\documentclass[a4paper,11pt]{article}
\pdfoutput=1 

\usepackage{jheppub} 

\usepackage[T1]{fontenc}

\usepackage{amsmath, amsthm, amssymb, mathrsfs, bm} 
\usepackage{graphics,bm}
\usepackage{epsfig}
\usepackage{graphicx}
\usepackage{amsmath}
\usepackage{wrapfig}
\usepackage{color}
\usepackage{tikz}
\RequirePackage{mathtools}
\RequirePackage{subfigure}
\usetikzlibrary{arrows,shapes}
\usetikzlibrary{trees}
\usetikzlibrary{matrix,arrows} 			
\usepackage[bottom]{footmisc}											
\usetikzlibrary{positioning}				
\usetikzlibrary{calc,through}				
\usepackage{pgffor}	
\usepackage[]{hyperref}

\usetikzlibrary{decorations.pathmorphing}	
\usetikzlibrary{decorations.markings}
\tikzset{
    sigmaCT/.style={draw=black, postaction={decorate},
        decoration={markings,mark=at position .99 with {\arrow[draw=black]{>}},mark=at 		 position .99 with {\arrow[
draw=black]{<}}}},
    pionCT/.style={dashed,draw=black, postaction={decorate},
        decoration={markings,mark=at position .99 with {\arrow[draw=black]{>}},mark=at position .99 with {\arrow[draw=black]{<}}}},    
    fermionCT/.style={draw=black, postaction={decorate},
        decoration={markings,mark=at position .5 with {\arrow[draw=black]{>}},mark=at position .99 with {\arrow[draw=black]{>}},mark=at position .99 with {\arrow[draw=black]{<}}}},    
    fermion/.style={draw=black, postaction={decorate},
        decoration={markings,mark=at position .55 with {\arrow[draw=black]{>}}}},
    fermionbar/.style={draw=black, postaction={decorate},
        decoration={markings,mark=at position .55 with {\arrow[draw=black]{<}}}},
    pion/.style={dashed,draw=black, postaction={decorate}},
    sigma/.style={draw=black, postaction={decorate}}
}

\newcommand{\beq}{\begin{equation}}
\newcommand{\eeq}{\end{equation}}
\newcommand{\bqa}{\begin{eqnarray}}
\newcommand{\eqa}{\end{eqnarray}}

\newcommand{\os}{\text{\tiny OS}}
\newcommand{\ms}{\overline{\text{\tiny MS}}}

\title{\boldmath On-shell versus curvature mass parameter fixing schemes in the three flavor quark-meson model with vacuum fluctuations}

\author[]{Vivek Kumar Tiwari}

\affiliation[]{Department of Physics, University of Allahabad, Prayagraj, India-211002}

\emailAdd{vivekkrt@gmail.com}

\abstract{ The vacuum effective potential and phase diagram for the three (2+1) flavor quark-meson model have been computed and compared in an extended mean-field approximation (e-MFA) where the model parameters are fixed by using different renormalization prescriptions after including quark one-loop vacuum fluctuations. When the vacuum one-loop divergence is regularized in the minimal subtraction scheme and the curvature masses of the scalar and pseudo-scalar mesons are used for fixing the parameters, the setting of the quark-meson model with the vacuum term (QMVT) turns out to be  inconsistent as one notes that the curvature masses are defined by the evaluation of self-energy at zero momentum. This work constitutes the first application of the consistent on-shell parameter fixing scheme to the three flavor quark-meson (QM) model. In this setting of the renormalized quark-meson (RQM) model, the physical (pole) masses of the $\pi, K, \eta$ and $\eta^{\prime}$ 
pseudo-scalar mesons and the scalar $\sigma$ meson,the pion decay constant and kaon decay constant are put into the relation of the running mass parameter and couplings by using the on-shell and the minimal subtraction renormalization schemes. The nonstrange direction normalized vacuum effective potential  plots for both the RQM model and QMVT model, are exactly identical for the $m_\sigma=$ 658.8 MeV while the nonstrange direction order parameter temperature variations and phase diagrams for both the models RQM and QMVT are identical when the $m_\sigma $ value is smaller by 10 MeV i.e. 
$m_\sigma=$ 648 MeV. This happens because the normalized vacuum effective potential variation in the nonstrange direction
is somewhat influenced by its variation in the strange direction. When the $m_\sigma < $ 658.8 MeV, the nonstrange direction  
normalized vacuum effective potential is deepest for the QMVT model. One observes an interesting trend reversal for the $m_\sigma > $ 658.8 MeV when the nonstrange direction vacuum effective potential of the RQM model becomes deepest. Similar $m_{\sigma}$ dependent differences and similarities are noticed in the nature of the RQM and QMVT model phase diagrams and the nonstrange direction order parameter temperature variations. The normalized vacuum effective potential plots in the
strange direction for both the RQM model and QMVT model, are nearly coincident for the $m_\sigma=$ 785 MeV. The effective potential variation in the strange direction for the $m_\sigma < $ 785 MeV is deepest in the QMVT model while for the $m_\sigma > $ 785 MeV, it becomes deepest for the RQM model.}

\begin{document} 
\maketitle
\flushbottom

\section{Introduction}
\label{secI}

 The study of quantum chromodynamics (QCD) phase diagram in all its details has been a very active research area of strong interaction physics since 1970s when the first QCD schematic phase diagram appeared \cite{Cabibbo75}. It depicted a confined phase of hadrons at a low temperature (low baryonic density) and a deconfined phase of quarks and gluons \cite{SveLer,SveLer1,Mull,Ortms,Riske} at a high temperature (zero baryonic density) or high baryonic density (zero temperature). One gets important and valuable information for the QCD phase transition from the lattice QCD simulations \cite{AliKhan:2001ek,Digal:01,Karsch:02,Fodor:03,Allton:05,Karsch:05,Aoki:06,Cheng:06,Cheng:08} at zero chemical potential but for the non zero baryon densities/chemical potentials, the lattice QCD calculations get seriously compromised as the QCD action becomes complex on account of the fermion sign problem \cite{Karsch:02}. For mapping out the phase diagram regions where the lattice simulations do not work, one gets much help from the investigations carried out in the ambit of phenomenological models developed using the effective degrees of freedom \cite{Alf,Fukhat}. 

In the zero quark mass limit, the QCD Lagrangian with the three flavor of quarks has the $ SU_{L+R}(3) \times SU_{L-R}(3) $ symmetry. The axial ($A=L-R$) part of the symmetry is called the chiral symmetry. It gets spontaneously broken in the low energy hadronic vacuum of the QCD. This leads to the formation of  chiral condensate and one gets eight massless pseudoscalar bosons  as Goldstone bosons. The chiral symmetry gets explicitly broken as well due to the small mass of the light quarks $u,d$ and a relatively heavy $s$ quark. In the nature, we find light pions while kaons and eta are heavier due to the large mass of strange quark. Furthermore 't Hooft~\cite{tHooft:76prl,tHooft:76prl1} showed that the $U_A(1)$ axial symmetry
is explicitly broken to $Z_{A}(N_f)$ at the quantum level by the instanton effects. Even in the chiral limit of zero quark masses, the $\eta'$ meson is not a massless Goldstone boson as it acquires a mass of about 1 GeV due to the $U_A(1)$ axial anomaly. The framework of the three flavor linear sigma model \cite{Rischke:00,Rischke:001,Schaefer:09} is very conducive for  investigating the $ SU_{A}(3)$ chiral as well as the $U_{A}(1)$ axial symmetry breaking and restoration. It enables the construction of chiral invariant combinations using the chiral partners from the respective octet as well as the singlet of the scalar and pseudoscalar mesons. When the nine scalar and nine pseudo-scalar mesons are coupled to  the three flavor of quarks, one gets the QCD-like framework of the quark-meson (QM) model for computing and exploring the QCD phase diagram.

Furthermore, the QCD confinement at low temperatures and densities is mimicked in a statistical sense by coupling the chiral models to a  constant background $SU(N_{c})$ gauge field $A_{\mu}^{a}$ \cite{Polyakov:78plb,fuku,benji, Pisarski:00prd,Vkt:06}. Thus the confinement of quarks inside the hadrons gets implemented by the introduction of the Polyakov loop. When the free energy density from the gluons in the form of the phenomenological Polyakov loop potential \cite{ratti,fuku2} is added to the QM model, one gets the PQM model. Different QCD phase structure/phase diagram studies, have already been done in the chiral models \cite{scav,Roder,fuku11,grahl,jakobi,Herpay:05,Herpay:06,Herpay:07,Kovacs:2006ym,kahara, kahara1, kahara2, Bowman:2008kc, Bowman:2008kc1,Fejos,Jakovac:2010uy,koch,marko}, two and three flavor QM model \cite{mocsy,bj,Schaefer:2006ds} and PQM model \cite{SchaPQM2F,SchaPQM3F,Mao,TiPQM3F}.  

The quark one-loop vacuum fluctuations and the associated renormalization issues are neglected altogether \cite{scav,mocsy,bj,Schaefer:2006ds,SchaPQM2F,kahara,kahara1,kahara2,Schaefer:09,SchaPQM3F,Mao,TiPQM3F} in the QM model with the standard mean field approximation (s-MFA) under the assumption that the redefinition of the meson potential parameters would account for their effects.~The QM model in the chiral limit under the s-MFA, gives a first-order chiral phase transition at zero baryon densities. This result is inconsistent because the general theoretical arguments \cite{rob,hjss} predict that the abovementioned chiral phase transition should be of second order. In order to remedy the inconsistency, proper treatment of the Dirac sea was first proposed in the Ref.~\cite{vac}. Afterwards, the  quark one-loop vacuum corrections were included in the two and three flavor QM/PQM models and its detailed impact on the phase diagram and phase structure was investigated in several research papers  \cite{lars,guptiw,schafwag12,chatmoh1,TranAnd,vkkr12,chatmoh2,vkkt13,Herbst,Weyrich,kovacs,zacchi1,zacchi2,Rai}. For the fixing of the model parameters, these publications have used the curvature masses of the mesons while the pion decay constant is identified with the vacuum expectation value of the sigma mean field. In the three flavor QM/PQM model  with fermionic vacuum correction \cite{schafwag12,chatmoh1,chatmoh2,vkkt13,Rai}, one has the kaon decay constant also which is given by the combination of the vacuum expectation values of the strange and nonstrange condensate. The quark one-loop vacuum divergence in the abovementioned works has been properly regularized by using the minimal subtraction scheme. The curvature mass is akin to defining the meson mass by the evaluation of self-energy at zero momentum because the effective potential is the generator of the n-point functions of the theory at vanishing external momenta  \cite{laine,Adhiand1,BubaCar,Naylor,fix1}. This consideration makes the above parameter fixing procedure inconsistent. For making comparisons and quantifying the effect of the parameter fixing with the curvature meson masses, this model setting has been named as the quark-meson model with vacuum term (QMVT).

The radiative corrections to the physical quantities in most of the renormalization procedures, change their tree level relations to the parameters of the Lagrangian. Thus the effective potential calculation becomes inconsistent if one uses the tree level values of the parameters. The on-shell parameters have their tree-level values while the running parameters in the $\overline{\text{MS}}$ scheme depend on the renormalization scale $\Lambda$. According to the  correct renormalization prescription, one needs to calculate the counterterms both in the $\overline{\text{MS}}$ scheme and in the on-shell scheme and then connect the renormalized parameters of the two schemes. Afterwards the effective potential is calculated using the modified minimal subtraction procedure where the relations between the on-shell parameters (physical quantities) and the running parameters are used as the input \cite{Adhiand1}. In a series of papers, Adhikari and collaborators \cite{Adhiand1,Adhiand2,Adhiand3,asmuAnd} have used this renormalization prescription for the proper accounting of the effect of Dirac sea in the context of two flavor QM model which uses the $O(4)$ sigma model with the iso-singlet scalar $\sigma$ and the iso-triplet pseudoscalar $\pi$ meson. 

In a very recent research work, we have also applied \cite{RaiTiw} the on-shell parameter fixing scheme to that version of the quark-meson (QM) model in which the two flavor of quarks are coupled to the  eight mesons of the $ SU_{L}(2) \times SU_{R}(2)$ linear sigma model and then made a comparative study of the effective potentials as well as the phase diagrams when the QM model setting with the on-shell parameters is contrasted with QM model parameter fixing using the curvature masses. In the present work, we will apply the on-shell parameter fixing scheme to the three (2+1) flavor quark-meson (QM) model in which the three flavor of quarks are coupled to the octect and singlet scalar as well as pseudoscalar mesons of the $ SU_{L}(3) \times SU_{R}(3) $ linear sigma model. This model setting has been termed as the renormalized quark-meson (RQM) model which has the advantage of providing us the framework in which apart from the $ SU_{A}(3)$ chiral, we can investigate the $ U_{A}(1)$ axial symmetry breaking and restoration also together with the interplay of axial $ U_{A}(1)$ and $ SU_{A}(3)$ chiral symmetry.

The paper is arranged as follows. The brief formulation of the $ SU_{L}(3) \times SU_{R}(3) $ QM model is presented in the section~\ref{secII}. The section~\ref{subsec:Vterm} presents the calculation of the effective potential of the quark-meson model with vacuum term (QMVT) together with its parameter fixing procedure which uses the curvature masses of the scalar and pseudo-scalar mesons. The on-shell scheme counterterms and self-energy calculations are presented in the section~\ref{subsecIVA}. The relations between the physical quantities and the running parameters, are derived in the section~\ref{subsecIVB}, the derivation of the effective potential in the RQM model is presented in the section~\ref{subsecIVC}. The result and discussion is presented in the section~\ref{secV}. Finally summary and conclusion is presented in the section~\ref{secVI}.

\section{Model Formulation}
\label{secII}
We are presenting the formulation of the $ SU_{L}(3) \times SU_{R}(3) $ quark-meson model in this section. Three flavor of quarks in this model are coupled to the $SU_V(3) \times SU_A(3)$ symmetric meson fields. The model Lagrangian is written in terms of quarks, mesons and couplings as
\bqa
{\cal L_{QM}}&=&\bar{\psi}[i\gamma^\mu \partial_\mu- g\; T_a\big( \sigma_a 
+ i\gamma_5 \pi_a\big) ] \psi+\cal{L(M)},
\label{lag}
\eqa
where $\psi$ is a color $N_c$-plet, a four-component Dirac spinor as well as a flavor triplet 
\bqa
\psi&=&
\left(
\begin{array}{c}
u\\
d\\
s
\end{array}\right)\;.
\eqa
The flavor blind Yukawa coupling $g$ couples the three flavor of quarks with the nine scalar ($\sigma_a, J^{P}=0^{+}$) and nine pseudoscalar ($\pi_a, J^{P}=0^{-}$) mesons.
The massless quarks become massive due to the  spontaneous breaking of the chiral symmetry as the chiral condensate assumes non-zero vacuum expectation value. The Lagrangian for the meson fields has the following form \cite{Schaefer:09,Roder,TiPQM3F} 
\bqa
\nonumber
\label{lag11}
\hspace{-1.5 cm}\cal{L(M)}&=&\text{Tr} (\partial_\mu {\cal{M}}^{\dagger}\partial^\mu {\cal{M}}-m^{2}({\cal{M}}^{\dagger}{\cal{M}}))\\
\nonumber
&&-\lambda_1\left[\text{Tr}({\cal{M}}^{\dagger}{\cal{M}})\right]^2-\lambda_2\text{Tr}({\cal{M}}^{\dagger}{\cal{M}})^2\\
&&+c[\text{det}{\cal{M}}+\text{det}{\cal{M}}^\dagger]+\text{Tr}\left[H({\cal{M}}+{\cal{M}}^\dagger)\right]\;,
\label{lag1}
\eqa
here the field ${\cal{M}}$ is a complex $3\times3$ matrix which contains the nine scalars $\sigma_a$ and the nine pseudoscalar $\pi_a$ mesons.
\bqa
{\cal{M}}&=&T_a\xi_a=T_a(\sigma_a+i\pi_a)\;.
\eqa
Here the $T_a$ represent 9 generators of $U(3)$ with $T_a = \frac{\lambda_a}{2}$ where $a=0,1 ,\dots ~,8$. The $\lambda_a$ are standard Gell-Mann matrices with $\lambda_0=\sqrt{\frac{2}{3}}\  {\mathbb I}_{3\times3}$. 
The generators follow the $U(3)$ algebra $\left[T_a, T_b\right]  = if_{abc}T_c$ and 
$\left\lbrace T_a, T_b\right\rbrace  = d_{abc}T_c$ where $f_{abc}$ and $d_{abc}$ are the standard 
antisymmetric and symmetric structure constants respectively with $f_{ab0}=0$ and 
$d_{ab0}=\sqrt{\frac{2}{3}}\ \delta_{ab}$ and matrices are normalized as 
$\text{Tr}(T_a T_b)=\frac{\delta_{ab}}{2}$. The following term breaks the  $SU_L(3) \times SU_R(3)$ chiral symmetry explicitly.
\bqa
H = T_a h_a\;.
\eqa
Here $H$ is a $3 \times 3$ matrix with nine external parameters. On account of the spontaneous breaking of the chiral symmetry, the filed $\xi$ picks up the nonzero vacuum expectation value, $\bar{\xi}$. Only three possible nonzero parameters $h_0$, $h_3$ and $h_8$ might cause the explicit breakdown of the chiral symmetry because $\bar{\xi}$ must have the quantum numbers of the vacuum. We are choosing $h_0$, $h_8  \neq 0$ and isospin symmetry breaking is neglected. Thus having the nonzero condensates $\bar{\sigma_0}$ and $\bar{\sigma_8}$, one gets the $2+1$ flavor symmetry breaking scenario.
The model has five other parameters in addition to the $h_0$ and $h_8$. These are the tree-level  mass parameter squared $m^2$, quartic coupling constants $\lambda_1$ and $\lambda_2$, a Yukawa coupling $g$ and a cubic coupling constant $c$ which models the $U_A(1)$ axial anomaly of the QCD vacuum.
\begin{table*}[!htbp]
    \caption{Meson masses calculated from the second derivative of the grand potential at its minimum as given in Ref.\ \cite{Schaefer:09,Herpay:06}} 
    \label{tab:table1}
    \begin{tabular}{p{0.08\textwidth} p{0.40\textwidth}|p{0.08\textwidth} p{0.40\textwidth}}
             \\ 
             \hline
            & Scalar meson masses \ \ \ \ \ \ \ \ \ \ \ &             & Pseudo-scalar meson masses  \\
      \hline 
      $(m_{a_{0}})^2$ & $m^2 +\lambda_1(x^2+y^2)+\frac{3\lambda_2}{2} x^2+
      \frac{\sqrt{2}c}{2}y $&$ (m_{\pi})^{2}$ & $m^2 + \lambda_1 (x^2 + y^2) +\frac{\lambda_2}{2} x^2 -\frac{\sqrt{2} c}{2} y$\\
      $(m_{\kappa})^{2}$ & $m^2+\lambda_1(x^2+y^2)+\frac{\lambda_{2}}{2}(x^2+\sqrt{2}xy+2y^2)+\frac{c}{2}x$&$(m_{K})^{2}$ &$m^2 + \lambda_1 (x^2 + y^2) +\frac{\lambda_2}{2} (x^2 - \sqrt{2} x y +2 y^2) - \frac{c}{2} x$\\
      $(m_{s,00})^2$ &  $m^2+\frac{\lambda_1}{3}(7x^2+4\sqrt{2}xy+5y^2)+\lambda_2(x^2 + y^2)-\frac{\sqrt{2}c}{3} (\sqrt{2} x +y)$&$(m_{p,00})^{2}$ & $m^2 + \lambda_1(x^2 +y^2) + \frac{\lambda_2}{3}(x^2 +y^2) + \frac{c}{3} (2x + \sqrt{2} y)$\\
      $(m_{s,88})^{2}$ & $m^2 +\frac{\lambda_1}{3}(5x^2-4\sqrt{2}xy +7y^2)+\lambda_2(\frac{x^2}{2} +2y^2)+\frac{\sqrt{2}c}{3} (\sqrt{2}x-\frac{y}{2})$&$(m_{p,88})^{2}$ & $m^2 +\lambda_1(x^2 +y^2) +\frac{\lambda_2}{6}(x^2 +4y^2)-\frac{c}{6}(4x -\sqrt{2}y)$\\
   $(m_{s,08})^{2}$ &  $\frac{2\lambda_1}{3}(\sqrt{2}x^2 -xy -\sqrt{2}y^2) +\sqrt{2}\lambda_2(\frac{x^2}{2}-y^2) +\frac{c}{3\sqrt{2}}(x- \sqrt{2}y)$&$(m_{p,08})^{2}$&$\frac{\sqrt{2}\lambda_2}{6}(x^2-2y^2)-\frac{c}{6}(\sqrt{2}x -2y)$\\
   $(m_{s,xx})^{2}$&$ m^2+3(\lambda_1+\frac{\lambda_2}{2})x^2+\lambda_1 y^2-\frac{c}{\sqrt{2}}y  $ &$(m_{p,xx})^{2}$&$  m^2+(\lambda_1+\frac{\lambda_2}{2})x^2+\lambda_1 y^2+\frac{c}{\sqrt{2}}y $ \\
   $(m_{s,yy})^{2}$&$ m^2+\lambda_1 x^2+3(\lambda_1+\lambda_2)y^2  $ &$(m_{p,yy})^{2}$&$m^2+\lambda_1 x^2+(\lambda_1+\lambda_2)y^2 $ \\
   $(m_{s,xy})^{2}$&$ 2\lambda_1 xy-\frac{c}{\sqrt{2}}x  $ &$(m_{p,xy})^{2}$&$\frac{c}{\sqrt{2}}x $ \\ &&& \\
   $m_{\sigma}^2$ & $\frac{1}{2}(m_{s,00}^2 + m_{s,88}^2)$ & $m_{\eta}^2$ & $\frac{1}{2}(m_{p,00}^2+m_{p,88}^2) $ \\
   &$-\frac{1}{2}\sqrt{(m_{s,00}^2-m_{s,88}^2)^2+4m_{s,00}^4}$&&$-\frac{1}{2} \sqrt{(m_{p,00}^2-m_{p,88}^2)^2+4m_{p,00}^4}$\\ &&&\\
    $m_{f_{0}}^2$ & $\frac{1}{2}(m_{s,00}^2+m_{s,88}^2)$
   &$m_{\eta^{\prime}}^2$ &$\frac{1}{2}(m_{p,00}^2+m_{p,88}^2) $ \\
   &$+\frac{1}{2} \sqrt{(m_{s,00}^2-m_{s,88}^2)^2+4m_{s,00}^4}$&&$+\frac{1}{2} \sqrt{(m_{p,00}^2-m_{p,88}^2)^2 +4m_{p,00}^4 }$\\
   \hline 
\end{tabular}
\end{table*}

\subsection{Grand Potential in the Mean Field Approach}
\label{subsec:potmf}
The considered system is spatially uniform and it is in thermal equilibrium at temperature $T$ and quark chemical potential $\mu_{f} (f=u,d,s)$. The partition function is obtained by the path integral over the quark/antiquark and meson fields \cite{Schaefer:09,TiPQM3F}
\bqa
\nonumber
\mathcal{Z}&=& \mathrm{Tr\, exp}[-\beta (\hat{\mathcal{H}}-\sum_{f=u,d,s} 
\mu_{f} \hat{\mathcal{N}}_{f})]  \\ \nonumber
&=& \int\prod_a \mathcal{D} \sigma_a \mathcal{D} \pi_a \int
\mathcal{D}\psi \mathcal{D} \bar{\psi} \; \mathrm{exp} \bigg[- \int_0^{\beta}d\tau\int_Vd^3x   \\ \label{eq:partf}
&& \bigg(\mathcal{L_{QM}^{E}} 
 + \sum_{f=u,d,s} \mu_{f} \bar{\psi}_{f} \gamma^0 \psi_{f} \bigg) \bigg]\;. 
\eqa
Where $\beta= \frac{1}{T}$ and the three dimensional volume of the system is $V$. In general, the three quark chemical potentials will be different for the three quark flavors. It is  assumed that the $SU_V(2)$ symmetry is preserved in this work. Hence the small difference in the mass of $u$ and $d$ quark is neglected. Thus the quark chemical potential for the $u$ and $d$ quarks is equal $\mu_u = \mu_d$ and the strange quark chemical potential is $\mu_s$.  
   
In the standard mean-field approximation \cite{scav,Schaefer:09,TiPQM3F}, the partition function is calculated by replacing the meson fields with their vacuum expectation values
$\langle M \rangle =  T_0 \bar{\sigma_0} + T_8 \bar{\sigma_8}$ and neglecting the thermal as well as quantum fluctuations of the meson fields while retaining the quarks and antiquarks as quantum fields. 
Using the standard method given in Refs.~\cite{fuku,SchaPQM2F,Kapusta_Gale}, one can find the expression of grand potential as the  sum of meson and quark/antiquark contribution, 
\bqa
\label{eq:grandp}
\Omega_{\rm MF}(T,\mu)=-\frac{T\ln Z}{V} &= &U(\sigma_0,\sigma_8)+ \Omega_{\bar{q}q} (T, \mu)\;.
\eqa 
The 2 + 1 flavor case is studied by performing the following basis transformation of condensates and external fields from the original singlet octet (0, 8) basis to the nonstrange strange basis ($x$, $y$)
\bqa
\sigma_x &=&x=
\sqrt{\frac{2}{3}}\bar{\sigma}_0 +\frac{1}{\sqrt{3}}\bar{\sigma}_8, \\
\sigma_y &=&y=
\frac{1}{\sqrt{3}}\bar{\sigma}_0-\sqrt{\frac{2}{3}}\bar{\sigma}_8.
\eqa
The grand potential is written in $x$, $y$ basis as,
\bqa
\Omega_{\rm MF }(T,\mu) =U(x,y)+\Omega_{q\bar{q}}(T,\mu)\;.
\label{Grandpxy}
\eqa
The external fields ($h_x$, $h_y$) are written in terms of the ($h_0$, $h_8$) by similar expressions. Since the nonstrange and strange quark/antiquark decouple, the quark masses are written as,
\bqa
\label{mums}
m_{u} = g \frac{x}{2}, \qquad m_{s} = g \frac{y}{\sqrt{2}}\;.
\eqa 
The tree level effective potential in the nonstrange-strange basis is written as,
\bqa
\label{eq:mesop}
\nonumber
 U(x,y) & = &\frac{m^{2}}{2}\left(x^{2} +
  y^{2}\right) -h_{x} x -h_{y} y
 - \frac{c}{2 \sqrt{2}} x^2 y \\ \nonumber 
 && + \frac{\lambda_{1}}{2} x^{2} y^{2}+
  \frac{1}{8}\left(2 \lambda_{1} +
    \lambda_{2}\right)x^{4} \\
 && +\frac{1}{8}\left(2 \lambda_{1} +
    2\lambda_{2}\right) y^{4}\ ,
\eqa    
The stationarity conditions $ \frac{\partial U(x,y)}{\partial x}|_{x=\overline x, y=\overline y}=0=\frac{\partial U(x,y)}{\partial y}|_{x=\overline x, y=\overline y}$  for the effective potential (\ref{eq:mesop}) give 
\bqa
\label{hxhy}
{\hskip -0.75 cm}h_{x}&=&\overline x \ m^2_{\pi} \ \text{and} \ h_{y}={\biggl\{\frac{\sqrt{2}}{2}(m_{K}^2-m_{\pi}^2) \overline  x+m_{K}^2 \overline y\biggr\}}\;. \quad
\eqa
The tree level curvature masses of the pions, kaons and other mesons in the QM model are given by the mass matrix $(m_{\alpha,ab})^2$ evaluated in Ref.~\cite{Rischke:00,Rischke:001, Schaefer:09}. Here $\alpha=$ s, p; ``s'' stands for the scalar and ``p'' stands for the pseudoscalar mesons and $a,b=0,1,2,\cdots,8$. In the scalar sector, the $a_{0}$ meson mass is given by the 11 element (degenerate with the 22 and 33 elements) and the $\kappa$ meson mass is given by the 44 element (degenerate with the 55, 66 and 77 elements). The $\sigma$ and $f_{0}$ meson masses are found by diagonalizing the (00)-(88) sector of the scalar mass matrix. In exactly analogous manner for the pseudoscalar sector $m^2_{\text{p},11}=m^2_{\text{p},22}=m^2_{\text{p},33}\equiv m^2_{\pi}$ and $m^2_{\text{p},44}=m^2_{\text{p},55}=m^2_{\text{p},66}=m^2_{\text{p},77} \equiv m^2_{K}$. Diagonalization of the pseudoscalar (00)-(88) sector of the  mass matrix gives us the masses of the physical $\eta$ and $\eta^{\prime} $ mesons. All the meson masses are given in the Table~\ref{tab:table1}.

The  quark/antiquark contribution is given by,
\bqa
\label{vac1}
\hspace{-0.75 cm}\Omega_{q\bar{q}} (T,\mu;x,y) &=& \Omega_{q\bar{q}}^{vac}+\Omega_{q\bar{q}}^{T,\mu}\;,
\\
\label{vac2}
\Omega_{q\bar{q}}^{vac} &=&- 2 N_c\sum_{u,d,s}  \int \frac{d^3 p}{(2\pi)^3} E_{f} \theta( \Lambda_c^2 - \vec{p}^{2})\;,
\\
\Omega_{q\bar{q}}^{T,\mu}&=&- 2 N_c\sum_{u,d,s} \int \frac{d^3 p}{(2\pi)^3} T \left[ \ln \left(1+e^{-E_{f}^{+}/T}\right)+\ln\left(1+e^{-E_{f}^{-}/T}\right)\right]\;.
\label{vac3}
\eqa
The fermion vacuum contribution is given by the first term of the Eq.~(\ref{vac1}), where $\Lambda_c$ is the ultraviolet cutoff. $E_{f}^{\pm} =E_f \mp \mu_{f} $ and $E_f=\sqrt{p^2 + m{_f}{^2}}$ is the flavor dependent single particle energy of the quark/antiquark, $m_{u}=m_{d}=\frac{gx}{2}$ is the mass of the light quarks $u$, $d$ and strange quark mass is $m_{s}=\frac{gy}{\sqrt{2}}$. For the present work, it is assumed that $\mu_{u}=\mu_{d}=\mu_{s}=\mu$.

In the standard mean-field approximation (s-MFA), the quark one-loop vacuum term of the Eq.~(\ref{vac1}) is neglected and the QM model grand potential is written as,
\bqa
\Omega_{QM}(T,\mu,x,y)&=&U(x,y)+\Omega_{q\bar{q}}^{T,\mu}\;.
 \label{Omega_MF}
 \eqa
The chiral order parameters $ x$ for the nonstrange and $ y$  for the strange sector are obtained by minimizing the thermodynamic potential the Eq.~(\ref{Omega_MF}) in the nonstrange and strange
directions
\begin{equation}
\frac{\partial \Omega_{QM}(T,\mu,x,y)}{\partial
      x}|_{x,y}= \frac{\partial \Omega_{QM}(T,\mu,x,y)}{\partial
      y}|_{x,y}=0.
\label{EoMMF1}
\end {equation}

\subsection{Parameter fixing}
\label{subsec:paramfix}
The six model parameters $m^2$, $\lambda_1 $, $\lambda_2$, $c$, $h_x$ and $h_y$ are obtained using six experimentally known quantities in the vacuum. The pion, kaon mass, the average squared mass of the $\eta$ and $\eta^{\prime}$ mesons ($m_{\eta}^2+m_{\eta^{\prime}}^2$) from the pseudo-scalar side and the mass $m_{\sigma}$ of the scalar meson $\sigma$ together with the pion and kaon decay constants $f_{\pi}$ and $f_K$ are used as the input \cite{Rischke:00,Rischke:001,Schaefer:09} for determining the six model parameters.

    In accordance of the partially conserved axial-vector
current relation (PCAC), the vacuum condensates values are $\overline x=f_{\pi}$ and  $\overline y=(2f_{K}-f_{\pi})/\sqrt{2}$. The minimum of the effective potential in the Eq.~(\ref{EoMMF1}) for $T=0,\mu=0$ is located at the above values. The parameters $\lambda_2$ and $c$ in the vacuum are obtained as 
\bqa 
\label{para1}
\lambda_2&=&\dfrac{2  \ }{(x^2+4y^2)(\sqrt{2} \ y -x)}\left[(3\sqrt{2} \  y) m_{K}^2- (\sqrt{2} y+2x)m_{\pi}^2-(\sqrt{2} y-x)(m_{\eta}^2+m_{\eta^{'}}^2)\right]\;, \hspace{0.8 cm }\\
c&=&\dfrac{2(m^2_K-m^2_\pi)}{(\sqrt{2} \ y-x)}-\sqrt{2} \ y \ \lambda_2 \;.
\label{para2}
\eqa
The difference of the $\sigma$ and $\pi$ mass squares $(m_{\sigma}^2 - m_{\pi}^2)$ does not have any mass parameter $m^2$ dependence. It depends on the parameters $\lambda_1$, $\lambda_2$ and $c$. When the $\lambda_2$ and $c$ as obtained from the above two equations are put into the expression of  $(m_{\sigma}^2 - m_{\pi}^2)$ and $x=\overline x=f_{\pi}$ and  $y=\overline y=(2f_{K}-f_{\pi})/\sqrt{2}$, one gets the vacuum value of the parameter $\lambda_1$. Using the expression of $m_{\pi}^2$, the mass parameter $m^2$ can be written as
\bqa
\label{para3}
m^2&=&m^2_\pi-\lambda_1 (x^2 + y^2)-\frac{\lambda_2}{2} x^2 +\frac{c}{\sqrt{2} } y \;.
\eqa
Putting the vacuum values of $m_{\pi}^2$, $\lambda_1$, $\lambda_2$, $c$, $x$ and $y$ in the Eq. (\ref{para3}), one gets the value of the mass parameter $m^2$. Putting the $\overline x$ and $\overline y$ values in the Eq.~(\ref{hxhy}), one gets 
\bqa
\label{para4}
h_{x}&=&f_{\pi} m^2_{\pi}  \ \text{and} \ h_{y}=\sqrt{2} f_{K} \ m_{K}^2-\frac{1}{\sqrt{2}}f_{\pi} \ m_{\pi}^2\;.
\eqa
Finally, the Yukawa coupling is fixed from the nonstrange constituent quark mass $g=\frac{2m_{u}}{f_{\pi}}$. For the $f_{\pi}=92.4$ MeV and $m_{u} \sim 300.3$ MeV, the $g \sim 6.5$ and strange quark mass is predicted to be $m_{s} \sim 334.34$ MeV. The experimental value of $m_{\eta}=547.5$ MeV and $m_{\eta^{\prime}}=957.78$ MeV. In Ref.~\cite{Schaefer:09}, the parameter $\lambda_{2}$ is determined by taking the $m_{\eta}=539$ MeV and $m_{\eta^{\prime}}=963$ MeV as input because 
the sum of the squared masses $m_{\eta}^2+m_{\eta^{\prime}}^2=(539)^2+(963)^2$ is almost equal to the $(547.5)^2+(957.78)^2$ and the calculated parameters reproduce $m_{\eta}=539$ MeV and $m_{\eta^{\prime}}=963$ MeV in the output.    
\section{QM Model with Vacuum Term }
\label{subsec:Vterm}
This section contains a brief description of the effective potential calculation when 
the scalar and pseudo-scalar mesons curvature masses are used for the 
parameter fixing and the vacuum  value of the nonstrange condensate is put equal to 
the pion decay constant while the strange condensate vacuum value is a combination of the 
pion and kaon decay constant. The quark one-loop vacuum divergence given by the first term 
of the Eq.~(\ref{vac1}) is regularized under the minimal subtraction scheme using the 
dimensional regularization as done for the two flavor case in Ref.~\cite{vac,guptiw,vkkr12} and 
the three flavor case in Ref.~\cite{schafwag12,chatmoh1,chatmoh2,vkkt13}. The quark one-loop 
vacuum term is written as 
\bqa
\label{vt_one_loop}
\Omega_{q\bar{q}}^{\rm vac} &=&-2 N_c \sum_{f=u,d,s} \int \frac{d^3
  p}{(2\pi)^3} E_f \;.
\eqa
The dimensional regularization of the Eq.~(\ref{vt_one_loop}) near three 
dimensions, $d=3-2\epsilon$ yields the $\epsilon$ zeroth order potential as
\begin{equation}
{\hskip -0.1 cm}\Omega_{q\bar{q}}^{\rm vac} =\sum_{f=u,d,s}\frac{N_c\ m_f^4}{16 \pi^2}\left[
  \frac{1}{\epsilon} -\frac{ 
\{-3+2\gamma_E +4\ln(\frac{m_f}{2\sqrt{\pi} \Lambda})\}}{2} \right],
\label{Omega_DR}
\end{equation}
Here $\Lambda$ is the arbitrary renormalization scale.
When the following counter term  $\delta \mathcal{L}$ is added to the QM model Lagrangian, 
\begin{equation}
\delta \mathcal{L} =\sum_{f=u,d,s} \frac{N_c}{16 \pi^2} m_f^4 \left[ \frac{1}{\epsilon} - \frac{1}{2}
 \left\{ -3 + 2 \gamma_E - 4 \ln (2\sqrt{\pi})\right\} \right], 
\label{counter}
\end{equation}
one gets the renormalized fermion vacuum loop contribution as: 
\begin{equation}
\Omega_{q\bar{q}}^{\rm vac} =  -\sum_{f=u,d,s} \frac{N_c}{8 \pi^2} m_f^4  \ln\left(\frac{m_f}{\Lambda}\right)\;,
\label{Omega_reg} 
\end{equation}
The vacuum grand potential becomes the renormalization scale $\Lambda$ dependent when the quark
one-loop contribution in the first term of the Eq.~(\ref{vac1}) is replaced by the Eq.~(\ref{Omega_reg}) and one writes :
\bqa
\Omega^{\Lambda} (x,y) =U(x,y)+\Omega_{q\bar{q}}^{\rm vac}
\label{Omeg_rel}
\eqa
Here the six unknown parameters $m^2$, $\lambda_1, \lambda_2$, $h_x, h_y$ and $c$ of the meson 
potential U($x,y$), are obtained from the $x$ and $y$ dependent curvature masses of the 
mesons. The procedural details  of finding the different parameters are presented in the Appendix (\ref{appenA}). When the parameter $\lambda_2$ is determined, the logarithmic $\Lambda$ dependence in the term 
$\Omega_{q\bar{q}}^{\rm vac}$ generates a renormalization scale $\Lambda$ dependent part 
$\lambda_{2\Lambda}$ and one gets $\lambda_2=\lambda_{2s}+n+\lambda_{2+}+\lambda_{2\Lambda}$. 
$\lambda_{2s}$ is the same old $\lambda_2$ parameter of the QM/PQM model in the 
Ref.~\cite{Rischke:00,Rischke:001,Schaefer:09,TiPQM3F}. Here, $n=\frac{N_cg^4}{32\pi^2}$, $\lambda_{2+}=\frac{n{f_{\pi}}^2}{f_{K} \left(f_{K}-f_{\pi}\right)}\log\{\frac{2 f_K-f_{\pi}}{f_{\pi}}\}$ and $\lambda_{2\Lambda}= 4n\log\{ \frac{g\left( 2f_K-f_{\pi}\right)}{2 \Lambda}\}$. When this value of the $\lambda_2$ is substituted in the expression of U($x,y$) and all the terms of the summation in $\Omega_{q\bar{q}}^{\rm vac}$  expression are written explicitly, the Eq. (\ref{Omeg_rel}) takes the form : 
\bqa
\label{eq:mesVop}
\nonumber
{\hskip -1.0 cm}\Omega^{\Lambda} (x,y)&=&\frac{m^{2}}{2} \left( x^{2} +
y^{2}\right) -h_{x} x-h_{y} y
-\frac{c}{2 \sqrt{2}} x^2 y+\frac{\lambda_{1}}{4}\left(x^{4} +y^{4}+2 x^{2} y^{2}\right) \\ 
&&+\frac{\left(\lambda_{2\text{v}}+n+\lambda_{2\Lambda} \right)}{8} \left(x^{4} + 2 y^{4} \right)-\frac{n x^4}{2}\log\left(\frac{g \ x}{2\Lambda}\right)-\ n y^4\log\left(\frac{g \ y}{\sqrt{2}\Lambda}\right)\;.   
\eqa 

The $\lambda_{2\text{v}}=\lambda_{2s}+\lambda_{2+}$. When the terms are rearranged, one finds that 
the scale dependence of all the terms in $\Omega_{q\bar{q}}^{\rm vac}$ gets completely cancelled by
the logarithmic $\Lambda$ dependence of the $\lambda_{2}$ contained in $\lambda_{2\Lambda}$. The scale independent vacuum effective potential expression is
\bqa
\nonumber
{\hskip -1.0 cm} \Omega(x,y) &=& \frac{m^{2}}{2}\left(x^{2} +
  y^{2}\right) -h_{x} x -h_{y} y
 - \frac{c}{2 \sqrt{2}} x^2 y +\frac{\lambda_{1}}{4} \left( x^{4} + y^{4} +2 x^{2} y^{2} \right)  \\
 &&+ \frac{ \left( \lambda_{2\text{v}}+n \right) }{8}\left( x^{4} + 2 y^{4} \right)-\frac{n x^4}{2} \log \left( \frac{x}{\left(2f_K-f_{\pi} \right) } \right)- n y^4 \log \left( \frac{\sqrt 2\ y}{\left(2f_K-f_{\pi}\right)}\right)\;.
\label{eq:mesVop}
\eqa
One notes that the parameters  $m^2$, $\lambda_1$ and $\lambda_2$ are modified by the fermionic vacuum correction in this parameter fixing scheme while the parameters $h_x$, $h_y $ and $c$ are not affected.

The thermodynamic grand potential with the renormalized fermionic vacuum correction in the quark meson model with vacuum term (QMVT) is written as 
\bqa
\Omega_{\rm QMVT}(T,\mu;x,y) & =&  \Omega(x,y) + \Omega_{q\bar{q}}^{\rm T}(T,\mu;x,y)\;.
\label{OmegaMFPQMVT}
\eqa
The nonstrange and strange quark condensates $x$ and $y$ are found by searching the global minimum of the grand potential for a given temperature $T$ and chemical potential $\mu$.
\bqa 
\label{eq:gapeq}
\nonumber
\frac{ \partial \Omega_{\rm QMVT} (T,\mu;x,y)}{\partial x}|_{x, y}  &=& \frac{ \partial \Omega_{\rm QMVT} (T,\mu;x,y)}{\partial y}|_{x,y} = 0 \\ .
\eqa 
 
Here it is relevant to remind that the dressing of the meson propagator is not considered in the curvature mass scheme of parameter fixing.
Hence the pion and kaon decay constants $f_\pi $ and $f_{K} $ do not get renormalized. The quark one-loop vacuum correction to the effective potential modifies the parameters in such a way that the stationarity conditions in the nonstrange and strange directions for the $T=0$ give the same result for $h_x$ and $h_y$ as in the QM model. The modified curvature masses of the pion and kaon as presented in the Appendix (B) of Ref.~\cite{vkkt13} remain the same as their pole masses. The minimum of the vacuum effective potential remains at $ \overline{x}=f_\pi $ and $ \overline{y}=\frac{2f_K-f_\pi)}{\sqrt{2}} $.
\section{Renormalized Quark Meson Model}
\label{secIV}
Model parameters in several of the  recent research works were fixed by taking the  $\pi, K, \eta, \eta^{\prime}$ and $\sigma$ meson masses equal to the their curvature (or screening) masses \cite {lars,guptiw, schafwag12,chatmoh1,TranAnd,vkkr12,chatmoh2,vkkt13,Herbst,Weyrich,kovacs,zacchi1,zacchi2,Rai} while the nonstrange condensate is put equal to the pion decay constant and the strange condensate is related to the pion and kaon decay constant. However, we know that the poles of the meson propagators give their physical masses and the residue of the pion propagator at its pole is related to the pion decay constant \cite{BubaCar,Naylor,fix1}. Furthermore, the curvature masses are akin to defining the meson masses by evaluating their self-energies at zero momentum \cite{laine,Adhiand1,Adhiand2,Adhiand3} as it is known that the effective potential is the generator of the n-point functions of the theory at zero external momenta. It is also to be noted that the pole definition is the physical and gauge invariant one \cite{Kobes,Kobes1,Rebhan}. In the absence of the Dirac sea contributions, the  pole mass prescription is equivalent to the curvature mass prescription for the parameter fixing of the model but when the quark one-loop vacuum correction is taken into account, the pole masses of the mesons start to differ from their screening masses \cite{BubaCar,fix1}. The above arguments necessitate the  use of the exact on-shell parameter fixing method for the renormalized quark-meson (RQM) model where the physical (pole) masses of the mesons, the pion and kaon decay constants are put into the relation of the running mass parameter and couplings by using the on-shell and the minimal subtraction renormalization prescriptions \cite{Adhiand2,asmuAnd,RaiTiw}.

\subsection{Self-energies and counterterms}
\label{subsecIVA}
When the quark one-loop vacuum corrections are included, the tree level parameters of the Eqs.~(\ref{para1})--(\ref{para4}) become inconsistent unless one uses the on-shell renormalization 
scheme. The divergent loop integrals in the on-shell scheme are also regularized by the dimensional regularization but the counterterm choices  are different from the minimal subtraction scheme. The suitable choice of counterterms in the on-shell scheme leads to the exact cancellation of the loop corrections to the self-energies. Since the couplings are evaluated on-shell, the renormalized parameters become renormalization scale independent. The parameters and wave functions/fields of the Eq.~(\ref{lag}) are bare quantities. The counterterms $\delta m^{2} $, $\delta \lambda_{1} $,  $\delta \lambda_{2} $, $\delta c$, $\delta h_{x}$, $\delta h_{y}$ and $\delta g^{2} $, for the parameters and the counterterms $\delta Z_\pi $, $\delta Z_{K} $, $\delta Z_\sigma $, $\delta Z_\eta $, $\delta Z_\eta^{\prime} $, 
$\delta Z_\psi $ $\delta Z_{x}$ and $\delta Z_{y} $ for the wave functions/fields are introduced in the Lagrangian (\ref{lag}) where the couplings and renormalized fields are defined as,
\bqa
\label{ctrm1}
\pi^i_b&=&\sqrt{Z_\pi} \ \pi^i, \ K_b=\sqrt{Z_K} \ K , \ \eta_b=\sqrt{Z_\eta} \ \eta, \\ 
 \eta^{\prime}_b&=&\sqrt{Z_\eta^{\prime}} \ \eta^{\prime}, \ \sigma_b=\sqrt{Z_\sigma} \ \sigma,\ m^2_b=Z_m \ m^2\\
\psi_b&=&\sqrt{Z_\psi} \ \psi, \
\lambda_{1b}=Z_{\lambda_{1}} \ \lambda_{1}, \ \lambda_{2b}=Z_{\lambda_{2}} \ \lambda_{2}, \\ \
g_b&=&\sqrt{Z_g} \ g, \
 h_{xb}=Z_{h_{x}} \ h_{x},\ h_{yb}=Z_{h_{y}} \ h_{y},  \\ \label{ctrm5} c_b&=&Z_c \ c,\ x_b=\sqrt{Z_{x}} \ x, \ y_b=\sqrt{Z_{y}} \ y 
\;.
\eqa
Here the $Z_ {(\pi,K,\eta,\eta^{\prime},\sigma,\psi,x,y)}=1+\delta Z_{(\pi,K,\eta,\eta^{\prime},\psi,x,y)} $, denote  the field strength renormalization constants while $Z_ {(m,\lambda_{1},\lambda_{2},g,h_{x},h_{y},c )}=1+\delta Z_{((m,\lambda_{1},\lambda_{2},g,h_{x},h_{y},c ) )} $ denote the mass and coupling renormalization constants. One loop correction to the quark fields and the quark masses is zero because in the large $N_{c}$ limit, the $\pi$ and $\sigma$ loops that may renormalize the quark propagators are of the order $N_{c}^0$. Hence the $Z_\psi=1$ and the respective quark self energy corrections for the nonstrange quarks and the strange quarks are $\delta m_{u}=0$ and  $\delta m_{s}=0$. Also, the one-loop correction at the pion-quark $\pi \overline{\psi} \psi$ vertex is of order $N_{c}^0$, hence get neglected. In consequence, we get $Z_{\psi}  \ \sqrt{Z_{g} \ g^2}\sqrt{Z_{\pi}} \approx g(1+\frac{1}{2}\frac{\delta g^2}{g^2}+\frac{1}{2}  \delta Z_\pi )=g$. Thus $\frac{\delta g^2}{g^2} \ + \delta Z_\pi=0$. Furthermore the $\delta m_{u}=0$ and $\delta m_{s}=0$ implies that $\delta g \ x/2 + g \ \delta x/2 =0 $ and  $\delta g \ y/\sqrt{2} + g \ \delta y/\sqrt{2} =$ 0. This gives $\delta x/x= \delta y/y=-\delta g/g$ which is written as
\bqa
\label{Zpi}
\frac{\delta  x^2}{ x^2}&=&\frac{\delta  y^2}{ y^2}=-\frac{\delta g^2}{g^2}=\delta Z_\pi\;.
\eqa
Following the Refs.~\cite{Adhiand1,Adhiand2,Adhiand3,asmuAnd,RaiTiw} and using the Eqs.~(\ref{ctrm1})-(\ref{ctrm5}) together with the Eqs.~(\ref{para1}) and (\ref{para2}), the counterterm $\delta \lambda_{2} $ can be expressed in terms of the counterterms  $\delta m^{2}_{\pi} $, $\delta m^{2}_{K}$, $\delta m^{2}_{\eta}$ , $\delta m^{2}_{\eta^{\prime}}$ and $\delta Z_{\pi}$ while the $\delta c $ is expressed in terms of the $\delta m^{2}_{\pi} $, $\delta m^{2}_{K}$, $\delta Z_{\pi}$ and the preceding $\delta \lambda_{2}$. The resulting expressions of the $\delta \lambda_{2}$ and $\delta c$ are the following.

\bqa
\nonumber
\label{delta:lambda_2}
\delta\lambda_{2}&=&\frac{2}{(x^2+4y^2)(\sqrt{2} \ y -x)}\biggl[(3\sqrt{2}  y)\delta m_{K}^2 - (\sqrt{2} y+2x)     \; \\ 
&&\delta m_{\pi}^2-(\sqrt{2} y-x)(\delta m_{\eta}^2+\delta m_{\eta^{'}}^2) \biggr]-\lambda_{2} \delta Z_{\pi} \;,
\\
\delta c&=&\dfrac{2(\delta m^2_K-\delta  m^2_\pi)}{(\sqrt{2} \ y-x)}-\sqrt{2} \  y \ \delta \lambda_{2}-(2\sqrt{2} \ y \ \lambda_{2}+c) \ \frac{\delta Z_{\pi}}{2} \;. 
\label{deltac}
\eqa 
Once the $\delta \lambda_{2}$ and $\delta c$ are written, using the expression of ($\delta m_{\sigma}^2-\delta m_{\pi}^2$) and doing some algebraic manipulations, one can write the counter term  $\delta \lambda_{1}$ as follows : 
\begin{align}
\label{lam1}
\delta\lambda_{1}=\frac{\delta \lambda_{1\text{{\tiny NUMI}}}}{\lambda_{1\text{{\tiny DENOM}}}}-\lambda_{1} \ \delta  Z_{\pi}.
\end{align}

\begin{align}
\label{lam1de}
\nonumber
\lambda_{1\text{{\tiny DENOM}}}= \biggl(\sqrt{(m^2_{s,00}-m^2_{s,88})^2+4m^4_{s,08}}\biggr) (x^2+y^2)-
\frac{(m^2_{s,00}-m^2_{s,88})}{3}(x^2+4\sqrt{2}xy-y^2)\\
 - \frac{4m^2_{s,08}}{3} (\sqrt{2} x^2-xy-\sqrt{2} y^2)
 \end{align}
\begin{align} 
\nonumber
&\delta\lambda_{1\text{{\tiny NUMI}}}=\sqrt{(m^2_{s,00}-m^2_{s,88})^2+4m^4_{s,08}}\biggl(\delta m_{\sigma}^2-\delta m_{\pi}^2\biggr)-\biggl\{\delta\lambda_{2}\frac{(x^2+6y^2)}{4} + \delta c \frac{\sqrt{2}\ y}{4}\biggr\} \\ \nonumber
&\sqrt{(m^2_{s,00}-m^2_{s,88})^2+4m^4_{s,08}} + \biggl\{\delta\lambda_{2}\frac{(x^2-2y^2)}{4}
-\delta c \frac{\sqrt{2}\ (4\sqrt{2}x+y)}{12}\biggr\}(m^2_{s,00}-m^2_{s,88}) \\ \nonumber
&+\biggl\{\delta\lambda_{2}\sqrt{2}(x^2-2y^2)+\delta c \frac{\sqrt{2}\ (x-\sqrt{2}y)}{3}\biggr\}m^2_{s,08}\; \\ \nonumber
&-\lambda_{2} \ \delta  Z_{\pi} \Biggl\{\frac{(x^2+6y^2)}{4}\sqrt{(m^2_{s,00}-m^2_{s,88})^2+4m^4_{s,08}}-\frac{(x^2-2y^2)}{4}(m^2_{s,00}-m^2_{s,88})-\sqrt{2}(x^2-2y^2)m^2_{s,08}\Biggr\}\; \\
\label{lam2}
&-c\frac{\delta  Z_{\pi}}{2}\Biggl\{\frac{\sqrt{2} y}{4}\sqrt{(m^2_{s,00}-m^2_{s,88})^2+4m^4_{s,08}} +\frac{\sqrt{2}\ (4\sqrt{2}x+y)}{12}(m^2_{s,00}-m^2_{s,88})+ \frac{\sqrt{2}\ (\sqrt{2}y-x)}{3}m^2_{s,08}\Biggr\}\;
\\ \nonumber 
&\text{Finally the  counterterm $\delta m^2$ is written in terms of  the $\delta m_{\pi}^2$, $\delta \lambda_{1}$, $\delta \lambda_{2}$, $\delta c$ and $\delta Z_{\pi} $} \\  
& \delta m^2=\delta  m^2_\pi-\delta \lambda_{1} \ (x^2+y^2)- \ \frac{(\delta \lambda_{2}) \ x^2}{2}+\frac{\delta c \ y}{\sqrt{2}} -\delta  Z_{\pi} \biggl\{\lambda_{1} \ (x^2+y^2)+ \ \frac{\lambda_{2} \ x^2}{2}-
 \frac{ c \ y}{2\sqrt{2}} \biggr\}\;.
\end{align}

\begin{figure*}[htb]
\subfigure[\ One-loop self energy and tadpole diagrams.]{
\label{sclarse} 
\begin{minipage}[b]{0.48\textwidth}
\centering \includegraphics[width=\linewidth]{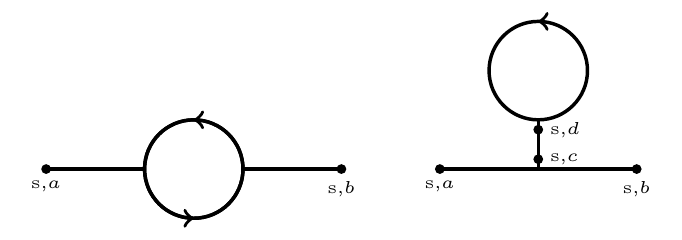}
\end{minipage}}
\hfill
\subfigure[\ One-loop self energy and tadpole counterterm diagrams.]{
\label{counts1b} 
\begin{minipage}[b]{0.48\textwidth}
\centering \includegraphics[width=\linewidth]{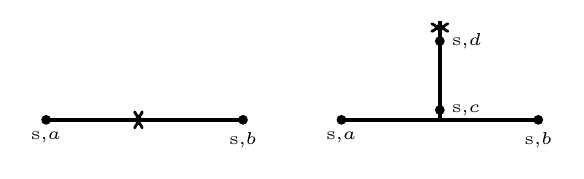}
\end{minipage}}
\caption{The solid line represents scalar particles  and an arrow on the solid line
denotes a quark.}
\label{fig:mini:fig1} 
\end{figure*}

\begin{figure*}[htb]
\subfigure[\ One-loop self energy and tadpole diagrams.]{
\label{psclarse} 
\begin{minipage}[b]{0.48\textwidth}
\centering \includegraphics[width=\linewidth]{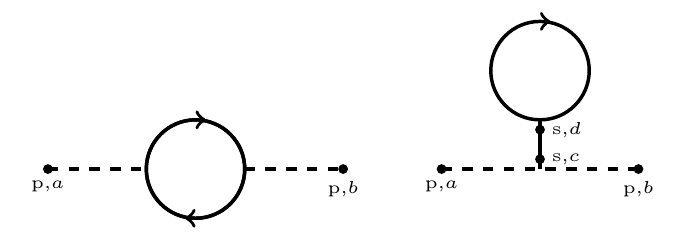}
\end{minipage}}
\hfill
\subfigure[\ One-loop self energy and tadpole counterterm diagrams.]{
\label{countp2b} 
\begin{minipage}[b]{0.48\textwidth}
\centering \includegraphics[width=\linewidth]{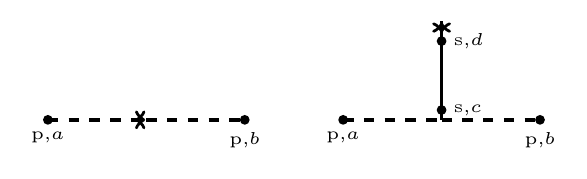}
\end{minipage}}
\caption{The dash line represents pseudo-scalar particles  and an arrow on the solid line
denotes a quark.}
\label{fig:mini:fig2} 
\end{figure*}

 The Fig.~(\ref{sclarse}) depicts the Feynman diagrams of the self energy and tadpole contributions for the scalar particles while the Fig.~(\ref{counts1b}) depicts the corresponding counter term diagrams. The Feynman diagrams of the self energy and tadpol contributions for the pseudo-scalar particles are given in Fig.~(\ref{psclarse}) and the corresponding diagrams for the counter terms are presented in the Fig.~(\ref{countp2b}). The self energies of the scalar sigma $\sigma$, pseudo-scalar eta~($\eta$), eta-prime~($\eta^{\prime}$), pion~($\pi$) and kaon~($K$) are required for the on-shell parameter fixing. The scalar $\sigma$ self energy correction is obtained in terms of the self energy corrections $\Sigma_{\text{s},00}(p^2)$, $\Sigma_{\text{s},88}(p^2)$ and $\Sigma_{\text{s},08}(p^2)$ while pseudo-scalar $\eta$ and $\eta^{\prime}$ self energy corrections are obtained in terms of self energy corrections $\Sigma_{\text{p},00}(p^2)$, $\Sigma_{\text{p},88}(p^2)$ and $\Sigma_{\text{p},08}(p^2)$. The expressions of scalar and pseudo-scalar self energies are written below:
 
\begin{align}
&\Sigma_{\text{s},00}(p^2)=-\frac{2}{3}N_cg^2\left[2\mathcal{A}(m^2_u)-(p^2-4m^2_u)\mathcal{B}(p^2,m_{u})\right]-\frac{1}{3}N_cg^2\left[2\mathcal{A}(m^2_s)-(p^2-4m^2_s)\mathcal{B}(p^2,m_{s})\right]+\Sigma^{tad}_{\text{s},00} \quad \;,\\
&\Sigma_{\text{s},11}(p^2)=-N_cg^2\left[2\mathcal{A}(m^2_u)-(p^2-4m^2_u)\mathcal{B}(p^2,m_{u})\right]+\Sigma^{tad}_{\text{s},11}\;,\\
&\Sigma_{\text{s},44}(p^2)=-N_cg^2\left[\mathcal{A}(m^2_u)+\mathcal{A}(m^2_s)-(p^2-(m_u+m_s)^2)\mathcal{B}(p^2,m_u,m_s)\right]+\Sigma^{tad}_{\text{ s},44}\;,\\ 
&\Sigma_{\text{s},88}(p^2)=-\frac{1}{3}N_cg^2\left[2\mathcal{A}(m^2_u)-(p^2-4m^2_u)\mathcal{B}(p^2,m_{u})\right]-\frac{2}{3}N_cg^2\left[2\mathcal{A}(m^2_s)-(p^2-4m^2_s)\mathcal{B}(p^2,m_{s})\right]+\Sigma^{tad}_{\text{s},88}\;,\\ \nonumber
&\Sigma_{\text{s},08}(p^2)=-\frac{\sqrt{2}}{3}N_cg^2\left[2\mathcal{A}(m^2_u)-(p^2-4m^2_u)\mathcal{B}(p^2,m_{u})\right]+\frac{\sqrt{2}}{3}N_cg^2\left[2\mathcal{A}(m^2_s)-(p^2-4m^2_s)\mathcal{B}(p^2,m_{s})\right]\\ 
&+\Sigma^{tad}_{\text{s},08},
\end{align}


\begin{align}
&\Sigma_{\text{p},00}(p^2)&=&-\frac{2}{3}N_cg^2\left[2\mathcal{A}(m^2_u)-p^2\mathcal{B}(p^2,m_{u})\right]-\frac{1}{3}N_cg^2\left[2\mathcal{A}(m^2_s)-p^2\mathcal{B}(p^2,m_{s})\right]+\Sigma^{tad}_{\text{p},00}\;,\\
\label{selfenpi}
&\Sigma_{\text{p},11}(p^2)&=&\Sigma_{\pi}(p^2)=-N_cg^2\left[2\mathcal{A}(m^2_u)-p^2\mathcal{B}(p^2,m_{u})\right]+\Sigma^{tad}_{\text{ p},11}\;,\\
\label{selfenK}
&\Sigma_{\text{ p},44}(p^2)&=&\Sigma_{K}(p^2)=-N_cg^2\left[\mathcal{A}(m^2_u)+\mathcal{A}(m^2_s)-\left\{p^2-(m_u-m_s)^2\right\} \mathcal{B}(p^2,m_u,m_s)\right]+\Sigma^{tad}_{\text{p},44}\;, \\
&\Sigma_{\text{p},88}(p^2)&=&-\frac{1}{3}N_cg^2\left[2\mathcal{A}(m^2_u)-p^2\mathcal{B}(p^2,m_{u})\right]-\frac{2}{3}N_cg^2\left[2\mathcal{A}(m^2_s)-p^2\mathcal{B}(p^2,m_{s})\right]+\Sigma^{tad}_{\text{ p},88}\;,\\
&\Sigma_{\text{ p},08}(p^2)&=&-\frac{\sqrt{2}}{3}N_cg^2\left[2\mathcal{A}(m^2_u)-p^2\mathcal{B}(p^2,m_{u})\right]+\frac{\sqrt{2}}{3}N_cg^2\left[2\mathcal{A}(m^2_s)-p^2\mathcal{B}(p^2,m_{s})\right]+\Sigma^{tad}_{\text{ p},08}.
\end{align}

\begin{figure}[htb]
\begin{center}
\includegraphics[width=0.5\textwidth]{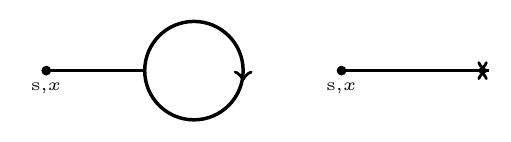}
\end{center}
\caption{One point diagram for the nonstrange scalar and its counterterm.}
\label{count33}
\end{figure}

\begin{figure}[htb]
\begin{center}
\includegraphics[width=0.5\textwidth]{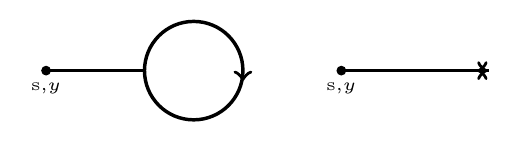}
\end{center}
\caption{One point diagram for the strange scalar and its counterterm.}
\label{count44}
\end{figure}
The one-point function diagram for the quark one-loop correction to the 
nonstrange component of the scalar $\sigma$ and its counterterm
is shown in the Fig.~(\ref{count33}). It is written as, 
\bqa
\delta\Gamma_x^{(1)}=-4 \ N_c \ g \ m_u \ \mathcal{A}(m_u^2)+i\delta t_{x}\;.
\eqa
The Fig.~(\ref{count44}) presents the one-point function diagram for the quark one-loop correction to the 
strange component of the scalar $\sigma$ and its counterterm. It can be written as,
\bqa
\delta\Gamma_y^{(1)}=-2\sqrt{2} \ N_c \ g \ m_s \ \mathcal{A}(m_s^2)+i\delta t_{y}\;.
\eqa
\subsection{Parameters with  Renormalization}
\label{subsecIVB}
The one-point functions $\Gamma^{(1)}_{x}=it_{x}=i(h_{x}-m_{\pi}^2 x)$ for the nonstrange and  $\Gamma^{(1)}_{y}=it_{y}=i{\tiny{\biggl\{h_{y}-\frac{\sqrt{2}}{2}(m_{K}^2-m_{\pi}^2) \  x-m_{K}^2y\biggr\}}}$ for the strange degree of freedom become zero and we get two tree level equations of motion $t_{x}=0$ and $t_{y}=0$. Thus the classical minimum of the effective potential gets fixed. The first renormalization condition for the nonstrange  $<\sigma_{x}>=0$ and the strange degree of freedom $<\sigma_{y}>=0$ requires that the respective one-loop corrections $\delta \Gamma^{(1)}_{x}$ and $\delta \Gamma^{(1)}_{y}$ to the one 
point functions, are put to zero such that the minimum of the effective potential does not change. Thus the $\delta\Gamma^{(1)}_{x}=0 $ and  $\delta\Gamma^{(1)}_{y}=0 $ give us 
\bqa
\delta t_{x}
&=&-4i \ N_c \ g \ m_u \ \mathcal{A}(m_u^2)  \;,
\\
\delta t_{y}
&=&-2\sqrt{2}i \ N_c \ g \ m_s \ \mathcal{A}(m_s^2)  \;. 
\eqa
Using the equation $h_{x}=t_{x}+m_{\pi}^2 \ x $ and $h_{y}=t_{y}+{\tiny{\biggl\{\frac{\sqrt{2}}{2}(m_{K}^2-m_{\pi}^2) \  x+m_{K}^2 \ y\biggr\}}} $, one can write the
counterterms $\delta h_{x}$ and $\delta h_{y}$ in terms of the corresponding tadpole counterterms  $\delta t_{x}$ and $\delta t_{y}$ as the following
\bqa
\label{delta:hx}
{\hskip -2.0 cm}\delta h_{x}&=&m^2_\pi \ \delta x +\delta m^2_\pi \ x+\delta t_{x},\qquad \qquad \qquad \qquad \; \\
\label{delta:hy}
{\hskip -2.0 cm}\delta h_{y}&=&{\biggl\{\frac{\sqrt{2}}{2}(m_{K}^2-m_{\pi}^2) \ \delta  x+\frac{\sqrt{2}}{2}(\delta m_{K}^2-\delta m_{\pi}^2) \ x \biggr\}}+m_{K}^2 \ \delta y +\delta m_{K}^2 \ y +\delta t_{y}\;. 
\eqa

Using the Eq.~(\ref{Zpi}), one can write
\bqa
\label{delta:hxn}
\delta h_x&=&\frac{1}{2}m^2_\pi \ x \ \delta Z_\pi+\delta m^2_\pi \ x +\delta t_x, \qquad \qquad \qquad \qquad \;
\\ \nonumber
\label{delta:hny}
{\hskip -1.5 cm}\delta h_{y}&=&{\biggl\{\frac{\sqrt{2}}{2}(m_{K}^2-m_{\pi}^2)\ x \ \frac{\delta  Z_{\pi}}{2}+\frac{\sqrt{2}}{2}(\delta m_{K}^2-\delta m_{\pi}^2) \  x \biggr\}}+m_{K}^2 \ y \ \frac{\delta Z_{\pi}}{2} +\delta m_{K}^2 \ y +\delta t_{y}.\; \\
\eqa
The inverse propagator for the pseudo-scalar
$\pi,K $ mesons can be written as 
\bqa
p^2-m_{\pi,K}^2-i\Sigma_{\pi,K}(p^2)
{\rm +counterterms}
\;.
\label{definv}
\eqa
The mixing in the $00$ and $88$ components for the scalar (s) and pseudo-scalar (p)  particles, gives us the physical states of the $\sigma$ and $f_{0}$ as the scalar particles and the $\eta$ and $\eta^{\prime}$ as the pseudo-scalar particles. The inverse propagator is given by the $2 \times 2$ matrix showing the mixing of the $00$ and $88$ components. When the determinant of this matrix is put to zero, negative root of the resulting equation gives the inverse propagator of the physical  $\sigma$ in the scalar and $\eta$ in the pseudo-scalar channel. The positive root gives the inverse propagator of the physically observed particles $f_{0}$ and  $\eta^{\prime}$ in the respective scalar and  pseudo-scalar channel.

\begin{equation}
\textnormal{Det}\left.
\begin{pmatrix} 
p^2-m^2_{\text{s(p)},00}-i\Sigma_{\text{s(p)},00}(p^2) \,\,& 
-m^2_{\text{s(p)},08}-i\Sigma_{\text{s(p)},08}(p^2) \\ 
-m^2_{\text{s(p)},08}-i\Sigma_{\text{s(p)},08}(p^2) & 
p^2-m^2_{\text{s(p)},88}-i\Sigma_{\text{s(p)},88}(p^2) \,\,
\end{pmatrix}
\right.=0\; .
\end{equation}
We obtain two solutions for the $p^{2}$ 
\begin{align}
\nonumber
& p^2=\frac{1}{2}\Biggl[{\tiny{\Biggl(\biggl\{ m^2_{\text{s(p)},00}+
i\Sigma_{\text{s(p)},00}(p^2)\biggr\}+\biggl\{m^2_{\text{s(p)},88}+i\Sigma_{\text{s(p)},88}(p^2)\biggr\} \Biggr)}} \mp  \\ 
& \qquad  {\tiny{ \sqrt{\Biggl( \biggl\{ m^2_{\text{s(p)},00}+
i\Sigma_{\text{s(p)},00}(p^2)\biggr\}-\biggl\{m^2_{\text{s(p)},88}+i\Sigma_{\text{s(p)},88}(p^2)\biggr\}\Biggr)^2+4\Biggl(m^2_{\text{s(p)},08}+i\Sigma_{\text{s(p)},08}(p^2) \Biggr)^2 }}}\Biggr]\;. \\ \nonumber
\end{align}
Neglecting the higher order ($N_{c}^2$) terms like $\tiny{ \biggl\{\Sigma_{\text{s(p)},00}(p^2))-\Sigma_{\text{s(p)},88}(p^2)\biggr\}^2 }$ and $ \Sigma_{\text{s(p)},08}^2(p^2)$ in self energy corrections, the above expression is written as,  

\begin{align}
\nonumber
& p^2=\frac{1}{2}\Biggl[\biggl( m^2_{\text{s(p)},00}+m^2_{\text{s(p)},88}\biggr)\mp {\tiny{ \sqrt{\Biggl( m^2_{\text{s(p)},00}-m^2_{\text{s(p)},88}\Biggr)^2+4\Biggl(m^2_{\text{s(p)},08}\Biggr)^2 }}}  \Biggr]+\frac{1}{2}\Biggl[\Biggl( i\Sigma_{\text{s(p)},00}(p^2)+ \\ \nonumber 
& i\Sigma_{\text{s(p)},88}(p^2)\Biggr)  \mp \frac{1}{\sqrt{(m^2_{\text{s(p)},00}-m^2_{\text{s(p)},88})^2+4m^4_{\text{s(p)},08}}} \biggl\{\biggl(i\Sigma_{\text{s(p)},00}(p^2)-i\Sigma_{\text{s(p)},88}(p^2)\biggr)(m^2_{\text{s(p)},00}-m^2_{\text{s(p)},88})  \\ \nonumber
&+4i\Sigma_{\text{s(p)},08}(p^2)  m^2_{\text{s(p)},08}  \biggr\}  \Biggr]\label{detroot}\;.\\ 
\end{align}
Negative root of the Eq.~(\ref{detroot}) gives the sum of the mass and self energy correction for the scalar $\sigma$ (pseudoscalar $\eta$)
\begin{align}
\nonumber
& p^2=m_{\sigma(\eta)}^2 + i\Sigma_{\sigma(\eta)}(p^2) \  \text{where}  \\ 
&m_{\sigma(\eta)}^2=\frac{1}{2}\Biggl[\biggl( m^2_{\text{s(p)},00}+m^2_{\text{s(p)},88}\biggr)-{\tiny{ \sqrt{\Biggl( m^2_{\text{s(p)},00}-m^2_{\text{s(p)},88}\Biggr)^2+4\Biggl(m^2_{\text{s(p)},08}\Biggr)^2 }}}  \Biggr] \ \text{and} \\ \nonumber
&\Sigma_{\sigma(\eta)}(p^2)=\frac{1}{2}\Biggl[\Sigma_{\text{s(p)},00}(p^2)+\Sigma_{\text{s(p)},88}(p^2)-\frac{1}{\sqrt{(m^2_{\text{s(p)},00}-m^2_{\text{s(p)},88})^2+4m^4_{\text{s(p)},08}}} \\ & \qquad \quad  \qquad \quad \biggl\{\biggl(\Sigma_{\text{s(p)},00}(p^2)-\Sigma_{\text{s(p)},88}(p^2)\biggr)(m^2_{\text{s(p)},00}-m^2_{\text{s(p)},88})  +4\Sigma_{\text{s(p)},08}(p^2) \   m^2_{\text{s(p)},08}  \biggr\}  \Biggr]\;.\label{selfensign}
\end{align}
Positive root of the Eq.~(\ref{detroot}) gives the sum of the mass and self energy correction for the scalar $f_{0}$ (pseudoscalar $\eta^{\prime}$)
\begin{align}
& p^2=m_{f_{0}(\eta^{\prime})}^2 + i\Sigma_{f_{0}(\eta^{\prime})}(p^2)  \text{where}\\ \nonumber 
&m_{f_{0}(\eta^{\prime})}^2=\frac{1}{2}\Biggl[\biggl( m^2_{\text{s(p)},00}+m^2_{\text{s(p)},88}\biggr)+{\tiny{ \sqrt{\Biggl( m^2_{\text{s(p)},00}-m^2_{\text{s(p)},88}\Biggr)^2+4\Biggl(m^2_{\text{s(p)},08}\Biggr)^2 }}}  \Biggr] \ \text{and} \\ \nonumber
&\Sigma_{f_{0}(\eta^{\prime})}(p^2)=\frac{1}{2}\Biggl[\Sigma_{\text{s(p)},00}(p^2)+\Sigma_{\text{s(p)},88}(p^2)+\frac{1}{\sqrt{(m^2_{\text{s(p)},00}-m^2_{\text{s(p)},88})^2+4m^4_{\text{s(p)},08}}} \\  & \qquad \quad  \qquad \quad \biggl\{\biggl(\Sigma_{\text{s(p)},00}(p^2)-\Sigma_{\text{s(p)},88}(p^2)\biggr)(m^2_{\text{s(p)},00}-m^2_{\text{s(p)},88})  +4\Sigma_{\text{s(p)},08}(p^2) \   m^2_{\text{s(p)},08}  \biggr\}  \Biggr]\;.\label{selfenf0n}
\end{align}

Thus the inverse propagator for the scalar $\sigma$ and the pseudo-scalar $ \eta, \ \eta^{\prime} $ mesons can be written as
\bqa
p^2-m_{\sigma,\eta,\eta^{\prime}}^2-i\Sigma_{\sigma,  \eta,  \eta^{\prime}}(p^2)
{\rm +counterterms}
\;.
\label{propmix}
\eqa
The renormalized mass in the Lagrangian is put equal to the physical mass, i.e.\ $m=m_{\rm pole}$ \footnote{The contributions of the imaginary parts of the self-energies for defining the mass are neglected.} when the on-shell scheme gets implemented and one can write
\bqa
\Sigma(p^2=m_{\sigma,\eta,  \eta^{\prime},\pi,K}^2)
{\rm +counterterms}
&=&0
\label{pole}
\;.
\eqa
Since the propagator residue is put to unity in the on-shell scheme, one gets 
\bqa
\label{res}
\nonumber
{\partial\over\partial p^2}\Sigma_{\sigma,\eta,  \eta^{\prime},\pi,K}(p^2)\Big|_{p^2=m_{\sigma,\eta,  \eta^{\prime},\pi,K}^2} \\
{\rm +counterterms}
&=&0\;.
\eqa
Using the diagrams of the Fig.~\ref{counts1b} and Fig.~\ref{countp2b}, the counterterms of the two point functions of the  scalar  and pseudo-scalar  mesons can be written as 
\bqa
\label{count1}
\Sigma_{\sigma}^{\rm ct1}(p^2)&=&i\left[\delta Z_{\sigma}(p^2-m_{\sigma}^2)-\delta m_{\sigma}^2\right]\;,
\\
\label{count2}
\Sigma_{\pi}^{\rm ct1}(p^2)&=&i\left[\delta Z_{\pi}(p^2-m_{\pi}^2)-\delta m_{\pi}^2\right]\;,
\\
\label{count3}
\Sigma_{K}^{\rm ct1}(p^2)&=&i\left[\delta Z_{K}(p^2-m_{K}^2)-\delta m_{K}^2\right]\;,
\\
\label{count4}
\Sigma_{\eta}^{\rm ct1}(p^2)&=&i\left[\delta Z_{\eta}(p^2-m_{\eta}^2)-\delta m_{\eta}^2\right]\;,
\\
\label{count5}
\Sigma_{\eta^{\prime}}^{\rm ct1}(p^2)&=&i\left[\delta Z_{\eta^{\prime}}(p^2-m_{\eta^{\prime}}^2)-\delta m_{\eta^{\prime}}^2\right]\;.
\eqa

The tadpole contributions to the scalar and pseudo-scalar self energies, contain two independent terms  proportional to $N_{c} g m_{u}\mathcal{A}(m_{u}^2)$  and $N_{c} g m_{s}\mathcal{A}(m_{s}^2)$ respectively as presented in the Appendix (\ref{appenB}). The tadpole counterterms $\Sigma^{\rm ct2}$ for the scalar and pseudo-scalar particles are chosen (negative of the respective tadpole contributions to the scalar and pseudo-scalar self energies) such that they completely cancel the respective tadpole contributions to the self-energies. The evaluation of the self-energies and their derivatives in the on-shell conditions, give all the renormalization constants. When the Eqs. (\ref{pole}), (\ref{res}) and (\ref{count1})--(\ref{count5}) are combined, we obtain the following set of equations :  
\bqa
\delta m_{\pi}^2&=&-i\Sigma_{\pi}(m_{\pi}^2)\;;\delta Z_\pi= i{\partial\over\partial p^2}\Sigma_\pi(p^2)\Big|_{p^2=m_\pi^2}
\eqa
\bqa
\delta m_{K}^2&=&-i\Sigma_{K}(m_{K}^2)\;;\delta Z_{K} =
i{\partial\over\partial p^2}\Sigma_{K}(p^2)\Big|_{p^2=m_{K}^2}\;,
\eqa
\bqa
\delta m_{\eta}^2&=&-i\Sigma_{\eta}(m_{\eta}^2)\;;\delta Z_\eta =i{\partial\over\partial p^2}\Sigma_\eta(p^2)\Big|_{p^2=m_\eta^2},
\eqa
\bqa
\delta m_{\eta^{\prime}}^2&=&-i\Sigma_{\eta^{\prime}}(m_{\eta^{\prime}}^2)\;;\delta Z_{\eta^{\prime}} =i{\partial\over\partial p^2}\Sigma_{\eta^{\prime}}(p^2)\Big|_{p^2=m_{\eta^{\prime}}^2},
\eqa
\bqa
\delta m_{\sigma}^2&=&-i\Sigma_{\sigma}(m_{\sigma}^2)\;;\delta Z_\sigma =
i{\partial\over\partial p^2}\Sigma_\sigma(p^2)\Big|_{p^2=m_\sigma^2}.
\eqa
When the self energy (neglecting the tadpole contributions) expressions from the Eqs.~(\ref{selfenpi}), (\ref{selfenK}), (\ref{selfensign}) and (\ref{selfenf0n}) are used, we get the following set of equations.

\bqa
&\delta m_{\pi}^2=2ig^2N_c[\mathcal{A}(m_u^2)-\mbox{$1\over2$}{m_{\pi}^2}\mathcal{B}(m_{\pi}^2,m_u)]\;, \\
&\delta m_{K}^2=ig^2N_c\biggl[\mathcal{A}(m_u^2)+\mathcal{A}(m_s^2)-\{m_{K}^2-(m_{u}-m_{s})^{2}\} 
\mathcal{B}(m_{K}^2,m_{u},m_{s})\biggr]\;,
\eqa

\begin{align}
\nonumber
&\delta m_{\eta}^2=\frac{-i}{2}\Biggl[\Sigma_{\text{p},00}(m_{\eta}^2)+\Sigma_{\text{p},88}(m_{\eta}^2)-\frac{1}{\sqrt{(m^2_{\text{p},00}-m^2_{\text{p},88})^2+4m^4_{\text{p},08}}} \\ 
&\qquad \quad \biggl\{\biggl(\Sigma_{\text{p},00}(m_{\eta}^2)-\Sigma_{\text{p},88}(m_{\eta}^2)\biggr)(m^2_{\text{p},00}-m^2_{\text{p},88})  +4\Sigma_{\text{p},08}(m_{\eta}^2) \   m^2_{\text{p},08}  \biggr\}  \Biggr]       \;, \\  \nonumber 
&\delta m_{\eta}^2=ig^2N_c\Biggl[ \biggl\{\mathcal{A}(m_u^2)+\mathcal{A}(m_s^2)-\mbox{$1\over2$}{m_{\eta}^2}\mathcal{B}(m_{\eta}^2,m_u)-\mbox{$1\over2$}{m_{\eta}^2}\mathcal{B}(m_{\eta}^2,m_s)\biggr\}  \\
&\qquad \quad -
\frac{\tiny{\biggl\{(m^2_{\text{p},00}-m^2_{\text{p},88})+4\sqrt{2}m^2_{\text{p},08}\biggr\}}}{3\sqrt{(m^2_{\text{p},00}-m^2_{\text{p},88})^2+4m^4_{\text{p},08}}} \biggl\{\mathcal{A}(m_u^2)-\mathcal{A}(m_s^2)-\mbox{$1\over2$}{m_{\eta}^2}\mathcal{B}(m_{\eta}^2,m_u)+\mbox{$1\over2$}{m_{\eta}^2}\mathcal{B}(m_{\eta}^2,m_s)\biggr\}\Biggr] \; , \\  \nonumber  
&\delta m_{\eta^{\prime}}^2=
\frac{-i}{2}\Biggl[\Sigma_{\text{p},00}(m_{\eta^{\prime}}^2)+\Sigma_{\text{p},88}(m_{\eta^{\prime}}^2)+\frac{1}{\sqrt{(m^2_{\text{p},00}-m^2_{\text{p},88})^2+4m^4_{\text{p},08}}} \\  
&\qquad \quad \biggl\{\biggl(\Sigma_{\text{p},00}(m_{\eta^{\prime}}^2)-\Sigma_{\text{p},88}(m_{\eta^{\prime}}^2)\biggr)(m^2_{\text{p},00}-m^2_{\text{p},88})  +4\Sigma_{\text{p},08}(m_{\eta^{\prime}}^2) \   m^2_{\text{p},08}  \biggr\}  \Biggr] \;,
\\
\nonumber
&\delta m_{\eta^{\prime}}^2=ig^2N_c\Biggl[ \biggl\{\mathcal{A}(m_u^2)+\mathcal{A}(m_s^2)-\mbox{$1\over2$}{m_{\eta^{\prime}}^2}\mathcal{B}(m_{\eta^{\prime}}^2,m_u)-\mbox{$1\over2$}{m_{\eta^{\prime}}^2}\mathcal{B}(m_{\eta^{\prime}}^2,m_s)\biggr\}         \\ 
&\qquad \quad + \frac{\tiny{\biggl\{(m^2_{\text{p},00}-m^2_{\text{p},88})+4\sqrt{2}m^2_{\text{p},08}\biggr\}}}{3\sqrt{(m^2_{\text{p},00}-m^2_{\text{p},88})^2+4m^4_{\text{p},08}}}  \biggl\{\mathcal{A}(m_u^2)-\mathcal{A}(m_s^2)-\mbox{$1\over2$}{m_{\eta^{\prime}}^2}\mathcal{B}(m_{\eta^{\prime}}^2,m_u)+\mbox{$1\over2$}{m_{\eta^{\prime}}^2}\mathcal{B}(m_{\eta^{\prime}}^2,m_s)\biggr\}\Biggr] \; , \\  \nonumber 
&\delta m_{\sigma}^2=
\frac{-i}{2}\Biggl[\Sigma_{\text{s},00}(m_{\sigma}^2)+\Sigma_{\text{s},88}(m_{\sigma}^2)-\frac{1}{\sqrt{(m^2_{\text{s},00}-m^2_{\text{s},88})^2+4m^4_{\text{s},08}}} \\ 
&\qquad \quad \biggl\{\biggl(\Sigma_{\text{s},00}(m_{\sigma}^2)-\Sigma_{\text{s},88}(m_{\sigma}^2)\biggr)(m^2_{\text{s},00}-m^2_{\text{s},88})  +4\Sigma_{\text{s},08}(m_{\sigma}^2) \  m^2_{\text{s},08}  \biggr\}  \Biggr] \;, \\
\nonumber  
&\delta m_{\sigma}^2=ig^2N_c\Biggl[ \biggl\{\mathcal{A}(m_u^2)+\mathcal{A}(m_s^2)-\mbox{$1\over2$}{(m_{\sigma}^2-4m_{u}^2)}
\mathcal{B}(m_{\sigma}^2,m_u)-\mbox{$1\over2$}{(m_{\sigma}^2-4m_{s}^2)}\mathcal{B}(m_{\sigma}^2,m_s)\biggr\}-\\ \nonumber
&\qquad \quad  \frac{\tiny{\biggl\{(m^2_{\text{s},00}-m^2_{\text{s},88})+4\sqrt{2}m^2_{\text{s},08}\biggr\}}}{3\sqrt{(m^2_{\text{s},00}-m^2_{\text{s},88})^2+4m^4_{\text{s},08}}} \biggl\{\mathcal{A}(m_u^2)-\mathcal{A}(m_s^2)-\frac{(m_{\sigma}^2-4m_{u}^2)}{2}\mathcal{B}(m_{\sigma}^2,m_u)\\ 
&+\frac{(m_{\sigma}^2-4m_{s}^2)}{2}\mathcal{B}(m_{\sigma}^2,m_s)\biggr\}\Biggr],
\\ 
&\delta Z_{\pi}=
ig^2N_c\left[\mathcal{B}(m_{\pi}^2,m_u)+m_{\pi}^2\mathcal{B}^{\prime}(m_{\pi}^2,m_u)
\right]\;,\\
&\delta Z_{K}=
ig^2N_c\left[\mathcal{B}(m_{K}^2,m_{u},m_{s})+(m_{K}^2-(m_{u}-m_{s})^2)\mathcal{B}^{\prime}(m_{K}^2,m_{u},m_{s})\right]\;,
\\
\nonumber
&\delta Z_{\eta}=\frac{ig^2N_c}{2}\Biggl[\biggl\{
\mathcal{B}(m_{\eta}^2,m_{u})+
\mathcal{B}(m_{\eta}^2,m_{s})+m_{\eta}^2  \  \mathcal{B}^{\prime}(m_{\eta}^2,m_{u}) + m_{\eta}^2 \ \mathcal{B}^{\prime}(m_{\eta}^2,m_{s})\biggr\}  \\ 
&\qquad +\frac{\tiny{\biggl\{(m^2_{\text{p},00}-m^2_{\text{p},88})+4\sqrt{2}m^2_{\text{p},08}\biggr\}}}{3\sqrt{(m^2_{\text{p},00}-m^2_{\text{p},88})^2+4m^4_{\text{p},08}}}\biggl\{-\mathcal{B}(m_{\eta}^2,m_{u})+\mathcal{B}(m_{\eta}^2,m_{s})-m_{\eta}^2 \ \mathcal{B}^{\prime}(m_{\eta}^2,m_{u}) + m_{\eta}^2 \ \mathcal{B}^{\prime}(m_{\eta}^2,m_{s})\biggr\}\Biggr]\;,
\end{align}

\begin{align}
\nonumber
&\delta Z_{\eta^{\prime}}=\frac{ig^2N_c}{2}\Biggl[\biggl\{
\mathcal{B}(m_{\eta^{\prime}}^2,m_{u})+
\mathcal{B}(m_{\eta^{\prime}}^2,m_{s})+m_{\eta^{\prime}}^2  \  \mathcal{B}^{\prime}(m_{\eta^{\prime}}^2,m_{u}) + m_{\eta^{\prime}}^2 \ \mathcal{B}^{\prime}(m_{\eta^{\prime}}^2,m_{s})\biggr\}  \\ 
& -\frac{\tiny{\biggl\{(m^2_{\text{p},00}-m^2_{\text{p},88})+4\sqrt{2}m^2_{\text{p},08}\biggr\}}}{3\sqrt{(m^2_{\text{p},00}-m^2_{\text{p},88})^2+4m^4_{\text{p},08}}}\biggl\{-\mathcal{B}(m_{\eta^{\prime}}^2,m_{u})+\mathcal{B}(m_{\eta^{\prime}}^2,m_{s})-m_{\eta^{\prime}}^2 \ \mathcal{B}^{\prime}(m_{\eta^{\prime}}^2,m_{u}) + m_{\eta^{\prime}}^2 \ \mathcal{B}^{\prime}(m_{\eta^{\prime}}^2,m_{s})\biggr\}\Biggr]\;,
\\
\nonumber
&\delta Z_{\sigma}=\frac{ig^2N_c}{2}\Biggl[\biggl\{
\mathcal{B}(m_{\sigma}^2,m_{u})+
\mathcal{B}(m_{\sigma}^2,m_{s})+(m_{\sigma}^2-4m_u^2)\mathcal{B}^{\prime}(m_{\sigma}^2,m_{u}) + (m_{\sigma}^2-4m_s^2)\mathcal{B}^{\prime}(m_{\sigma}^2,m_{s})\biggr\}+
\\ \nonumber
& \quad  \frac{\tiny{\biggl\{(m^2_{\text{s},00}-m^2_{\text{s},88})+4\sqrt{2}m^2_{\text{s},08}\biggr\}}}{3\sqrt{(m^2_{\text{s},00}-m^2_{\text{s},88})^2+4m^4_{\text{s},08}}}\biggl\{\mathcal{B}(m_{\sigma}^2,m_{s})-\mathcal{B}(m_{\sigma}^2,m_{u})-(m_{\sigma}^2-4m_u^2)\mathcal{B}^{\prime}(m_{\sigma}^2,m_{u}) \\
&+ (m_{\sigma}^2-4m_s^2)\mathcal{B}^{\prime}(m_{\sigma}^2,m_{s})\biggr\}\Biggr].
\end{align}
The field renormalization constant expressions are given above for the $\pi$, $K$, $\eta$, $\eta^{\prime}$ and $\sigma$. However in the calculations below, one needs to have the simplified expression of $\delta Z_\pi$ only. Substituting the expressions of $\delta Z_{\pi},  \delta m_{K}^2, \ \delta m_{\pi}^2,\ \delta m_{\eta}^2$ and $ \delta m_{\eta^{\prime}}^2$ from the above in the Eq.(\ref{delta:lambda_2}), the $\delta \lambda_{2}$ is written as,
\begin{align}
\nonumber
&\delta\lambda_{2\os}=\dfrac{2 \ iN_{c} \ g^2 \ }{(x^2+4y^2)(\sqrt{2} \ y -x)}\Biggl[(3\sqrt{2} \  y) \biggl\{A(m_{u}^2)+A(m_{s}^2)-\biggl( m_{K}^2-(m_{s}-m_{u})^2 \biggr) \mathcal{B}(m^2_K,m_u,m_s)  \biggr\}\; \\ \nonumber
&\qquad \quad-(\sqrt{2} y+2x) \biggl\{2 A(m_{u}^2)  -m_{\pi}^2\mathcal{B}(m^2_\pi,m_u) \biggr\}-(\sqrt{2} y-x)\Biggl\{2 A(m_{u}^2)+2 A(m_{s}^2)-\frac{m_{\eta}^2}{2}\biggl\{ \mathcal{B}(m^2_\eta \ , m_u) \; \\ \nonumber 
&\qquad \quad + \mathcal{B}(m^2_\eta \ , \ m_s)\biggr\}-\frac {m_{\eta^{'}}^2}{2}\biggl\{\mathcal{B}(m^2_{\eta^{'}} \ , \ m_u)+\mathcal{B}(m^2_{\eta^{'}}\ , \ m_s)    \biggr\}-\frac{\biggl(m^2_{p,00}-m^2_{p,88}+4\sqrt{2}m^2_{p,08}\biggr)}{6 \ \biggl( \sqrt{(m^2_{p,00}-m^2_{p,88})^2+4m^4_{p,08}\biggr) }} \\ 
&\qquad \quad\Biggl\{ m_{\eta}^2\biggl\{- \mathcal{B}(m^2_\eta \ , m_u)+\mathcal{B}(m^2_\eta \ , \ m_s)\biggr\} -m_{\eta^{'}}^2\biggl\{-\mathcal{B}(m^2_{\eta^{'}} \ , \ m_u)+\mathcal{B}(m^2_{\eta^{'}}\ , \ m_s)\biggr\} \Biggr\}\Biggr]\;,
\\
\label{lam2os}
&\delta\lambda_{2\os}=\delta \lambda_{2\text{div}}+\lambda_{2\text{{\tiny FIN}}}\;,
\\
\nonumber
&\lambda_{2\text{{\tiny FIN}}}= \dfrac{N_{c} g^2}{(4\pi)^2}(\lambda_2-g^2)\ln\biggl(\frac{\Lambda^2}{m_u^2}\biggr) +  \dfrac{N_cg^2}{(4\pi)^2} \dfrac{2}{(x^2+4y^2)}\Biggl[\dfrac{(\sqrt{2} y+2x)}{(\sqrt{2}y -x)}\Biggl\{ m^2_{u}-m^2_{s} \biggl\{1-2\ln\biggl(\frac{m_s}{m_u}\biggr)\biggr\} \\ \nonumber
& \quad-m_{\pi}^2\mathcal{C}(m^2_\pi,m_u)\Biggr\}+\dfrac{3\sqrt{2} \ y}{(\sqrt{2}y -x)} \biggl\{ m_{K}^2-(m_{s}-m_{u})^2 \biggr\} \mathcal{C}(m^2_K,m_u,m_s)- \frac{ m_{\eta}^2}{2}\biggl\{ \mathcal{C}(m^2_\eta \ , m_u)+\mathcal{C}(m^2_\eta \ , \ m_s)\;
\\
\nonumber
&-2\ln\biggl(\frac{m_s}{m_u}\biggr)\biggr\}-\frac {m_{\eta^{'}}^2}{2}\biggl\{\mathcal{C}(m^2_{\eta^{'}} \ , \ m_u)+\mathcal{C}(m^2_{\eta^{'}}\ , \ m_s)-2\ln\biggl(\frac{m_s}{m_u}\biggr)\biggr\} +\frac{\biggl(m^2_{p,00}-m^2_{p,88}+4\sqrt{2}m^2_{p,08}\biggr)}{6 \ \biggl( \sqrt{(m^2_{p,00}-m^2_{p,88})^2+4m^4_{p,08}}\biggr) }
\end{align}
\begin{align}
& \quad \Biggl\{ m_{\eta}^2\biggl\{ \mathcal{C}(m^2_\eta \ , m_u)-\mathcal{C}(m^2_\eta \ , \ m_s)+2\ln\biggl(\frac{m_s}{m_u}\biggr)\biggr\} -m_{\eta^{'}}^2\biggl\{\mathcal{C}(m^2_{\eta^{'}} \ , \ m_u)-\mathcal{C}(m^2_{\eta^{'}}\ , \ m_s)+2\ln\biggl(\frac{m_s}{m_u}\biggr)\biggr\} \Biggr\}\Biggr]\;,
\\
\nonumber
& \text{Substituting the expressions of $\delta Z_{\pi},\  \delta m_{K}^2 ,\  \delta m_{\pi}^2, \ \delta \lambda_{2}$ in the Eq.~(\ref{deltac}), the $\delta c$ is written as} \\ \nonumber
&\delta c_{\os}= \dfrac{2 \ i \ N_{c} \ g^{2}}{(\sqrt{2} \ y-x)}\biggl\{A(m_{u}^2)+A(m_{s}^2)-\biggl( m_{K}^2-(m_{s}-m_{u})^2 \biggr) \mathcal{B}(m^2_K,m_u,m_s)-2 A(m_{u}^2)  + m_{\pi}^2\mathcal{B}(m^2_\pi,m_u) \biggr\} \\  \label{cos}
& \quad \ \  -\sqrt{2} \ y \ \delta \lambda_{2\os} - (2\sqrt{2}\ y \ \lambda_{2}+c)\frac{\delta Z_{\pi}}{2} \ ;  \ \ \ \ \   \delta c_{\os}=\delta c_{\text{div}}+c_{\text{\tiny{FINTOT}}} \ \ \ ; \ \ \ \ \ c_{\text{\tiny{FINTOT}}}=-\sqrt{2} \ y \ \lambda_{2\text{{\tiny FIN}}}+c_{\text{{\tiny FIN}}}\;,
\\ \nonumber 
&c_{\text{{\tiny FIN}}}
= \dfrac{N_cg^2}{(4\pi)^2}\Biggl[\biggl\{ c+\sqrt{2} y (\lambda_2-g^2) \biggr\}  \ln\biggl(\frac{\Lambda^2}{m_u^2}\biggr)   +     \dfrac{2}{(\sqrt{2}y -x)} \Biggl\{ \biggl\{ m_{K}^2-(m_{s}-m_{u})^2 \biggr\} \mathcal{C}(m^2_K,m_u,m_s) 
 \\ 
&\qquad-m_{\pi}^2\mathcal{C}(m^2_\pi,m_u) \Biggr\} -\dfrac{ g^2}{2}(\sqrt{2} y+x)+\dfrac{2 g^2 y^2 }{(\sqrt{2}y -x)}\ln\biggl(\frac{m_s}{m_u}\biggr)  \Biggr]\;.
\end{align}
Using the Eq.~(\ref{lam1}) and substituting the expressions of $\delta Z_{\pi},\  \delta m_{\sigma}^2 ,\ \delta m_{\pi}^2, \ \delta \lambda_{2}$ and $\delta c$ in the Eq.~(\ref{lam2}), the $\delta \lambda_{1}$ is written as
\begin{align}
&\delta\lambda_{1\os}=\frac{\lambda_{1\text{{\tiny NUMOS}}}}{\lambda_{1\text{{\tiny DENOM}}}}-\lambda_{1} \delta Z_{\pi}\;, \\ \nonumber
&\lambda_{1\text{{\tiny NUMOS}}}=iN_{c}g^2\Biggl[\biggl(\sqrt{(m^2_{s,00}-m^2_{s,88})^2+4m^4_{s,08}}\biggr)\biggl\{A(m_{u}^2)+A(m_{s}^2)-\biggl( \frac{m_{\sigma}^2-4m_{u}^2}{2} \biggr) \mathcal{B}(m^2_{\sigma},m_u) \\ \nonumber 
&-\biggl( \frac{m_{\sigma}^2-4m_{s}^2}{2} \biggr) \mathcal{B}(m^2_{\sigma},m_s)-2A(m_{u}^2)+m_{\pi}^2 \mathcal{B}(m^2_{\pi},m_u) \biggr\} -\biggl(\frac{m^2_{s,00}-m^2_{s,88} +4\sqrt{2}m^2_{s,08}}{3}\biggr) \\ \nonumber
&  \biggl\{A(m_{u}^2)-A(m_{s}^2)-\biggl( \frac{m_{\sigma}^2-4m_{u}^2}{2} \biggr)  \mathcal{B}(m^2_{\sigma},m_u)+\biggl( \frac{m_{\sigma}^2-4m_{s}^2}{2} \biggr) \mathcal{B}(m^2_{\sigma},m_s) \biggr\}  \Biggr] \\ \nonumber
&+\frac{(m^2_{s,00}-m^2_{s,88})}{12}\biggl\{ (3x^2-6y^2)\delta \lambda_{2\os}-\sqrt{2}(4\sqrt{2}x+y)\delta c_{\os}\biggr\}  -\frac{\sqrt{2}m^2_{s,08}}{3}\\  \nonumber
& \biggl\{(2y^2-x^2) \delta \lambda_{2\os}+(\sqrt{2} y-x) \delta c_{\os} \biggl\}-\frac{1}{4}\sqrt{(m^2_{s,00}-m^2_{s,88})^2+4m^4_{s,08}}\biggl\{(x^2+6y^2) \delta \lambda_{2\os}+\sqrt{2} y \delta c_{\os} \biggl\} \\ \nonumber 
& + \delta Z_{\pi} \Biggl[ \frac{(m^2_{\sigma_{00}}-m^2_{s,88})}{12}\biggl\{ (3x^2-6y^2) \lambda_{2}-\sqrt{2}(4\sqrt{2}x+y) \frac{c}{2}\biggr\} -\frac{\sqrt{2}m^2_{s,08}}{3}\biggl\{(2y^2-x^2)  \lambda_2+(\sqrt{2} y-x)  \frac{c}{2} \biggl\} \\  
&\qquad \quad \ \  - \frac{1}{4}\biggl(\sqrt{(m^2_{s,00}-m^2_{s,88})^2+4m^4_{s,08}}\biggr)\biggl\{(x^2+6y^2) \lambda_2+\sqrt{2} y \frac {c}{2} \biggl\}\Biggr]\;, \\ 
\label{lam1os}
&\delta\lambda_{1\os}=\delta \lambda_{1\text{div}}+\lambda_{1\text{{\tiny FIN}}} ;\qquad \lambda_{1\text{{\tiny FIN}}}=\frac{\lambda_{1\text{{\tiny NUMF}}}}{\lambda_{1\text{{\tiny DENOM}}}} ;\qquad \lambda_{1\text{{\tiny NUMF}}}=\lambda_{1\text{{\tiny NUMF-I}}}+\lambda_{1\text{{\tiny NUMF-II}}} \;, \\   \nonumber
&\text{expression of }\ \lambda_{1\text{{\tiny DENOM}}} \ \text{is given in the Eq.~(\ref{lam1de})}\;,
\end{align}
\begin{align}
\nonumber
&\lambda_{1\text{{\tiny NUMF-I}}}=\frac{(m^2_{s,00}-m^2_{s,88})}{12}\biggl\{ (3x^2+8\sqrt{2}xy-4y^2)\lambda_{2\text{{\tiny FIN}}}-\sqrt{2}(4\sqrt{2}x+y)c_{\text{{\tiny FIN}}}\biggr\}-\frac{\sqrt{2}m^2_{s,08}}{3}\biggl\{(\sqrt{2}y-x) c_{\text{{\tiny FIN}}} \; \\ 
&\qquad \quad \ \ \  + (4y^2+\sqrt{2}xy-3x^2)\lambda_{2\text{{\tiny FIN}}} \biggr\}-\biggl(\frac{1}{4}\sqrt{(m^2_{s,00}-m^2_{s,88})^2+4m^4_{s,08}}\biggr) \ \biggl\{ (x^2+4y^2)\lambda_{2\text{{\tiny FIN}}}+\sqrt{2}y c_{\text{{\tiny FIN}}}\biggr\}\;,
\\
\nonumber 
&\lambda_{1\text{{\tiny NUMF-II}}}=\dfrac{N_cg^2}{(4\pi)^2}\Biggl[\biggl(\sqrt{(m^2_{s,00}-m^2_{s,88})^2+4m^4_{s,08}}\biggr)\biggl\{\dfrac{g^2}{4}(x^2-2 y^2)+(m^2_{\sigma}-m^2_{\pi}-m^2_{u}-3m^2_{s}) \ln\biggl(\frac{\Lambda^2}{m_q^2}\biggr)\;\\ \nonumber
& +2m^2_{s}\ln\biggl(\frac{m_s}{m_u}\biggr) + \frac{ (m_{\sigma}^2-4m^2_{u})}{2} \ \mathcal{C}(m^2_\sigma \ , m_u)+\frac{ (m_{\sigma}^2-4m^2_{s})}{2} \ \biggl( \mathcal{C}(m^2_\sigma \ , m_s)-2\ln\biggl(\frac{m_s}{m_u}\biggr)\biggr)-m_{\pi}^2 \mathcal{C}(m^2_\pi \ , m_u)\biggr\}\; \\ \nonumber
&\qquad \quad \ \ - \biggl(\dfrac{(m^2_{s,00}-m^2_{s,88})+4\sqrt{2}m^2_{s,08}}{3}\biggr) \biggl\{ \dfrac{g^2}{4}(2 y^2-x^2)\biggl(1+3\ln\biggl(\frac{\Lambda^2}{m_q^2}\biggr)\biggr)+(m^2_{\sigma}-6 \ m^2_{s} \ ) \ln\biggl(\frac{m_s}{m_u}\biggr)\; \\ 
&\qquad \quad \ \ + \frac{ (m_{\sigma}^2-4m^2_{u})}{2} \ \mathcal{C}(m^2_\sigma \ , m_u)-\frac{ (m_{\sigma}^2-4m^2_{s})}{2} \  \mathcal{C}(m^2_\sigma \ , m_s) \biggr\} \Biggl] \;,  
\\ \nonumber
&\delta m^2_{\os}=iN_{c}g^2\biggl\{2A(m_{u}^2)-m_{\pi}^2 \mathcal{B}(m^2_{\pi},m_u)  \biggl\}-\delta \lambda_{1\os}  (x^2+y^2)-\delta \lambda_{2\os}  \frac{  x^2}{2}+\frac{\delta c_{\os} y}{\sqrt{2}}\\ 
&\qquad \quad \ \ -\delta  Z_{\pi} \biggl\{\lambda_{1} (x^2+y^2)+\lambda_{2}  \frac{  x^2}{2}-\frac{ c  y}{2\sqrt{2}} \biggr\}\;, \\
\label{m2os}
&\delta m^2_{\os}=\delta m^2_{\text{div}}+ m^2_{\text{\tiny{FIN}}} \;, \\ \nonumber
&m^2_{\text{\tiny{FIN}}}=\dfrac{N_cg^2}{(4\pi)^2} \Biggl[-2 m_{u}^2+(m_{\pi}^2-2 m_{u}^2)\ln\biggl(\frac{\Lambda^2}{m_u^2}\biggr)+m_{\pi}^2 \mathcal{C}(m^2_\pi \ , m_u)\Biggr] \\
&-\Biggl[\lambda_{1\text{\tiny{FIN}}}(x^2+y^2)+\lambda_{2\text{\tiny{FIN}}}\frac{x^2}{2}-c_{\text{\tiny{FINTOT}}}\frac{y}{\sqrt{2}}\Biggr]\;, \\ 
&\delta h_{x\os}=-\frac{i}{2}N_cg^2 m^2_\pi \ x \left[\mathcal{B}(m^2_\pi,m_u)-m^2_\pi \mathcal{B}^\prime(m^2_\pi,m_u)\right]\;,
\\
\label{hxos}
&\delta h_{x\os}=\delta h_{x\text{div}}+ h_{x\text{\tiny{FIN}}} \;, \\ 
&h_{x\text{\tiny{FIN}}}=\dfrac{N_cg^2}{2(4\pi)^2}h_x\left[\ln\left(\frac{\Lambda^2}{m_u^2}\right)+\mathcal{C}(m^2_\pi,m_u)-m^2_\pi \mathcal{C}^{\prime}(m^2_\pi,m_u)\right]\;,
\\
\nonumber
&\delta h_{y\os}=iN_cg^2\Biggl[ \left(\frac{\sqrt{2}}{2}x-y\right)\left\{\mathcal{A}(m^2_s)-\mathcal{A}(m^2_u)\right\}-\left(\frac{\sqrt{2}}{2}x+y\right)\left\{m^2_{K}-(m_u-m_s)^2\right\}\mathcal{B}(m^2_K,m_u,m_s) \\ 
&\qquad \quad \ + \left(\frac{\sqrt{2}}{2}x+y\right)\frac{m^2_K}{2}\left[\mathcal{B}(m^2_\pi,m_u) +m^2_\pi \mathcal{B}^\prime(m^2_\pi,m_u)\right] +
\frac{\sqrt{2}}{4}x \ m^2_\pi\left[\mathcal{B}(m^2_\pi,m_u)-m^2_\pi \mathcal{B}^\prime(m^2_\pi,m_u)\right] \Biggr]\;,
\\
\label{hyos}
&\delta h_{y\os}=\delta h_{y\text{div}}+ h_{y\text{\tiny{FIN}}} \;, \\ \nonumber
&h_{y\text{\tiny{FIN}}}=\dfrac{N_cg^2}{(4\pi)^2}\Biggl[ \dfrac{h_y}{2} \biggl\{  \ln\left(\frac{\Lambda^2}{m_u^2}\right) - \mathcal{C}(m^2_\pi,m_u)-m^2_\pi \mathcal{C}^{\prime}(m^2_\pi,m_u) \biggr\}-\dfrac{\sqrt{2} \  h_x}{2}\mathcal{C}(m^2_\pi,m_u) \\
& \qquad \quad +\left(\dfrac{\sqrt{2} \  x}{2}-y\right) \ \biggl\{m_{u}^2-m_{s}^2+2m_{s}^2 \ln(\frac{m_s}{m_u})\biggr\}+ \left( \dfrac{\sqrt{2} \  x}{2}+y \right)  \biggl\{ m_{K}^2-(m_{s}-m_{u})^2 \biggr\} \mathcal{C}(m^2_K,m_u,m_s) \Biggr]\;, \\
\label{zpi}
&\delta Z^{\os}_{\pi}=\delta Z_{\pi,\rm div}-
\frac{N_cg^2}{(4\pi)^2}\left[\ln\left(\frac{\Lambda^2}{m_u^2}\right)+\mathcal{C}(m_{\pi}^2,m_u)+m_{\pi}^2\mathcal{C}^{\prime}(m_{\pi}^2,m_u)
\right]\;,
\end{align}
\begin{align}
&\delta g^2_{\os}=-iN_cg^4\left[m^2_\pi \mathcal{B}^\prime(m^2_\pi,m_u)+\mathcal{B}(m^2_\pi,m_u)\right]=\delta g^2_{\text{div}}+\dfrac{N_cg^4}{(4\pi)^2}\left[\ln\left(\frac{\Lambda^2}{m_u^2}\right)+\mathcal{C}(m^2_\pi,m_u)+m^2_\pi \mathcal{C}^{\prime}(m^2_\pi,m_u)\right]  \;,\\ 
&\delta x^2_{\os}=iN_cg^2 x^2\left[m^2_\pi \mathcal{B}^\prime(m^2_\pi,m_u)+\mathcal{B}(m^2_\pi,m_u)\right]=\delta x^2_{\text{div}}-\dfrac{N_cg^2 x^2}{(4\pi)^2}\left[\ln\left(\frac{\Lambda^2}{m_u^2}\right)+\mathcal{C}(m^2_\pi,m_u)+m^2_\pi \mathcal{C}^{\prime}(m^2_\pi,m_u)\right]\;,\\
\label{y2os}
&\delta y^2_{\os}=iN_cg^2 y^2\left[m^2_\pi \mathcal{B}^\prime(m^2_\pi,m_u)+\mathcal{B}(m^2_\pi,m_u)\right]=\delta y^2_{\text{div}}-\dfrac{N_cg^2 y^2}{(4\pi)^2}\left[\ln\left(\frac{\Lambda^2}{m_u^2}\right)+\mathcal{C}(m^2_\pi,m_u)+m^2_\pi \mathcal{C}^{\prime}(m^2_\pi,m_u)\right]\;.\\ \nonumber
&\text{The common factor in the r.h.s. of the above four equations is defined as} \\ 
&\text{SCF}=\left[\ln\left(\frac{\Lambda^2}{m_u^2}\right)+\mathcal{C}(m^2_\pi,m_u)+m^2_\pi \mathcal{C}^{\prime}(m^2_\pi,m_u)\right]\;.
\end{align}

The $\mathcal{A}(m^2_f), \ \mathcal{B}(m^2,m_f), \ \mathcal{B}(m^2,m_u,m_s), \ \mathcal{B}^{\prime}(m^2,m_f)$, $\mathcal{C}(m^2,m_f)$, $\mathcal{C}(m^2,m_u,m_s)$, $\mathcal{C}^{\prime}(m^2,m_f)$ and $\mathcal{C}^{\prime}(m^2,m_u,m_s)$ are defined in the Appendix~(\ref{appenC}). The divergent part of the counterterms are  $ \delta \lambda_{2\text{div}}=\frac{N_cg^2}{(4\pi)^2\epsilon}(2\lambda_2-g^2)\;$,\ $\delta c_{\text{div}}=\frac{3N_cg^2 c}{2(4\pi)^2\epsilon}\;$, \ $\delta \lambda_{1\text{div}}=\frac{N_cg^22\lambda_1}{(4\pi)^2\epsilon}\;$, \ $\delta m^2_{\text{div}}=\frac{N_cg^2 m^2}{(4\pi)^2\epsilon}\;$, \ $ \delta h_{x\text{div}}=\frac{N_cg^2 h_x}{2(4\pi)^2\epsilon}\;$,\ $ \delta h_{y\text{div}}=\frac{N_cg^2 h_y}{2(4\pi)^2\epsilon}\;$, \ $\delta g^2_{\text{div}}=\frac{N_cg^4}{(4\pi)^2\epsilon}\; $, \ $\delta x^2_{\text{div}}=-\frac{N_cg^2x^2}{(4\pi)^2\epsilon}\; $, $\delta y^2_{\text{div}}=-\frac{N_cg^2y^2}{(4\pi)^2\epsilon}\;$, $\delta Z_{\pi,\rm div}=-\frac{N_cg^2}{(4\pi)^2\epsilon}$ . For both, the on-shell and the $\overline{\text{MS}}$ schemes, the divergent part of the counterterms are the same, i.e. $\delta \lambda_{1\text{div}}=\delta \lambda_{1\ms}$, $\delta \lambda_{2\text{div}}=\delta \lambda_{2\ms}$ etc.


Since the bare parameters are independent of the renormalization scheme, we can immediately write down the relations between the renormalized parameters in the on-shell and $\overline{\text{MS}}$ schemes as the following
\bqa
\lambda_{2\ms}&=&\lambda_2+\delta \lambda_{2\os}-\delta \lambda_{2\ms}\;,
\eqa
\bqa
c_{\ms}&=&c+\delta c_{\os}-\delta c_{\ms}\;,
\eqa
\bqa
\lambda_{1\ms}&=&\lambda_1+\delta \lambda_{1\os}-\delta \lambda_{1\ms}\;,
\eqa
\bqa
m^2_{\ms}&=&m^2+\delta m^2_{\os}-\delta m^2_{\ms}\;,
\eqa
\bqa
h_x{\ms}&=&h_{x}+\delta h_x{\os}-\delta h_x{\ms}\;,
\eqa
\bqa
h_y{\ms}&=&h_{y}+\delta h_y{\os}-\delta h_y{\ms}\;,
\eqa
\bqa
g^2_{\ms}&=&g^2+\delta g^2_{\os}-\delta g^2_{\ms}\;,
\eqa
\bqa
x^2_{\ms}&=&x^2+\delta x^2_{\os}-\delta x^2_{\ms}\;,
\eqa
\bqa
y^2_{\ms}&=&y^2+\delta y^2_{\os}-\delta y^2_{\ms}\;.
\eqa

The minimum of the vacuum effective potential is at $\overline{x}=f_\pi$ and $\overline{y}=\frac{(2f_{K}-f_\pi)}{\sqrt{2}}$. Using the above set of equations together with the Eqs.~(\ref{lam2os}), (\ref{cos}), (\ref{lam1os}), (\ref{m2os}), (\ref{hxos}), (\ref{hyos}) and (\ref{zpi})--(\ref{y2os}), one can write the scale $\Lambda$ dependent running parameters in the $\overline{\text{MS}}$ scheme as the following

\bqa
\label{params1}
\lambda_{2\ms}(\Lambda)
&=&\lambda_2+\lambda_{2\text{\tiny{FIN}}}\;,
\eqa
\bqa
c_{\ms}(\Lambda)
&=&c+c_{\text{\tiny{FINTOT}}}\;,
\eqa
\bqa
\lambda_{1\ms}(\Lambda)
&=&\lambda_1+\lambda_{2\text{\tiny{FIN}}}\;,
\eqa
\bqa
m^2_{\ms}(\Lambda)
&=&m^2+m^2_{\text{\tiny{FIN}}}\;,
\eqa
\bqa
h_x{\ms}(\Lambda)&=&h_{x}+h_{x\text{\tiny{FIN}}}\;,
\eqa
\bqa
\label{params5}
h_y{\ms}(\Lambda)&=&h_{y}+h_{y\text{\tiny{FIN}}}\;,
\eqa
\bqa
\label{params6}
g^2_{\ms}(\Lambda)
&=&g^2+\frac{N_cg^4}{(4\pi)^2} \text{SCF} \;,
\eqa
\bqa
\label{params7}
x^2_{\ms}(\Lambda)
&=&f_\pi^2-\dfrac{4N_cm^2_u}{(4\pi)^2} \text{SCF} \;,
\eqa
\bqa
\label{params8}
y^2_{\ms}(\Lambda)
&=&\left(\frac{2f_{K}-f_\pi}{\sqrt{2}}\right)^2-\dfrac{2N_cm^2_s}{(4\pi)^2} \text{SCF}\;.
\eqa
The parameters $\lambda_2$, $c$, $\lambda_1$, $m^2$, $h_x$, $h_y$ and $g^2$ in the Eqs.~(\ref{params1})--(\ref{params6}) and also in the earlier expressions, have the same tree level values of the QM model that one obtains after putting the $x=f_\pi$ and $y=\frac{(2f_{K}-f_\pi)}{\sqrt{2}}$ in the expressions  of the parameters described in the section \ref{subsec:paramfix}.

In the large-$N_c$ limit the parameters $\lambda_{2\ms}$, $c_{\ms}$, $\lambda_{1\ms}$, $m^2_{\ms}$, $h_{x\ms}$, $h_{y\ms}$ and $g^2_{\ms}$ are running with the scale $\Lambda$ and satisfy a set of the following simultaneous renormalization group equations 
\bqa
\label{diffpara1}
\dfrac{d\lambda_{2\ms}(\Lambda)}{d\ln(\Lambda)}&=&\dfrac{2N_c}{(4\pi)^2}\left[2\lambda_{2\ms}g^2_{\ms}-g^4_{\ms}\right]\;,
\eqa
\bqa
\dfrac{d c_{\ms}(\Lambda)}{d\ln(\Lambda)}&=&\dfrac{2N_c}{(4\pi)^2}g^2_{\ms}c_{\ms}\;,
\eqa
\bqa
\dfrac{d\lambda_{1\ms}(\Lambda)}{d\ln(\Lambda)}&=&\dfrac{4N_c}{(4\pi)^2}g^2_{\ms}\lambda_{1\ms}\;,
\eqa
\bqa
\dfrac{dm^2_{\ms}(\Lambda)}{d\ln(\Lambda)}&=&\dfrac{2N_c}{(4\pi)^2}g^2_{\ms}m^2_{\ms}\;,
\eqa
\bqa
\dfrac{d h_{x\ms}(\Lambda)}{d\ln(\Lambda)}&=&\dfrac{N_c}{(4\pi)^2}g^2_{\ms}h_{x\ms}\;,
\eqa
\bqa
\dfrac{d h_{y\ms}(\Lambda)}{d\ln(\Lambda)}&=&\dfrac{N_c}{(4\pi)^2}g^2_{\ms}h_{y\ms}\;,
\eqa
\bqa
\dfrac{d g^2_{\ms}}{d\ln(\Lambda)}&=&\dfrac{2N_c}{(4\pi)^2}g^4_{\ms}\;,
\eqa
\bqa
\label{diffpara7}
\dfrac{d x^2_{\ms}}{d\ln(\Lambda)}&=&-\dfrac{2N_c}{(4\pi)^2}g^2_{\ms}x^2_{\ms}\;,
\eqa
\bqa
\label{diffpara8}
\dfrac{d y^2_{\ms}}{d\ln(\Lambda)}&=&-\dfrac{2N_c}{(4\pi)^2}g^2_{\ms}y^2_{\ms}\;.
\eqa

Solving the differential the Eqs.~(\ref{diffpara1})--(\ref{diffpara8}), we get the following solutions

\bqa
\label{para01}
{\hskip -0.5 cm}\lambda_{2\ms}(\Lambda)&=&\frac{\lambda_{20}-\dfrac{N_c g^4_0}{(4\pi)^2}\ln\left(\dfrac{\Lambda^2}{\Lambda^2_0}\right)}{\left(1-\dfrac{N_c g^2_0}{(4\pi)^2}\ln\left(\dfrac{\Lambda^2}{\Lambda^2_0}\right)\right)^2}\;,  
\\
{\hskip -0.5 cm}c_{\ms}(\Lambda)&=&\frac{c_0}{\sqrt{\left[1-\dfrac{N_c g^2_0}{(4\pi)^2}\ln\left(\dfrac{\Lambda^2}{\Lambda^2_0}\right)\right]^3}}\;, \\
{\hskip -0.5 cm}\lambda_{1\ms}(\Lambda)&=&\frac{\lambda_{10}}{\left(1-\dfrac{N_c g^2_0}{(4\pi)^2}\ln\left(\dfrac{\Lambda^2}{\Lambda_0^2}\right)\right)^2}\;,
\eqa
\bqa
m^2_{\ms}(\Lambda)&=&\frac{m^2_0}{1-\dfrac{N_c g^2_0}{(4\pi)^2}\ln\left(\dfrac{\Lambda^2}{\Lambda^2_0}\right)}\;,
\\
\label{para05}
{\hskip -0.5 cm}h_{x\ms}(\Lambda)&=&\frac{h_{x0}}{\sqrt{1-\dfrac{N_c g^2_0}{(4\pi)^2}\ln\left(\dfrac{\Lambda^2}{\Lambda^2_0}\right)}}\;, \\ 
\label{para06}
{\hskip -0.5 cm}h_{y\ms}(\Lambda)&=&\frac{h_{y0}}{\sqrt{1-\dfrac{N_c g^2_0}{(4\pi)^2}\ln\left(\dfrac{\Lambda^2}{\Lambda^2_0}\right)}}\;, \\
\label{para07}
{\hskip -0.5 cm}g^2_{\ms}(\Lambda)&=&\frac{g^2_0}{1-\dfrac{N_c g^2_0}{(4\pi)^2}\ln\left(\dfrac{\Lambda^2}{\Lambda^2_0}\right)}\;,
\\
\label{para08}
{\hskip -0.5 cm}x^2&=&f^2_\pi\left[1-\frac{N_cg^2_0}{(4\pi)^2}\ln\left(\frac{\Lambda^2}{\Lambda^2_0}\right)\right]\;, \\
\label{para09}
{\hskip -0.5 cm}y^2&=&\frac{(2f_{K}-f_\pi)^{2}}{2}\left[1-\frac{N_cg^2_0}{(4\pi)^2}\ln\left(\frac{\Lambda^2}{\Lambda^2_0}\right)\right]\;.
\eqa
Where the parameters $\lambda_{10}$,\ $\lambda_{20}$,\ $g^2_0$,\ $m^2_0$,\ $c_0$, $h_{x0}$ and $h_{y0}$ are the running parameter values at the scale $\Lambda_0$. We can choose the $\Lambda_0$ to satisfy the following relation
\bqa
\ln\left(\frac{\Lambda^2_0}{m_u^2}\right)+\mathcal{C}(m^2_\pi)+m^2_\pi \mathcal{C}^{\prime}(m^2_\pi)&=&0\;.
\eqa
Now, we can calculate the parameters of the Eqs.~(\ref{params1})--(\ref{params8}) at the scale $\Lambda=\Lambda_0$ and find $\lambda_{10}$,\ $\lambda_{20}$,\ $g^2_0$,\ $m^2_0$,\ $c_0$, $h_{x0}$ and $h_{y0}$.
\\

\subsection{Effective Potential}
\label{subsecIVC}
Using the values of the parameters from the Eqs.~(\ref{para01})--(\ref{para07}), the vacuum effective potential in the $\overline{\text{MS}}$ scheme can be written as
\bqa
\label{omegarqm}
\Omega_{vac}&=&U(x_{\ms},y_{\ms})+\Omega^{q,vac}_{\ms}+\delta U(x_{\ms},y_{\ms})\;,
\eqa
where
\bqa
\label{omegams1}
\nonumber
U(x_{\ms},y_{\ms})&=&\frac{m_{\ms}^{2}}{2}\left(x_{\ms}^{2} +
  y_{\ms}^{2}\right) -h_{x \ms} \ x_{\ms} \\ \nonumber
&&-h_{y \ms} y_{\ms} 
- \frac{c_{\ms}}{2 \sqrt{2}} x_{\ms}^2 y_{\ms} 
  + \frac{\lambda_{1\ms}}{2} x_{\ms}^{2} y_{\ms}^{2} \\ \nonumber
&&+\frac{2 \lambda_{1\ms}+\lambda_{2\ms}}{8} \ x_{\ms}^{4}  
  +\frac{2 \lambda_{1\ms}+2\lambda_{2\ms}}{8} \ y_{\ms}^{4}\;, \\ 
\eqa
\bqa
\nonumber
\delta U(x_{\ms},y_{\ms})&=&\frac{\delta m_{\ms}^{2}}{2}\left(x_{\ms}^{2}+y_{\ms}^{2}\right)+\frac{m_{\ms}^{2}}{2}\left(
\delta x_{\ms}^{2} + \delta y_{\ms}^{2}\right)-\delta h_{x \ms} \ x_{\ms}
-h_{x \ms} \ \delta x_{\ms}-\delta h_{y \ms} \ y_{\ms}  \\ \nonumber
&&- h_{y \ms} \ \delta  y_{\ms}-\frac{\delta c_{\ms}}{2 \sqrt{2}} \ x_{\ms}^2 \ y_{\ms}- \frac{c_{\ms}}{2 \sqrt{2}} \ (\delta x_{\ms}^2  \ y_{\ms} +  x_{\ms}^2  \ \delta y_{\ms}) 
+ \frac{\delta \lambda_{1\ms}}{2} x_{\ms}^{2} \ y_{\ms}^{2} \\ \nonumber
&&+ \frac{\lambda_{1\ms}}{2} (\delta x_{\ms}^{2}y_{\ms}^{2}+x_{\ms}^{2} \ \delta y_{\ms}^{2})+(\frac{2 \delta \lambda_{1\ms}+\delta \lambda_{2\ms}}{8}) \ x_{\ms}^{4}
+(\frac{2 \lambda_{1\ms}+\lambda_{2\ms}}{8}) \ \delta x_{\ms}^{4} \\ \nonumber
 && +(\frac{2 \delta \lambda_{1\ms}+2 \delta \lambda_{2\ms}}{8}) \ y_{\ms}^{4} + (\frac{2 \lambda_{1\ms}+2\lambda_{2\ms}}{8}) \ \delta y_{\ms}^{4}\;, \\ 
\eqa
the $\mathcal{O}(N^2_c)$ terms are dropped as these are two-loop terms and one gets
\bqa
\delta U(x_{\ms},y_{\ms})&=&-\frac{N_cg^4_{\ms} (x^4_{\ms}+2y^4_{\ms})}{8(4\pi)^2}\frac{1}{\epsilon}=-\frac{N_c(2 \Delta_{x}^4+\Delta_{y}^4)}{(4\pi)^2}\frac{1}{\epsilon}\;.
\eqa
The quark one-loop  vacuum correction for the two nonstrange and one strange flavor is written as, 
\bqa
\nonumber
\label{omegavac}
\Omega^{q,vac}_{\ms}&=&\frac{N_cg^4_{\ms} x^4_{\ms}}{8(4\pi)^2}\left[\frac{1}{\epsilon}+\frac{3}{2}+\ln\left(\frac{4\Lambda^2}{g^2_{\ms}x^2_{\ms}}\right)\right]+\frac{N_cg^4_{\ms} 2 y^4_{\ms}}{8(4\pi)^2}\left[\frac{1}{\epsilon}+\frac{3}{2}+\ln\left(\frac{2\Lambda^2}{g^2_{\ms}y^2_{\ms}}\right)\right] \\ \nonumber \\
&=&\frac{2N_c\Delta_{x}^4}{(4\pi)^2}\left[\frac{1}{\epsilon}+\frac{3}{2}+\ln\left(\frac{\Lambda^2}{\Delta_{x}^2}\right)\right]+\frac{N_c\Delta_{y}^4}{(4\pi)^2}\left[\frac{1}{\epsilon}+\frac{3}{2}+\ln\left(\frac{\Lambda^2}{\Delta_{y}^2}\right)\right]\;.
\eqa
One can define the scale $\Lambda$ independent parameters $\Delta_{x}=\frac{g_{\ms} \ x_{\ms}}{2}$ and $\Delta_{y}=\frac{g_{\ms} \ y_{\ms}}{\sqrt{2}}$ using the Eqs.~(\ref{params6}), (\ref{params7}) and (\ref{params8}). It is instructive to write the Eq.~(\ref{omegams1}) in terms of the scale independent $\Delta_{x}$ and $\Delta_{y}$ as \\ 
\bqa
\nonumber
&&U(\Delta_{x},\Delta_{y})=\frac{m_{\ms}^2(\Lambda)}{g^2_{\ms}(\Lambda)}(2\Delta_{x}^2+\Delta_{y}^2)-2\frac{h_{x\ms}(\Lambda)}{g_{\ms}(\Lambda)}\Delta_{x}-\sqrt{2}\frac{h_{y\ms}(\Lambda)}{g_{\ms}(\Lambda)}\Delta_{y}-2\frac{c_{\ms}(\Lambda)}{g^3_{\ms}(\Lambda)}\Delta_{x}^2 \ \Delta_{y}+4\frac{\lambda_{1\ms}(\Lambda)}{g^4_{\ms}(\Lambda)}\Delta_{x}^2 \ \Delta_{y}^2 \\ 
&&{\hskip 2 cm}+2\frac{(2 \lambda_{1\ms}+\lambda_{2\ms})}{g^4_{\ms}(\Lambda)} \ \Delta_{x}^{4}+\frac{( \lambda_{1\ms}+\lambda_{2\ms})}{g^4_{\ms}(\Lambda)} \ \Delta_{y}^{4}\;,  \\ \nonumber
&&U(\Delta_{x},\Delta_{y})=\frac{m^2_0}{g^2_0}(2\Delta_{x}^2+\Delta_{y}^2)-2\frac{h_{x0}}{g_0}\Delta_{x}-\sqrt{2}\frac{h_{y0}}{g_0}\Delta_{y}-2\frac{c_{0}}{g^3_{0}}\Delta_{x}^2 \ \Delta_{y}+4\frac{\lambda_{10}}{g^4_{0}}\Delta_{x}^2 \ \Delta_{y}^2 \ \\ 
&&{\hskip 2 cm}+2\frac{(2 \lambda_{10}+\lambda_{20})}{g^4_{0}} \ \Delta_{x}^{4}+\frac{( \lambda_{10}+\lambda_{20})}{g^4_{0}} \ \Delta_{y}^{4}\;, 
\\ \nonumber
&&\Omega_{vac}(\Delta_{x},\Delta_{y})=\frac{m^2_0}{g^2_0}(2\Delta_{x}^2+\Delta_{y}^2)-2\frac{h_{x0}}{g_0}\Delta_{x}-\sqrt{2}\frac{h_{y0}}{g_0}\Delta_{y}-2\frac{c_{0}}{g^3_{0}}\Delta_{x}^2 \ \Delta_{y}+4\frac{\lambda_{10}}{g^4_{0}}\Delta_{x}^2 \ \Delta_{y}^2+2\frac{(2 \lambda_{10}+\lambda_{20})}{g^4_{0}}  \Delta_{x}^{4} \ \\ 
&&{\hskip 2 cm}+\frac{(\lambda_{10}+\lambda_{20})}{g^4_{0}}  \Delta_{y}^{4}+
\frac{2N_c\Delta_{x}^4}{(4\pi)^2}\left[\frac{3}{2}+\ln\left(\frac{\Lambda^2}{\Delta_{x}^2}\right)\right]+\frac{N_c\Delta_{y}^4}{(4\pi)^2}\left[\frac{3}{2}+\ln\left(\frac{\Lambda^2}{\Delta_{y}^2}\right)\right]\;.
\eqa
When the couplings and mass parameter are expressed in terms of the physical meson masses, pion decay constant, kaon decay constant and Yukawa coupling, one can write
\bqa
\label{rqmeff}
\nonumber
\Omega_{vac}(\Delta_{x},\Delta_{y})&=&\frac{(m^2+                                        
m^2_{\text{\tiny{FIN}}})}{2} \left\lbrace f^2_{\pi}\left(\frac{\Delta_{x}^2}{m^2_u}\right)+\frac{(2f_{K}-f_{\pi})^2}{2}\left(\frac{\Delta_{y}^2}{m^2_s}\right)\right\rbrace-(h_{x}+h_{x\text{\tiny{FIN}}})f_{\pi}\left(\frac{\Delta_{x}}{m_u}\right)  \\ \nonumber
&& -(h_{y}+h_{y\text{\tiny{FIN}}})\frac{(2f_{K}-f_{\pi})}{\sqrt{2}} \left(\frac{\Delta_{y}}{m_s}\right)-\frac{(c+c_{\text{\tiny{FINTOT}}})}{4}f^2_{\pi}(2f_{K}-f_{\pi})\left(\frac{\Delta_{x}^2}{m^2_u}\right)\left(\frac{\Delta_{y}}{m_s}\right)\\ \nonumber
&&+\frac{(\lambda_{1}+\lambda_{1\text{\tiny{FIN}}})}{4} \
f^2_{\pi} (2f_{K}-f_{\pi})^2\left(\frac{\Delta_{x}^2}{m^2_u}\right) 
\left(\frac{\Delta_{y}^2}{m^2_s}\right)+\frac{\lbrace 2(\lambda_{1}+\lambda_{1\text{\tiny{FIN}}})+(\lambda_{2}+\lambda_{2\text{\tiny{FIN}}})\rbrace}{8}f^4_\pi\left(\frac{\Delta_{x}^4}{m^4_u}\right)\\   \nonumber
&&+\frac{\lbrace (\lambda_{1}+\lambda_{1\text{\tiny{FIN}}})+(\lambda_{2}+\lambda_{2\text{\tiny{FIN}}})\rbrace}{16}(2f_{K}-f_{\pi})^4\left(\frac{\Delta_{y}^4}{m^4_s}\right) +\frac{2N_c\Delta_{x}^4}{(4\pi)^2} \\ \nonumber
&&\left[\frac{3}{2}-\ln\left(\frac{\Delta_{x}^2}{m^2_{u}}\right)-\mathcal{C}(m^2_\pi)-m^2_\pi \mathcal{C}^{\prime}(m^2_\pi)\right]+\frac{N_c\Delta_{y}^4}{(4\pi)^2}\left[\frac{3}{2}-\ln\left(\frac{\Delta_{y}^2}{m^2_{u}}\right)-\mathcal{C}(m^2_\pi)-m^2_\pi \mathcal{C}^{\prime}(m^2_\pi)\right]\;. \\
\eqa

It is to be noted that the pion decay constant, kaon decay constant and Yukawa coupling get renormalized in the vacuum because of the dressing of the meson propagator in the
on-shell scheme of the RQM model. But the  Eqs.~(\ref{params6}), (\ref{params7}) and (\ref{params8}) at the scale $\Lambda_0$ give us $g_{\ms}=g_{ren}=g$, $x_{\ms}=f_{\pi,ren}=f_\pi$ and $y_{\ms}=\frac{2f_{K,ren}-f_{\pi,ren}}{\sqrt{2}}=\frac{2f_K-f_\pi}{\sqrt{2}}$. Applying the stationarity condition $\frac{\partial \Omega_{vac}(\Delta_{x},\Delta_{y})}{\partial \Delta_{x}}=0$ to the Eq.~(\ref{rqmeff}) in  the nonstrange  direction, one gets  $h_{x0}=m_{\pi,c}^2 \ x_{\ms} =m^2_\pi \left\lbrace 1-\frac{N_cg^2}{(4\pi)^2}m^2_\pi\mathcal{C}^\prime(m^2_\pi)\right\rbrace f_\pi$.~Thus the pion curvature mass $ m_{\pi,c}^2 =m^2_\pi \left\lbrace 1-\frac{N_cg^2}{(4\pi)^2}m^2_\pi\mathcal{C}^\prime(m^2_\pi)\right\rbrace$. The stationarity condition $\frac{\partial \Omega_{vac}(\Delta_{x},\Delta_{y})}{\partial \Delta_{y}}=0$ in the strange direction, gives $\quad h_{0y}=\left({\frac{x_{\ms}}{\sqrt{2}}+y_{\ms}}\right)m^2_{K,c}-\frac{x_{\ms}}{\sqrt{2}}m^2_{\pi,c}$=$\sqrt{2}f_Km^2_{K,c}-\frac{f_{\pi}}{\sqrt{2}}m^2_{\pi,c}$. Using the expression of 
$h_{y\ms}(\Lambda_0)=h_{y0}$ in the Eq.~(\ref{params5}), one gets the expression of kaon curvature mass $m_{K,c}^2$ as written below in the Eq.~(\ref{mkcurve}). It is pointed out that the pion curvature mass $ m_{\pi,c}$ (as  in Ref.~\cite{fix1})  and the kaon curvature mass are different from their pole masses $m_{\pi} $ and $m_{K}$ due to the consistent on-shell parameter fixing. The minimum of the effective potential remains fixed at $x_{\ms}=f_\pi$ and 
$y_{\ms}=\frac{(2f_K-f_\pi)}{\sqrt{2}}$.

\begin{table*}[!htbp]
    \caption{Parameters of the different model scenarios. The RQM model parameters are obtained by putting the $\Lambda=\Lambda_0$ in the Eqs.~(\ref{params1})--(\ref{params5}).}
    \label{tab:table2}
    \begin{tabular}{p{0.123\textwidth} p{0.123\textwidth}  p{0.123\textwidth} p{0.123\textwidth} p{0.123\textwidth} p{0.123\textwidth} p{0.123\textwidth} p{0.123\textwidth} }
   \hline 
      Model&$m_{\sigma}(\text{MeV})$&$\lambda_2$&$c(\text{MeV}^2)$ &$\lambda_1$&$m^2(\text{MeV}^2)$& $h_x(\text{MeV}^3)$ & $h_y(\text{MeV}^3)$\\
      \hline 
      \hline
      &$400$&46.43 &4801.82 &-5.89 &$(494.549)^2$ &$(120.73)^3$ &$(336.43)^3$ \\
      &$500$&46.43 &4801.82 &-2.69 &$(434.305)^2$ &$(120.73)^3$ &$(336.43)^3$\\
      QM&$600$&46.43&4801.82 &1.141 &$(342.139)^2 $ &$(120.73)^3$ &$(336.43)^3$\\
      &$648$&46.43&4801.82 &3.75 &$(275.92)^2$&$(120.73)^3$ &$(336.43)^3$\\
      &$700$&46.43&4801.82 &6.63 &$(160.93)^2$&$(120.73)^3$ &$(336.43)^3$\\ \hline
      &$400$&46.43 &4801.82 &-8.17 &$(282.338)^2$ &$(120.73)^3$ &$(336.43)^3$ \\
      &$500$&46.43 &4801.82 &-5.28 &$(171.167)^2$&$(120.73)^3$ &$(336.43)^3$\\
      QMVT&$600$&46.43 &4801.82 &-1.66 &$-(184.28)^2 $&$(120.73)^3$ &$(336.43)^3$\\
      &$648$&46.43 &4801.82 &0.369 &$-(263.494)^2 $&$(120.73)^3$ &$(336.43)^3$\\
      &$700$&46.43 &4801.82 &2.82 &$-(334.918)^2$ &$(120.73)^3$ &$(336.43)^3$\\ \hline
      &$400$&34.88 &7269.20 &1.45 &$(442.447)^2 $&$(119.53)^3$ &$(323.32)^3$ \\
      &$500$&34.88 & 7269.20&3.676 &$(396.075)^2 $&$(119.53)^3$ &$(323.32)^3$\\
      RQM&$600$&34.88 &7269.20 &8.890 & $(256.506)^2$&$(119.53)^3$ &$(323.32)^3$\\
      &$648$&34.88 &7269.20 &13.905 &$-(147.619)^2$ &$(119.53)^3$ &$(323.32)^3$\\
      &$700$&34.88 & 7269.20&19.23 &$-(338.906)^2$ & $(119.53)^3$ &$(323.32)^3$\\
      \hline 
      \hline
    \end{tabular}
\end{table*}
\begin{equation}
\label{mkcurve}
\resizebox{1.1\hsize}{!}{
$m^2_{K,c}=m^2_K\left[1-\frac{N_cg^2}{(4\pi)^2}\left\{\mathcal{C}(m^2_\pi,m_u)+m^2_\pi\mathcal{C}^{\prime}(m^2_\pi,m_u)-\left(1-\frac{(m_s-m_u)^2}{m_K^2}\right)\mathcal{C}(m^2_K,m_u,m_s)+\left(1-\frac{f_\pi}{f_K}\right)\frac{m^2_u-m_s^2+2m_s^2\ln\left(\frac{m_s}{m_u}\right)}{m^2_K}\right\}\right]$}\;.
\end{equation}
The grand potential of the RQM model is written as,
\bqa
\nonumber
\label{rqmomega}
\Omega_{RQM}(\Delta_{x},\Delta_{y},T,\mu)&=&\frac{(m^2+                                        
m^2_{\text{\tiny{FIN}}})}{2} \left\lbrace f^2_{\pi}\left(\frac{\Delta_{x}^2}{m^2_u}\right)+\frac{(2f_{K}-f_{\pi})^2}{2}\left(\frac{\Delta_{y}^2}{m^2_s}\right)\right\rbrace-(h_{x}+h_{x\text{\tiny{FIN}}})f_{\pi}\left(\frac{\Delta_{x}}{m_u}\right)\; \\ \nonumber
&&{\hskip -2 cm}-(h_{y}+h_{y\text{\tiny{FIN}}})\frac{(2f_{K}-f_{\pi})}{\sqrt{2}} \left(\frac{\Delta_{y}}{m_s}\right)-\frac{(c+c_{\text{\tiny{FINTOT}}})}{4}f^2_{\pi}(2f_{K}-f_{\pi})\left(\frac{\Delta_{x}^2}{m^2_u}\right)\left(\frac{\Delta_{y}}{m_s}\right) \\ \nonumber
&&{\hskip -2 cm} +\frac{(\lambda_{1}+\lambda_{1\text{\tiny{FIN}}})}{4} 
f^2_{\pi} (2f_{K}-f_{\pi})^2 \left(\frac{\Delta_{x}^2}{m^2_u}\right) 
\left(\frac{\Delta_{y}^2}{m^2_s}\right)+\frac{\lbrace 2(\lambda_{1}+\lambda_{1\text{\tiny{FIN}}})+(\lambda_{2}+\lambda_{2\text{\tiny{FIN}}})\rbrace}{8}f^4_\pi\left(\frac{\Delta_{x}^4}{m^4_u}\right)\\   \nonumber
&&{\hskip -4 cm}+\frac{\lbrace (\lambda_{1}+\lambda_{1\text{\tiny{FIN}}})+(\lambda_{2}+\lambda_{2\text{\tiny{FIN}}})\rbrace}{16}(2f_{K}-f_{\pi})^4\left(\frac{\Delta_{y}^4}{m^4_s}\right)+\frac{2N_c\Delta_{x}^4}{(4\pi)^2}\left[\frac{3}{2}-\ln\left(\frac{\Delta_{x}^2}{m^2_{u}}\right)-\mathcal{C}(m^2_\pi)-m^2_\pi \mathcal{C}^{\prime}(m^2_\pi)\right] \\  \nonumber
&&{\hskip -2 cm} +\frac{N_c\Delta_{y}^4}{(4\pi)^2}\left[\frac{3}{2}-\ln\left(\frac{\Delta_{y}^2}{m^2_{u}}\right)-\mathcal{C}(m^2_\pi)-m^2_\pi \mathcal{C}^{\prime}(m^2_\pi)\right]\;-2N_cT \sum_{f=u,d,s} \\ 
&&\int \frac{d^3p}{(2\pi)^3}\left\lbrace \ln\left[1+e^{-E_{f}^{+}/T)}\right]+\ln\left[1+e^{-E_{f}^{-}/T)}\right]\right\rbrace.
\eqa
One gets the nonstrange condensate $\Delta_{x}$ and strange condensate $\Delta_{y}$ in the RQM model by searching the global minimum of the grand potential in the Eq.~(\ref{rqmomega}) for a given value of 
temperature T and chemical potential $\mu$

\begin{equation}
\frac{\partial \Omega_{RQM}(\Delta_{x},\Delta_{y},T,\mu)}{\partial
      \Delta_{x}}=\frac{\partial \Omega_{RQM}(\Delta_{x},\Delta_{y},T,\mu)}{\partial\Delta_{y}} =0
\label{EoMMF3}
\end {equation}

In our calculations, we have used the $m_\pi=138.0$ MeV, $m_K=496$ MeV.
Here in the RQM model, fixing the $m_{\eta}^2+m_{\eta^{\prime}}^2=(547.5)^2+(957.78)^2$ and then taking the $\eta$ mass as 527.58 MeV, one gets the ${\eta^{\prime}}$ mass equal to 968.89 MeV. The pole mass  $m_{\eta}=527.58$ MeV and $m_{\eta^{\prime}}=968.89$ MeV have been used for calculating the self energy corrections (for $\eta,\eta^{\prime}$) and fixing of the parameters in the on-shell scheme because it has been checked that when the masses are calculated with the new set of renormalized parameters and respective self energy corrections are added, the same pole masses are reproduced.  
\begin{table*}[!ht]
 \begin{tabular}{|c|c|c|c|c|c|c|}
\hline
$m_{\sigma}(\text{MeV})$&QM $\ T_{c}^{\chi}$ & RQM $\ T_{c}^{\chi}$ & QMVT $\  T_{c}^{\chi}$ & QM $\  T_{c}^{s}$ & RQM $\  T_{c}^{s}$ & QMVT $\ T_{c}^{s}$ \\ \hline
400 & 112.5 & 121.1  &  144.1  & 231.6 & 210.1  & 236.1 \\ \hline
500 & 129.0 & 133.6  &  156.8  & 238.6 & 213.3  & 241.1  \\ \hline
600 & 146.1 & 158.6  &  170.8  & 248.3 & 220.6  & 247.8   \\ \hline 
648 & 154.5 & 178.1  &  178.1  & 254.8 & 229.1  & 251.8    \\ \hline
700 & 163.9 & 195.8  &  186.3  & 261.7 & 240.3  & 256.7     \\ \hline
 \end{tabular}
\caption{Critical temperature in MeV for the nonstrange sector $\ T_{c}^{\chi}$ and the strange sector $\ T_{c}^{s}$ for the  $m_\sigma=$400, 500, 600, 648 and 700 MeV.}
\label{tab:Tc}
\end{table*}

\section{Results and Discussion}
\label{secV}

\begin{figure*}[htb]
\subfigure[\ Effective potential in Nonstrange direction for $m_\sigma=400 \ \text{MeV}$.]{
\label{fig:mini:fig5:a} 
\begin{minipage}[b]{0.30\textwidth}
\centering \includegraphics[width=\linewidth]{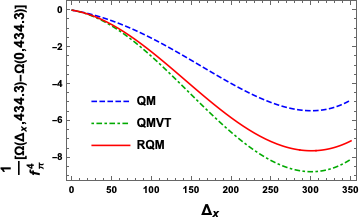}
\end{minipage}}%
\hfill
\subfigure[\ Nonstrange and strange order parameter  for $m_\sigma=400 \ \text{MeV}$.]{
\label{fig:mini:fig5:b} 
\begin{minipage}[b]{0.32\textwidth}
\centering \includegraphics[width=\linewidth]{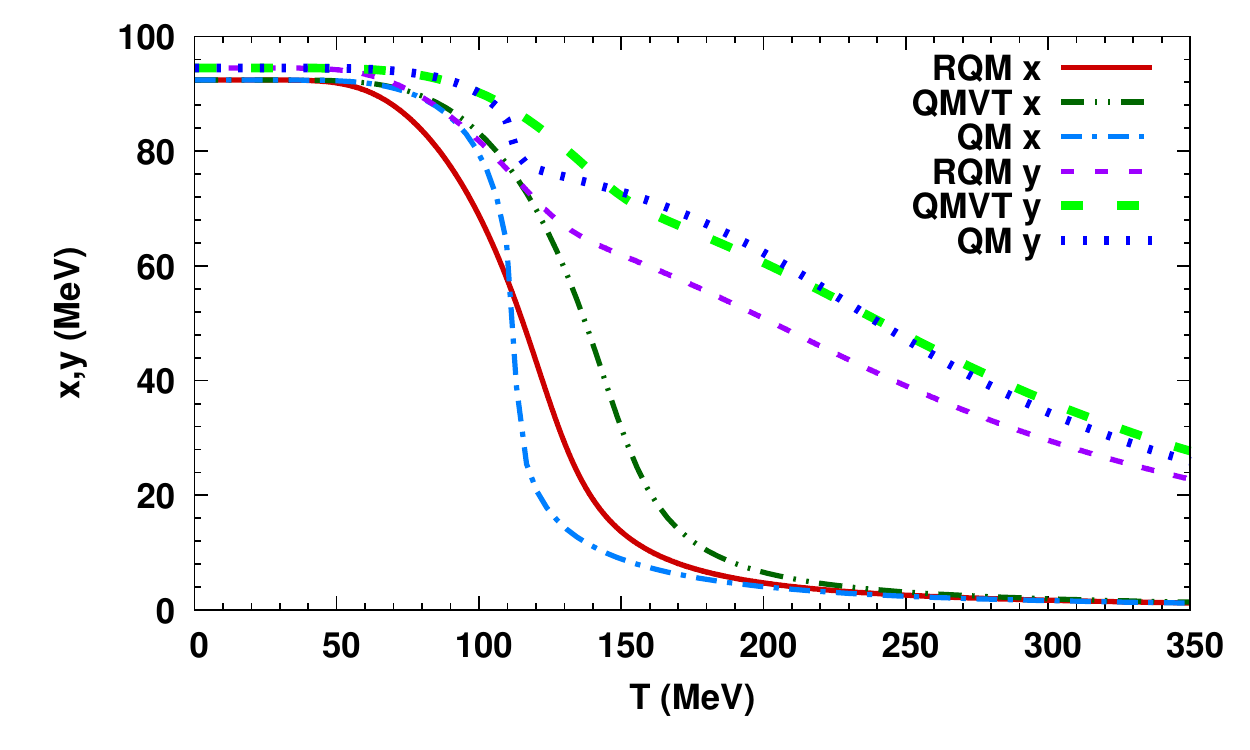}
\end{minipage}}
\hfill
\subfigure[\ Phase diagram for $m_\sigma=400 \ \text{MeV}$.]{
\label{fig:mini:fig5:c} 
\begin{minipage}[b]{0.32\textwidth}
\centering \includegraphics[width=\linewidth]{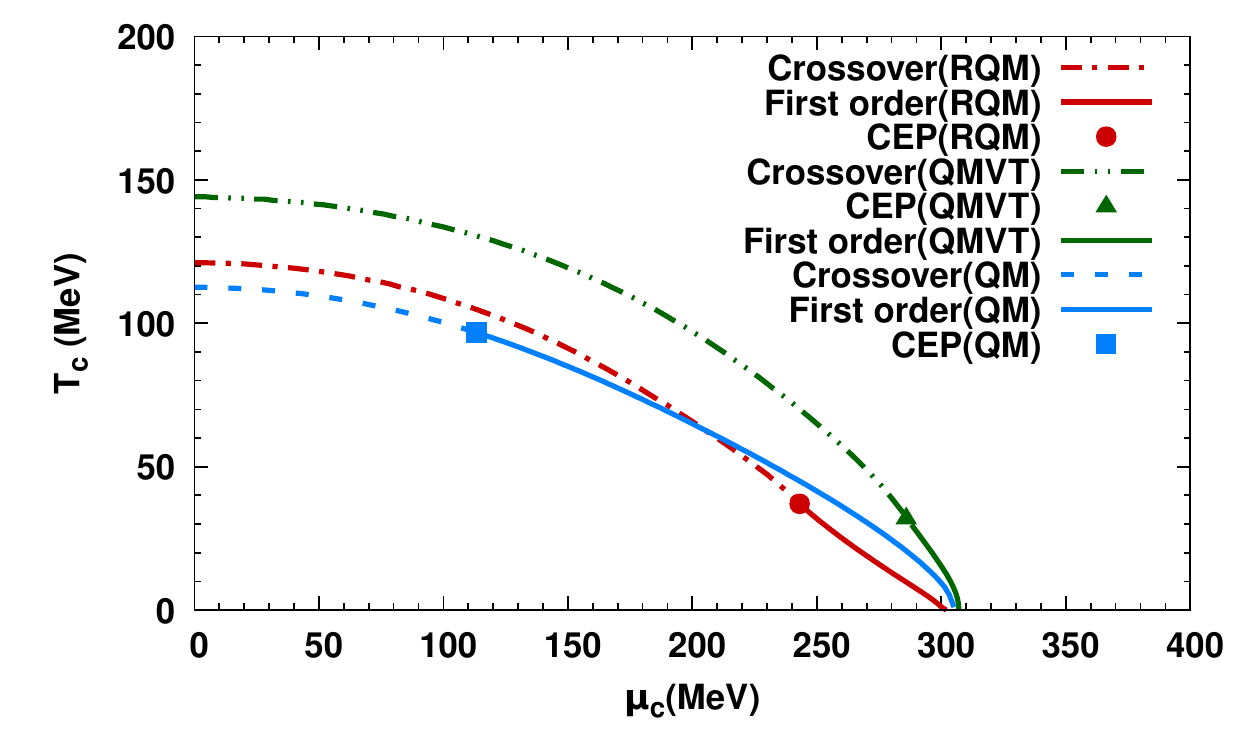}
\end{minipage}}
\caption{}
\label{fig:mini:fig5} 
\end{figure*} 


Parameters in the different model scenarios are presented in the Table~\ref{tab:table2}. The vacuum effective potential $\Omega$ is a function of the two variables $\Delta_{x}$ and $\Delta_{y}$. Its minimum is located at $\Delta_{x}=m_{u/d}=300.3$ MeV, $\Delta_{y}=m_{s}=433.34$ MeV in all the  2+1 flavor model scenarios, irrespective of the $\sigma$ meson mass values. In order to study the variation of the $\Omega$ in nonstrange direction, the $\Delta_{y}$ is fixed at 433.34 MeV and the normalized vacuum effective potential difference $\boldmath{\frac{\{ \Omega_{vac}(\Delta_{x}, 334.43 )-\Omega_{vac}(0,334.43 )\}}{f_{\pi}^4}}$ is plotted in Fig.(\ref{fig:mini:fig5:a}) with respect to the scale independent nonstrange quark mass parameter $\Delta_{x} $ for the $m_\sigma=$ 400 MeV. It is most shallow for the no-sea approximation of the QM model and deepest for the QMVT model while the on-shell parameter fixing of the RQM model when compared to the QMVT model, gives a shallower effective potential. The nonstrange and strange condensate temperature variations for the $\mu=0$, are presented in  Fig.(\ref{fig:mini:fig5:b}). Similar to the two flavor case \cite{RaiTiw}, the sharpest QM model temperature variation of the nonstrange quark condensate $x$ becomes quite smooth for the on-shell parameterization of the RQM model and the most smooth variation of the nonstrange condensate is witnessed for the QMVT model plot. The temperature derivative of the condensate in the nonstrange and strange direction when the $\mu$=0 MeV, defines the chiral crossover transition temperature  (called the pseudo-critical temperature) $T_{c}^{\chi}$ for the nonstrange direction and  $T_{c}^{s}$ for the strange direction. The early and sharpest crossover transition occurs  at a pseudo-critical temperature of $T_{c}^{\chi}= 112.5 $ MeV in the QM model and a smoother chiral transition is witnessed for the RQM model at $T_{c}^{\chi}=121.1 $ MeV while a most delayed and smooth chiral crossover occurs at $T_{c}^{\chi}=144.1 $ MeV in the QMVT model. The melting of the strange condensate $y$ is most significant in the RQM model when it is compared with its temperature variation in the QM and QMVT models. The Fig.(\ref{fig:mini:fig5:c}) depicts the phase diagram for the $m_\sigma=400$ MeV. The critical end point (CEP) location $\mu=$113.3 MeV, $T=$ 96.76 MeV of the QM model shifts significantly (similar to the results reported in earlier works \cite{guptiw,schafwag12,chatmoh1,vkkr12,chatmoh2}) to a far right position in the lower corner of the $\mu-T$ plane at $\mu=$285.91 MeV, $T=$32.23 MeV for the QMVT model setting. It is worthwhile to emphasize that due to the exact on-shell parameter fixing, the CEP in the RQM model moves to a higher position when compared with the QMVT model CEP and it gets located at a higher temperature $T=$37.03 MeV and a lower chemical potential $\mu=$ 243.12 MeV.

\begin{figure*}[htb]
\subfigure[\ $m_\sigma=500 \ \text{MeV}$.]{
\label{fig:mini:fig6:a} 
\begin{minipage}[b]{0.32\textwidth}
\centering \includegraphics[width=\linewidth]{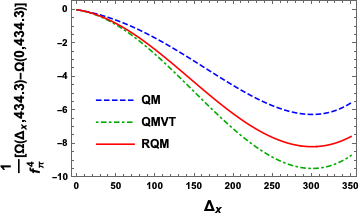}
\end{minipage}}%
\hfill
\subfigure[\ $m_\sigma=658.8 \ \text{MeV}$.]{
\label{fig:mini:fig6:b} 
\begin{minipage}[b]{0.32\textwidth}
\centering \includegraphics[width=\linewidth]{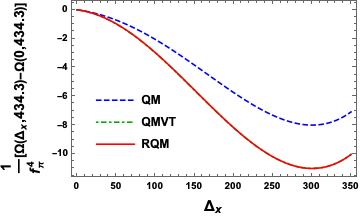}
\end{minipage}}
\hfill
\subfigure[ \ $m_\sigma=700 \ \text{MeV}$. ]{
\label{fig:mini:fig6:c} 
\begin{minipage}[b]{0.32\textwidth}
\centering \includegraphics[width=\linewidth]{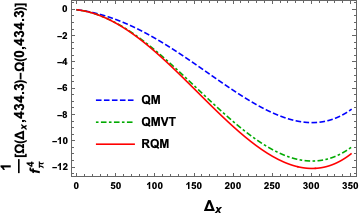}
\end{minipage}}
\caption{Normalized effective potential difference in the nonstrange direction for the QM,RQM and QMVT model.}
\label{fig:mini:fig6} 
\end{figure*}

In order to see the effect of the sigma meson mass on the nature of the vacuum effective potential, the respective plots of the normalized vacuum effective potential difference
$\boldmath{\frac{\{ \Omega_{vac}(\Delta_{x}, 334.43 )-\Omega_{vac}(0,334.43 )\}}{f_{\pi}^4}}$
with respect to the nonstrange quark mass parameter $\Delta_{x} $ for the $m_\sigma=$ 500 MeV, 658.8 MeV and 700 MeV, have been presented in the Figs.~(\ref{fig:mini:fig6:a}), (\ref{fig:mini:fig6:b}) and (\ref{fig:mini:fig6:c}). The effective potential difference is highest (i.e. it is most shallow) in the case of the QM model for every $m_\sigma$. Similar to the $m_\sigma=$ 400 MeV case, the effective potential is deepest for the QMVT model when the $m_\sigma=$ 500 MeV in the Fig.~(\ref{fig:mini:fig6:a}) while the on-shell parameterization in the RQM model gives a shallower effective potential in comparison. When the $m_\sigma$ value is increased, one notices that the effective potential difference in the RQM model becomes deeper and the QMVT model effective potential difference shows a rising trend and both the effective potential plots merge with each other for the $m_\sigma=658.8 \ \text{MeV}$, in the Fig.~(\ref{fig:mini:fig6:b}). As the $m_\sigma$ is increased beyond the 658.8 MeV, the plot of the effective potential difference turns deepest in the RQM model.The trend seen in the Fig~(\ref{fig:mini:fig6:a}) gets reversed in the plots of the Fig.~(\ref{fig:mini:fig6:c}) for the $m_\sigma=700 \text{MeV}$ case as the effective potential difference becomes shallower for the QMVT model and deepest for the RQM model. In our very recent work for the two flavor case \cite{RaiTiw}, we found similar trend reversal for the $m_\sigma>616$ MeV.

\begin{figure*}[htb]
\subfigure[\  $m_\sigma=500 \ \text{MeV}$.]{
\label{fig:mini:fig7:a} 
\begin{minipage}[b]{0.32\textwidth}
\centering \includegraphics[width=\linewidth]{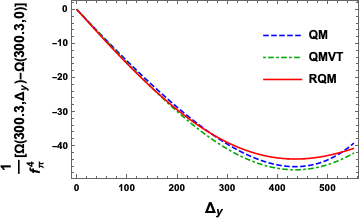}
\end{minipage}}%
\hfill
\subfigure[\  $m_\sigma=785 \ \text{MeV}$.]{
\label{fig:mini:fig7:b} 
\begin{minipage}[b]{0.32\textwidth}
\centering \includegraphics[width=\linewidth]{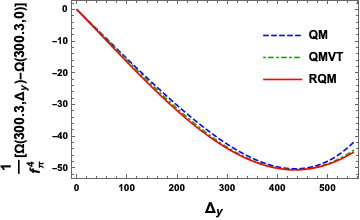}
\end{minipage}}
\hfill
\subfigure[\ $m_\sigma=850 \ \text{MeV}$. ]{
\label{fig:mini:fig7:c} 
\begin{minipage}[b]{0.32\textwidth}
\centering \includegraphics[width=\linewidth]{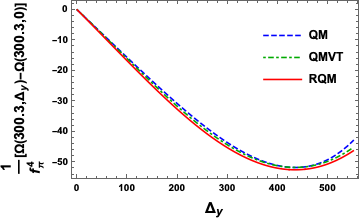}
\end{minipage}}
\caption{Normalized effective potential difference in the strange direction for the QM, RQM and QMVT model.}
\label{fig:mini:fig7} 
\end{figure*} 

The variation of the $\Omega$ in strange direction has also been investigated by fixing the $\Delta_{x}$ at 300.3 MeV. The normalized vacuum effective potential difference $\boldmath{\frac{\{ \Omega_{vac}(300.3,\Delta_{y})-\Omega_{vac}(300.3,0 )\}}{f_{\pi}^4}}$ is plotted with respect to the scale independent strange quark mass parameter $\Delta_{y} $ respectively in  the Fig.~(\ref{fig:mini:fig7:a}), (\ref{fig:mini:fig7:b}) and (\ref{fig:mini:fig7:c})  for the $m_\sigma=$ 500, 785 and 850 MeV. The strange direction effective potential difference is most shallow in the RQM model, deeper in the QM model and deepest in the QMVT model for the $m_\sigma=$ 500 MeV. It becomes deeper in the RQM model on increasing the $m_\sigma$ value and shows rising trend for the QMVT model. Both the effective potential plots nearly coincide with each other (similar to the nonstrange direction effective potential when the $m_\sigma=658.8 \ \text{MeV}$)  for the $m_\sigma=785 \ \text{MeV}$ while the QM model effective potential difference looks little shallow in comparison in the Fig.~(\ref{fig:mini:fig7:b}). The trend of plots in the Fig.~(\ref{fig:mini:fig7:a}) gets reversed in the Fig.~(\ref{fig:mini:fig7:c}) for the $m_\sigma=850 \ \text{MeV}$ case as the strange direction effective potential difference becomes shallower for the QMVT model and deepest for the RQM model. The trend reversal in the strange direction sets in for higher sigma mass when $m_{\sigma}>785$ MeV.

\begin{figure*}[htb]
\subfigure[\   $m_\sigma=500 \ \text{MeV}$.]{
\label{fig:mini:fig8:a} 
\begin{minipage}[b]{0.32\textwidth}
\centering \includegraphics[width=\linewidth]{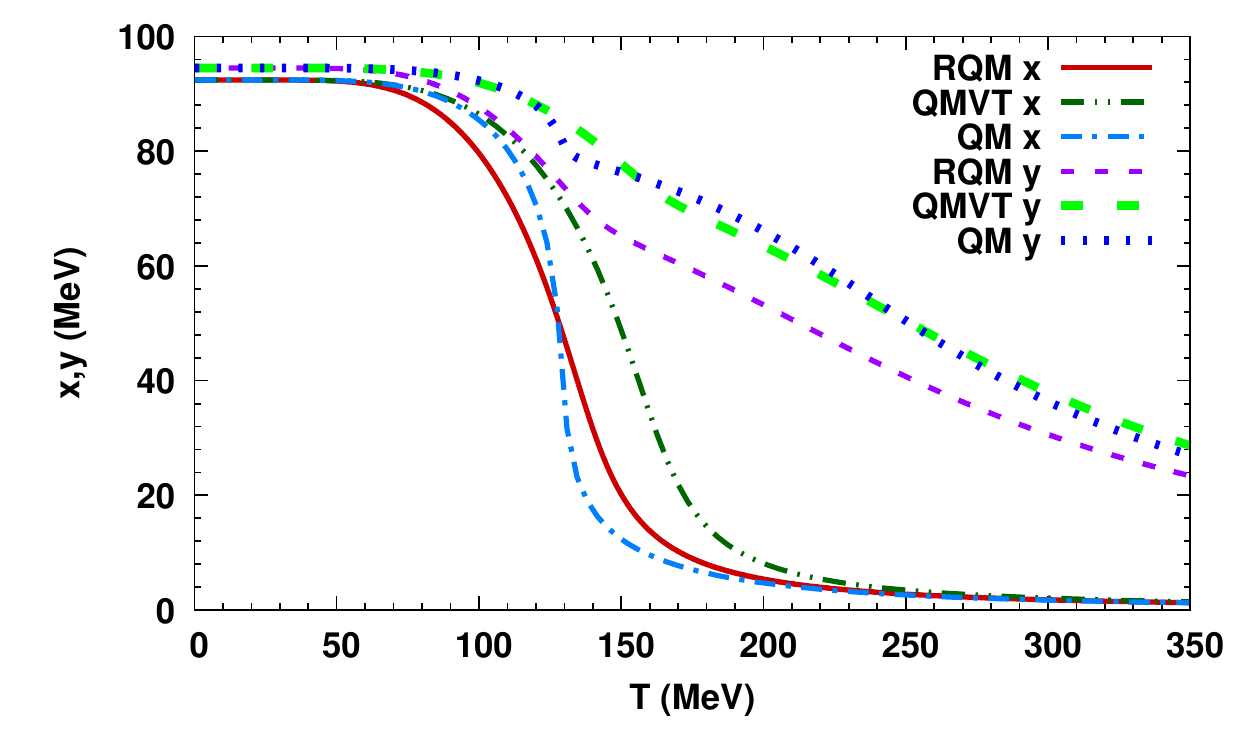}
\end{minipage}}%
\hfill
\subfigure[\  $m_\sigma=648 \ \text{MeV}$.]{
\label{fig:mini:fig8:b} 
\begin{minipage}[b]{0.32\textwidth}
\centering \includegraphics[width=\linewidth]{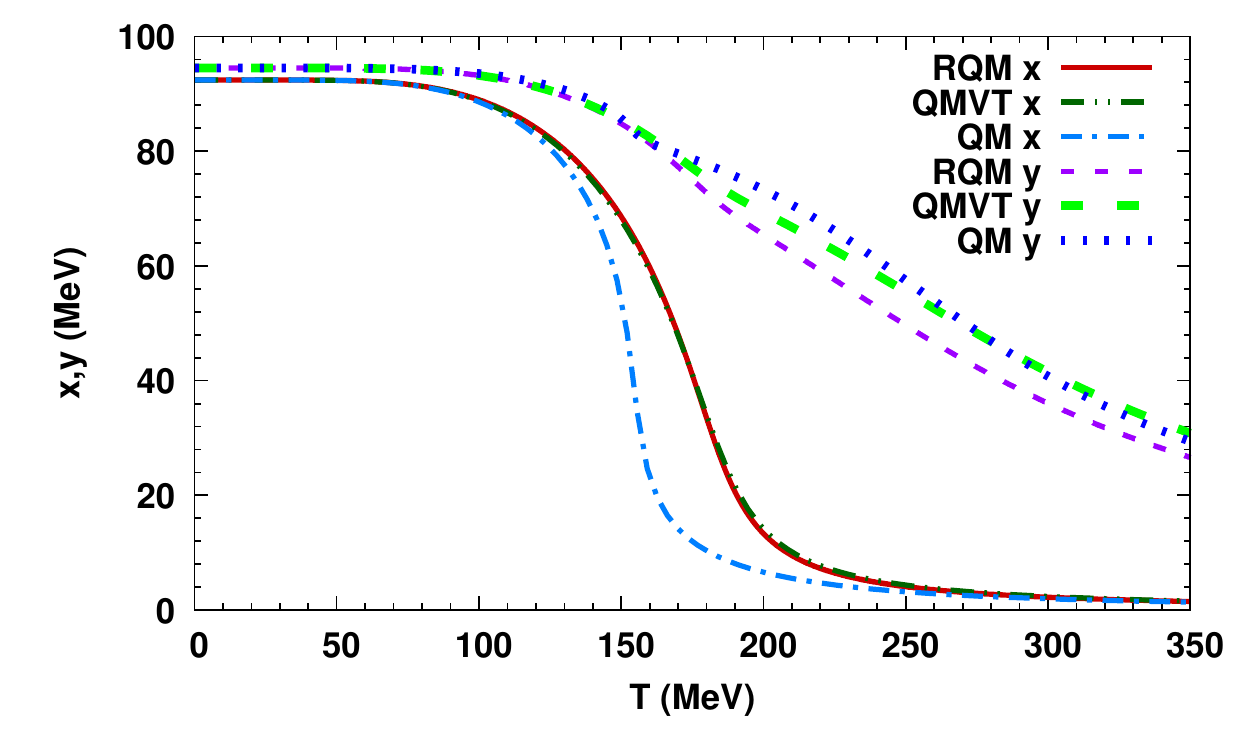}
\end{minipage}}
\hfill
\subfigure[\  $m_\sigma=700 \ \text{MeV}$.]{
\label{fig:mini:fig8:c} 
\begin{minipage}[b]{0.32\textwidth}
\centering \includegraphics[width=\linewidth]{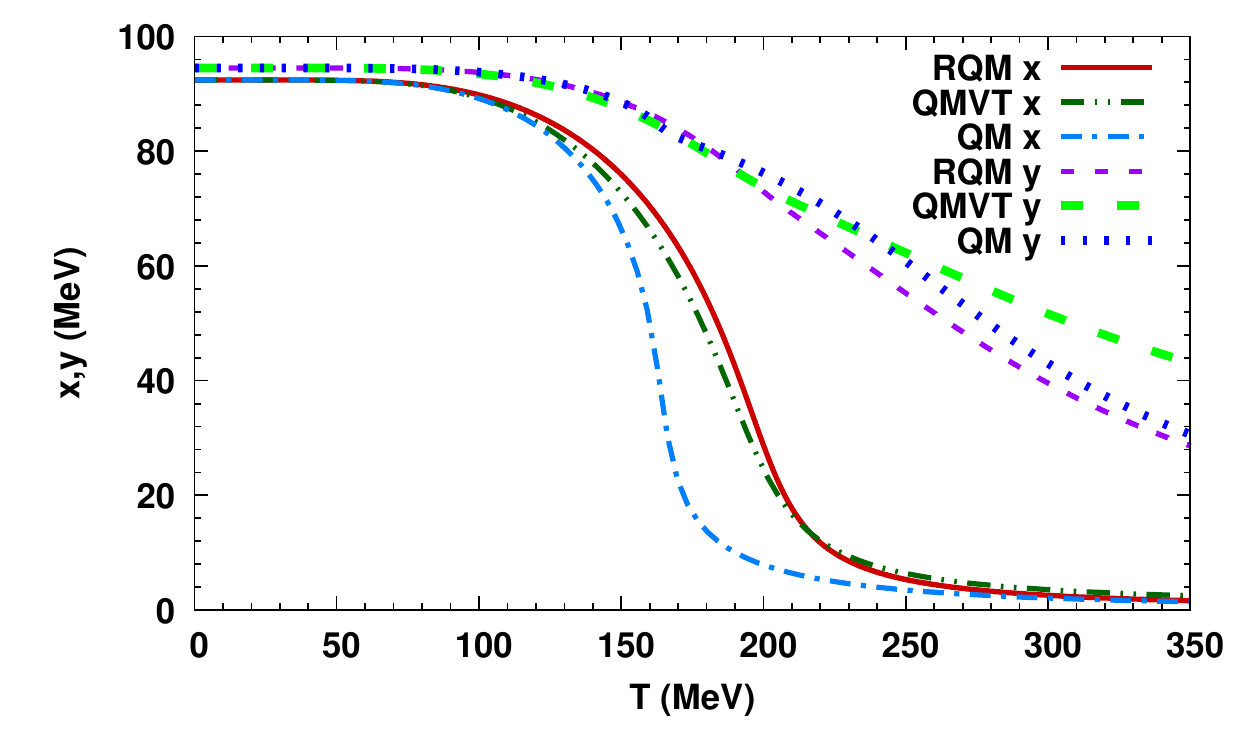}
\end{minipage}}
\caption{Temperature variation of the nonstrange and strange order parameter}
\label{fig:mini:fig8} 
\end{figure*}
The temperature variations of the nonstrange and strange quark condensates $x$ and $y$  (obtained from the $\Delta_{x}$ and $\Delta_{y} $ since the renormalization does not change the Yukawa coupling $g$) at the $\mu$=0 MeV, are plotted in Fig.~(\ref{fig:mini:fig8}) for the three values of $m_\sigma=$ 500, 648 and 700 MeV. The light blue dash dotted line presents the QM model while the solid red line depicts the RQM model and the dash double dotted deep green line shows the QMVT model plot for the nonstrange quark condensate $x$. The quark one-loop vacuum correction gives rise to a smoother chiral transition in general. The sharpest nonstrange condensate temperature variation of the QM model in Fig.~(\ref{fig:mini:fig8:a}) for the $m_\sigma=$ 500 MeV case, becomes smoother for the on-shell parameterization in the RQM model while the most smooth variation occurs in the QMVT model. The QM model nonstrange chiral crossover transition at $\mu=0$ MeV is sharpest and occurs early at a pseudo-critical temperature of $T_{c}^{\chi}=129.0$ MeV while a smoother chiral crossover for the RQM model occurs at $T_{c}=133.6$ MeV and the smoothest but delayed chiral crossover occurs in the QMVT model at $T_{c}^{\chi}=156.8$ MeV. The temperature variation of the strange quark condensate $y$ has been plotted by the thin dash line in purple for the RQM model, the dotted line in blue for the QM model and the thick dash line in green for the QMVT model.
 
 The RQM model temperature variation of the nonstrange quark condensate almost merges with the QMVT model result when the $m_{\sigma}=$ 648 MeV in the Fig.~(\ref{fig:mini:fig8:b}) and the $\mu$=0 chiral crossover   transition in the nonstrange sector occurs at $T_{c}^{\chi}=178.1$ MeV in both the models. This pattern is consequence of the coincidence of the RQM model plot of the normalized vacuum effective potential difference $\boldmath{\frac{\{ \Omega_{vac}(\Delta_{x}, 334.43 )-\Omega_{vac}(0,334.43 )\}}{f_{\pi}^4}}$ versus the $\Delta_{x}$ variation with the corresponding plot in the QMVT model in the Fig.~(\ref{fig:mini:fig6:b}) for the $m_\sigma=$ 658.8 MeV case. Here, it is relevant to point out that in our recent work \cite{RaiTiw} for the two flavor, we have shown that when the $m_\sigma=$ 616 MeV, the vacuum effective potential variation with respect to the constituent quark mass parameter ($ \Delta $) and the chiral order parameter temperature variation for the on-shell parameter fixing scheme in the RQM model, completely merge with the corresponding effective potential and the chiral order parameter variation computed with the curvature mass parameterization in the QMVT model. In the present 2+1 flavor work, since the vacuum effective potential depends on the nonstrange and strange constituent quark mass parameter $\Delta_{x}$ and $\Delta_{y}$ variations in the two independent directions, the $\Delta_{x}$ variation of the effective potential is  somewhat influenced by its $\Delta_{y}$ variation. Hence though the RQM model nonstrange direction $\Delta_{x}$ dependent variation of the vacuum effective potential difference $\boldmath{\frac{\{ \Omega_{vac}(\Delta_{x}, 334.43 )-\Omega_{vac}(0,334.43 )\}}{f_{\pi}^4}}$ coincides with the corresponding effective potential difference variation in the QMVT model for the $m_\sigma=$ 658.8 MeV, the consequential coincidence of the nonstrange order parameter temperature variations for both the model settings of the RQM and QMVT, occurs when the $\sigma$ mass is smaller by about 10 MeV i.e. $m_\sigma=$ 648 MeV. For the $m_\sigma=$ 648 MeV, the temperature variation of the strange condensate $y$ in the RQM model comes closer to the corresponding results in the QM model and QMVT models. The Fig.~(\ref{fig:mini:fig8:c}) when the $m_\sigma=$ 700 MeV, shows that the most smooth nonstrange quark condensate temperature variation takes place in the RQM model and the $\mu=0$  chiral crossover transition occurring  at  $T_{c}^{\chi}=195.8$ MeV, is very delayed while a less smooth chiral crossover transition is noticed in the QMVT model occurring earlier at $T_{c}^{\chi}=186.3$ MeV. Note that when the nonstrange quark condensate temperature variation in the RQM model is compared with the corresponding result in the QMVT model, one finds that the trend for the $m_\sigma=$ 700 MeV turns opposite of what one observes in the Fig.~(\ref{fig:mini:fig8:a}) for the $m_\sigma=$ 500 MeV case where the RQM model condensate variation is less smooth and occurs earlier than that in the QMVT model. This happens because the nonstrange direction  effective potential difference in the RQM model, becomes deepest for the $m_\sigma=$ 700 MeV ($>$ 658.8 MeV) case while it is shallower than that of the QMVT model for the $m_\sigma=$ 500 MeV ($<$658.8 MeV). The melting of the strange condensate ($y$) is most pronounced in the RQM model for the $m_\sigma=$ 500 MeV. Its melting in the QMVT model for the 160-240 MeV temperature range, is more than that of the QM model result. The strange condensate melting in the RQM model becomes less pronounced as the $\sigma$ meson mass is increased and its temperature variation becomes closer to the QM model result for the $m_\sigma=$ 648  and 700 MeV. Table~\ref{tab:Tc} gives the summary of the pseudo-critical temperatures $T_{c}^{\chi}$ and $T_{c}^{s}$ for the chiral crossover transition in the nonstrange and strange direction for different $m_{\sigma}$= 400, 500, 600, 648 and 700 MeV.   

 \begin{figure*}[htb]
\subfigure[\   $m_\sigma=500$ MeV]{
\label{fig:mini:fig9:a} 
\begin{minipage}[b]{0.32\textwidth}
\centering \includegraphics[width=\linewidth]{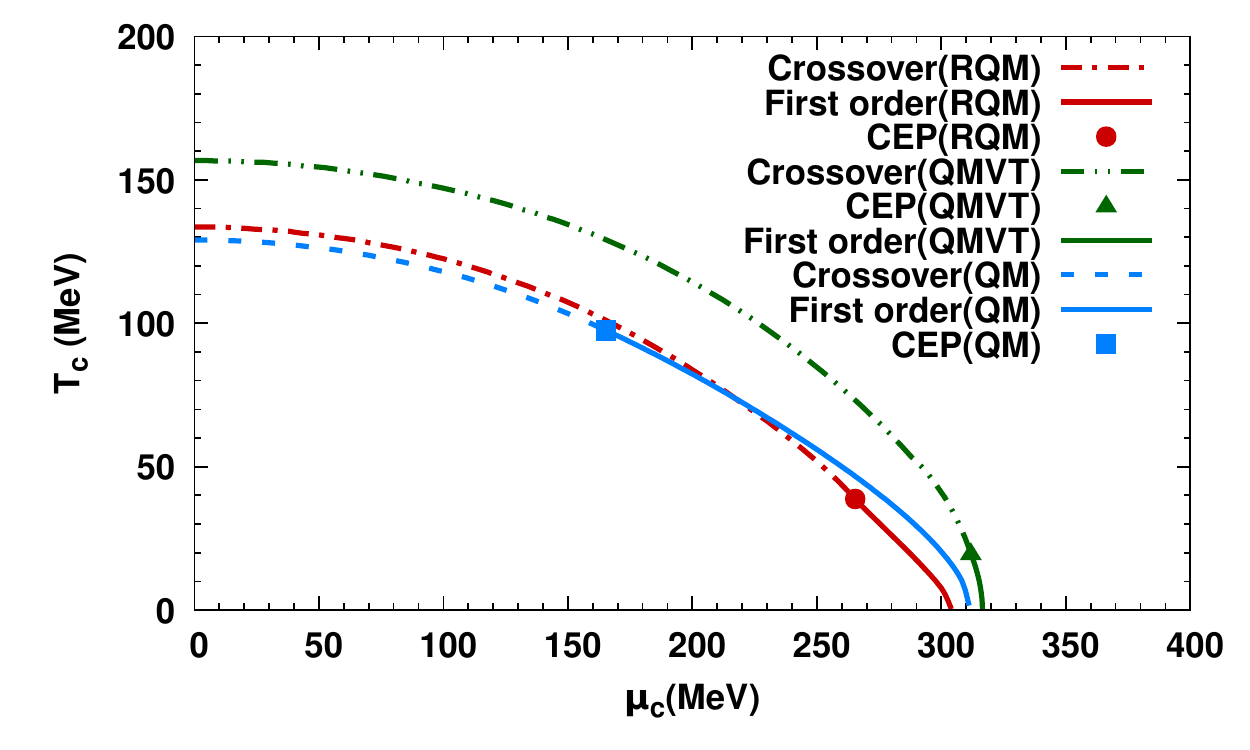}
\end{minipage}}%
\hfill
\subfigure[\  $m_\sigma=648$ MeV]{
\label{fig:mini:fig9:b} 
\begin{minipage}[b]{0.32\textwidth}
\centering \includegraphics[width=\linewidth]{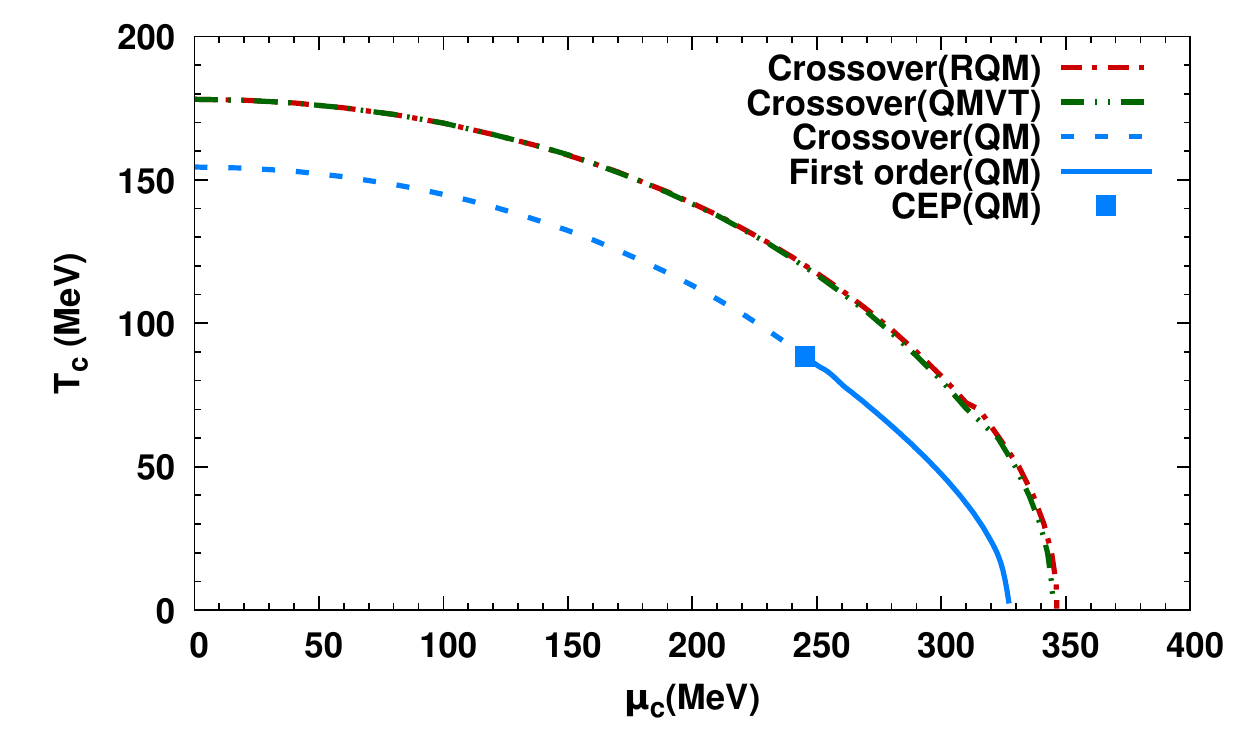}
\end{minipage}}
\hfill
\subfigure[\  $m_\sigma=700$ MeV]{
\label{fig:mini:fig9:c} 
\begin{minipage}[b]{0.32\textwidth}
\centering \includegraphics[width=\linewidth]{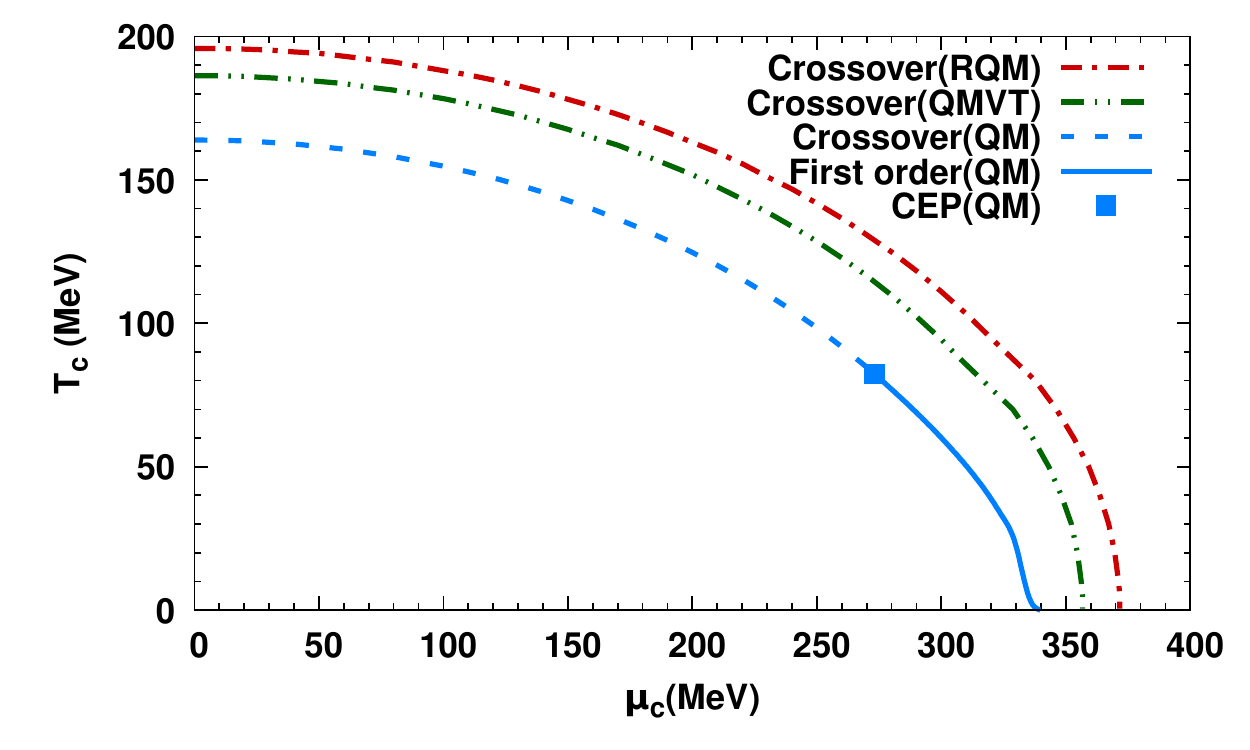}
\end{minipage}}
\caption{Phase diagrams for different $m_\sigma$}
\label{fig:mini:fig9} 
\end{figure*}

The Fig.~(\ref{fig:mini:fig9:a}) presents the phase diagram in the $\mu-T$ plane for the $m_\sigma=500$ MeV and  line types for different models are labelled. The QM model critical end point (CEP) at $\mu=$165.24 MeV, $T=$97.52 MeV, shifts to a far right lower corner of the phase diagram at $\mu=$ 311.85 MeV, $T=$19.8 MeV due to the curvature mass parameterization in the QMVT model. This robust shift in the present 2+1 flavor QMVT model is larger ($\mu=$ is larger by 11 MeV and $T$ is smaller by about 10 MeV) than the corresponding shift reported \cite{RaiTiw} for the two flavor QMVT model setting. It is important to  note that this significantly large shift of the CEP as also reported in several of the earlier studies \cite{guptiw,schafwag12,chatmoh1,vkkr12,chatmoh2}, becomes small due to the exact on-shell parameter fixing in the RQM model. The CEP in the  RQM model moves relatively higher up in the phase diagram and gets  located at a lower chemical potential $\mu=$265.42 MeV and higher temperature $T=$ 38.71 MeV when compared to the CEP location in the QMVT model. It is also noteworthy that the  QM model and RQM model phase diagrams stand in close proximity of each other.

Similar to the two flavor \cite{RaiTiw} result, here in the 2+1 flavor investigation also, the RQM model and QMVT model phase diagrams (crossover lines in the entire $\mu-T$ plane) merge with each other in the Fig.~(\ref{fig:mini:fig9:b}) for the $m_\sigma$=648 MeV. The above overlap is consequence of the coincidence that occurs in  the normalized vacuum effective potential difference and nonstrange order parameter plots respectively in the Fig.~(\ref{fig:mini:fig6:b}) and Fig.~(\ref{fig:mini:fig8:b})for both the models RQM and QMVT. The critical end point for the QM model gets located at $\mu=$  223.3 MeV, $T=$ 88.37 MeV in the Fig.~(\ref{fig:mini:fig9:b}). The phase diagrams for the RQM as well as the QMVT model, are crossover lines in the entire $\mu-T$ plane for the $m_\sigma=700$ MeV. One notices that when the $m_\sigma=700$ MeV in the Fig.~(\ref{fig:mini:fig9:c}), instead of the RQM model phase diagram proximity to the QM model phase boundary for the $m_\sigma=500$ MeV, the phase diagram of the QMVT model stands closer to the QM model phase boundary. The abovementioned nature of the plots correspond to the trend reversal seen respectively in the Fig.~(\ref{fig:mini:fig8:c}) and the Fig.~(\ref{fig:mini:fig6:c}) for the nonstrange direction order parameter and the normalized vacuum effective potential difference when the $m_\sigma=700$ MeV  plots are  compared with the corresponding $m_\sigma=500$ MeV  plots. In the QM model phase diagram, the first order line gets terminated at the critical end point (CEP) $\mu=$  273.12 MeV, $T=$ 82.30 MeV. 

\begin{figure}[!htb]
\centering \includegraphics[width=\linewidth]{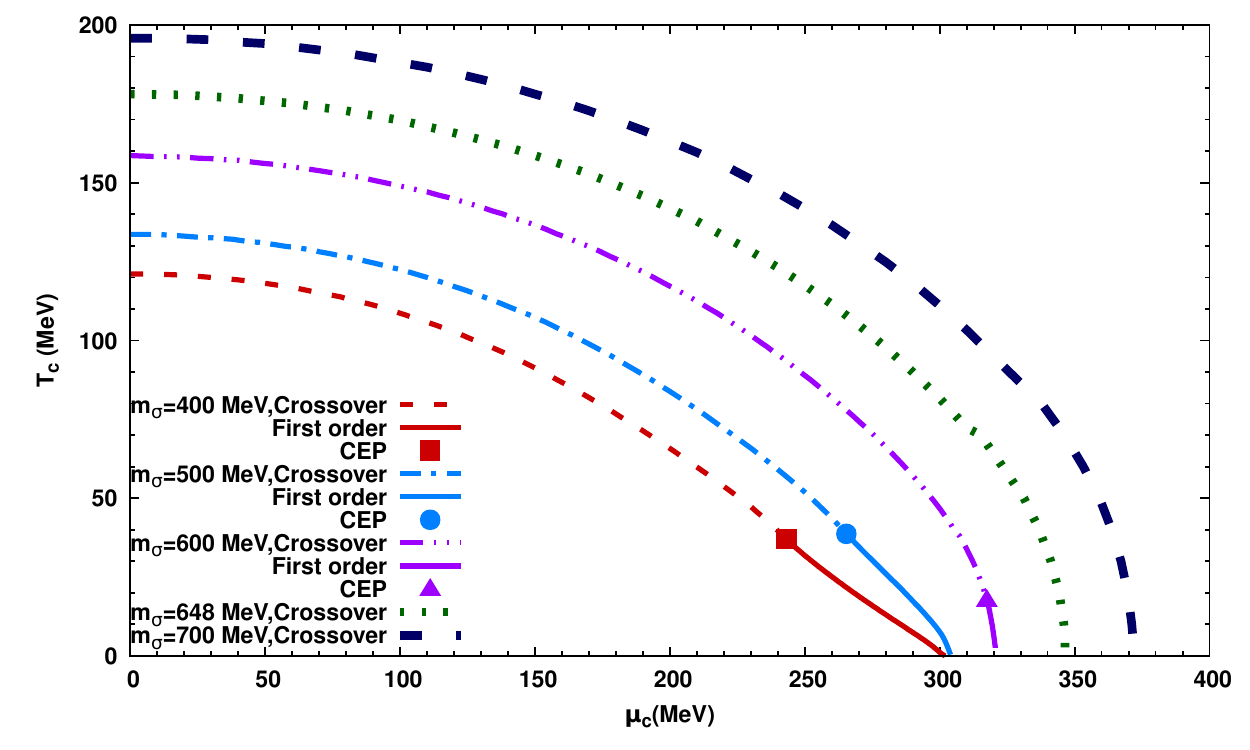}%
\caption{Phase diagram for the sigma masses of 400, 500, 600, 648 and 700 MeV in the RQM model.}
\label{fig:mini:fig10} 
\end{figure}
The comparison of the five phase diagrams for the $m_\sigma=$ 400 MeV, 500 MeV, 600 MeV, 648 MeV and 700 MeV in the RQM model and the influence of the sigma meson mass on phase diagrams  can be seen in the Fig.~(\ref{fig:mini:fig10}). The CEP position $\mu=$ 243.12 MeV, $T=$ 37.03 MeV for the $m_\sigma=$ 400 MeV, shifts rightwards to $\mu=$ 265.42 MeV, $T=$ 38.71 MeV when the $m_\sigma=$ 500 MeV. The CEP moves to the bottom right extreme of the phase diagram at $\mu$= 317.45 MeV $T=$ 17.7 MeV for the $m_\sigma=$ 600 MeV. The two crossover transition lines in the whole of the $\mu-T$ plane for the $m_\sigma=$ 648 MeV and 700 MeV are also shown.

\section{Summary and Conclusion}
\label{secVI}
This work presents the first application of the on-shell parameter fixing scheme to calculate the  effective potential after including the quark one-loop vacuum fluctuation and properly renormalizing it in the 2+1 flavor renormalized quark-meson (RQM) model. The seven running parameters of the model $\lambda_{1}$, $\lambda_{2}$, $c$, $m^{2}$, $h_{x}$, $h_{y}$ and $g$ are determined by relating the $\overline{\text{MS}}$, on-shell schemes and the experimental values of the pion decay constant $f_{\pi}$, the kaon decay constant $f_{K}$, the scalar $\sigma$ meson mass $m_{\sigma}$, the pseudo-scalar pion, kaon, eta and eta-prime meson masses  $m_{\pi},m_{K},m_{\eta},m_{\eta^{\prime}}$ and the nonstrange as well as the strange constituent quark masses. For comparison, the earlier calculation of the effective potential in the quark-meson model with the vacuum term (QMVT) has also been presented where the curvature meson masses have been used for fixing the model parameters. We have computed and compared the effective potentials, the nonstrange and strange order parameter temperature variations and the phase diagrams for the QM, RQM and QMVT model settings.

The $\sigma$ meson mass dependent similarities and differences are observed in the plots of the vacuum effective potentials in the nonstrange and strange directions for the RQM model and the QMVT model. The nonstrange direction normalized vacuum effective potential difference, is deepest for the QMVT model when the $m_\sigma=$ 400 and 500 MeV while it is shallower in the RQM model and most shallow in the QM model. For the $m_\sigma=$658.8 MeV, the nonstrange direction vacuum effective potential difference in both the plots for the RQM and QMVT models coincide with each other and for the higher $m_\sigma=$ 700 ($>$ 658.8) MeV, one finds that the trend of the plots for the $m_\sigma=$500 MeV  gets reversed and the  effective potential is deepest in the RQM model. The strange direction vacuum effective potential difference is most shallow in the RQM model, deeper in the QM model and deepest in the QMVT model for the $m_\sigma=$ 500 MeV. It becomes deeper in the RQM model on increasing the $m_\sigma$ value and shows rising trend for the QMVT model. Both the effective potential plots nearly merge with each other  for the $m_\sigma=785 \ \text{MeV}$. The trend of plots for the $m_\sigma=500$ MeV gets reversed for the $m_\sigma=850$ MeV as the strange direction effective potential difference becomes shallower for the QMVT model and deepest for the RQM model.

Comparing the $\mu=0$ MeV nonstrange order parameter ($x$), one finds that its sharpest QM model temperature variation for the $m_\sigma=$500 MeV, becomes smoother for the on-shell parameterization of the RQM model while the curvature mass parameterization of the QMVT model gives rise to a most smooth and delayed variation. This pattern also gets reversed for the higher $m_\sigma=$700 ($>$648) MeV as the nonstrange order parameter temperature variation becomes most smooth and delayed for the RQM model. We point out that our recent two flavor work \cite{RaiTiw} reported that, for the $m_\sigma=$ 616 MeV, the vacuum effective potential variation (dependent on the constituent quark mass parameter $ \Delta $) and the chiral order parameter temperature variation for the on-shell parameter fixing scheme in the RQM model, completely merge with the corresponding effective potential and the chiral order parameter temperature variation computed with the curvature mass parameterization in the QMVT model. In the present 2+1 flavor work, since the vacuum effective potential depends on the nonstrange and strange constituent quark mass parameter $\Delta_{x}$ and $\Delta_{y}$ variations in the two independent directions, the $\Delta_{x}$ variation of the effective potential is  somewhat influenced by its $\Delta_{y}$ variation. Hence though the RQM model nonstrange direction $\Delta_{x}$ dependent variation of the vacuum effective potential difference $\boldmath{\frac{\{ \Omega_{vac}(\Delta_{x}, 334.43 )-\Omega_{vac}(0,334.43 )\}}{f_{\pi}^4}}$ coincides with the corresponding effective potential difference variation in the QMVT model for the $m_\sigma=$ 658.8 MeV, the consequential coincidence of the nonstrange order parameter temperature variations for both the model settings of the RQM and QMVT, occurs when the $\sigma$ mass is smaller by about 10 MeV i.e. $m_\sigma=$ 648 MeV. 
The melting of the strange condensate ($y$) is most pronounced in the RQM model for the $m_\sigma=$ 400 and 500 MeV and its melting in the QMVT model is more than that of the QM model result in the temperature range of 160-240 MeV.The strange condensate melting in the RQM model becomes less pronounced as the $\sigma$ meson mass is increased and its temperature variation comes closer to the QM model result as the value of the $m_\sigma$ changes from 500 MeV to  648  and 700 MeV. 

It is well known that the CEP shifts to the far right side of the phase diagram in the $\mu-T$ plane for the  QMVT model  \cite{guptiw,schafwag12,chatmoh1,vkkr12,chatmoh2,RaiTiw}. Similar to the two flavor results \cite{RaiTiw}, here in the present 2+1 flavor work also, it is noticed that when the $m_\sigma=$ 400, 500 ($<$ 648) MeV, the  shift in the position of the CEP due to the on-shell parameterization in the RQM model, is smaller than what is found in the QMVT model. Furthermore, the phase boundaries depicting the crossover transition lines, for both the models RQM and QMVT, completely merge with each other for the $m_\sigma=$ 648 MeV. The crossover transition line of the QMVT model phase diagram, comes closer to the QM model phase boundary for the higher $m_\sigma=$ 700 ($>$ 648) MeV. This trend is opposite of what one notices for the $m_\sigma=$ 500 ($<$ 648) MeV case where the RQM model phase diagram stands in close proximity to the QM model phase boundary. 

\appendix
\section{THE QMVT PARAMETER FIXING}
\label{appenA}
This appendix presents a brief description of the parameter fixing procedure in the QMVT model as given in Ref.~\cite{vkkt13}. The vacuum meson mass matrix is written as 
\bqa
\label{eq:derVac}
 m_{\alpha,ab}^{2}&=&\frac{\partial^2 \Omega^{\Lambda} (x,y)}{\partial \xi_{\alpha,a}
 \partial \xi_{\alpha,b}} \bigg|_{min}=({m^{\text{m}}_{\alpha,ab}})^{2}+({\delta m^{\text{v}}_{\alpha,ab}})^{2} \quad
\eqa
Here, the $({m^{\text{m}}_{\alpha,ab}})^{2}$  expressions for $\alpha=\text{s, p}$ and $a,b=0,1,\cdots,8$ are same as the vacuum values of the meson masses $m_{\alpha,ab}^{2}$ which were originally evaluated for
the  QM model under s-MFA in Ref.~\cite{Rischke:00,Rischke:001,Schaefer:09} by taking the second derivatives of the pure mesonic potential at its minimum. The vacuum ($T=0$, $\mu=0$) mass modifications on account of the fermionic vacuum correction are given by

\begin{align}
\nonumber
&({\delta m^{\text{v}}_{\alpha,ab}})^{2}=\frac{\partial^2 \Omega_{q\bar{q}}^{\rm vac}}{\partial \xi_{\alpha,a}\partial \xi_{\alpha,b}} \bigg|_{min}=-\frac{N_c}{8\pi^2}\sum_f\biggl[\left(2\log\left(\frac{m_f}{\text{M}}\right)+\frac{3}{2} \right)\left(\frac{\partial m_f^2}{\partial \xi_{\alpha,a}}\right)\left(\frac{\partial m_f^2}{\partial \xi_{\alpha,b}}\right)\\ 
&{\hskip 5 cm}+\left(\frac{m_f^2}{2}+2m_f^2\log\left(\frac{m_f}{\text{M}} \right) \right) \frac{\partial^2
  m_f^2}{\partial \xi_{\alpha,a}\partial \xi_{\alpha,b}}\biggr]
\label{eq:dFVac}
\end{align}
Here $|_{min}$ denotes the global minimum of the full grand potential in the Eq.(\ref{Grandpxy}).
The first $m^{2}_{f,a} \equiv \partial m^{2}_{f}/ \partial \xi_{\alpha,a}$ and second  
$m^{2}_{f,ab} \equiv \partial m^{2}_{f,a}/ \partial \xi_{\alpha,b}$ partial derivatives of the squared quark mass
with respect to the different meson fields are evaluated in Ref.~\cite{Schaefer:09} in the non strange-strange basis. The values of these derivatives can be found from the Table III of the Ref. \cite{Schaefer:09,vkkt13}.
The mass modifications given in the Eq. (\ref{eq:dFVac}) due to the fermionic vacuum correction were evaluated in the Ref.~\cite{vkkt13} and different expressions of $({\delta m^{\text{v}}_{\alpha,ab}})^{2}$ are presented in the Table~\ref{tab*:MesonVM} for all the mesons of the scalar and pseudo-scalar nonet.

In the QMVT model calculations, the vacuum mass expressions in the Eq.(\ref{eq:derVac}) that determine 
$\lambda_2$ and c are $m^{2}_{\pi}=({m^{\text{m}}_{\pi}})^{2}+({\delta m^{\text{v}}_{\text{p},11}})^{2}$, 
$m^{2}_K=({m^{\text{m}}_K})^{2}+({\delta m^{\text{v}}_{p,44}})^{2}$ and $m^{2}_{\eta}+m^{2}_{\eta'}=m^{2}_{\text{p},00}+m^{2}_{\text{p},88}$ where
$m^{2}_{\text{p},00}=({m^{\text{m}}_{\text{p},00}})^{2}+({\delta m^{\text{v}}_{\text{p},00}})^{2}$ and 
$m^{2}_{\text{p},88}=({m^{\text{m}}_{\text{p},88}})^{2}+({\delta m^{\text{v}}_{\text{p},88}})^{2}$. We can write $m^{2}_{\eta}+m^{2}_{\eta'}=({m^{\text{m}}_{\eta}})^{2}+({m^{\text{m}}_{\eta'}})^{2}+({\delta m^{\text{v}}_{\text{p},00}})^{2}+({\delta m^{\text{v}}_{\text{p},88}})^{2}$ where $({m^{\text{m}}_{\eta}})^{2}+({m^{\text{m}}_{\eta'}})^{2}=
({m^{\text{m}}_{\text{p},00}})^{2}+({m^{\text{m}}_{\text{p},88}})^{2}$. Using mass modification  expressions
$({\delta m^{\text{v}}_{\alpha,ab}})^{2}$  given in the Table \ref{tab*:MesonVM}, one writes
\bqa
\label{PIKETA}
({m^{\text{m}}_K})^{2}&=&m^{2}_K+\frac{N_cg^4}{64\pi^2}\left(\frac{x-\sqrt{2}y}{x^2-2y^2}\right) 
\left( x^3X+2\sqrt{2}y^3Y\right) ; \ ({m^{\text{m}}_{\pi}})^{2}=m^{2}_{\pi}+
\frac{N_cg^4}{64\pi^2} \ x^2X  \ \ \ \\ \nonumber 
&&\text{and}\left({m^{\text{m}}_{\eta}})^{2}+({m^{\text{m}}_{\eta'}})^{2}\right)=\left( m^{2}_{\eta}+m^{2}_{\eta'}\right)+
\frac{N_cg^4}{192\pi^2}\left(3x^2X+6y^2Y\right)\;.
\eqa
\begin{table*}[!hbtp]
\caption{The superscript m in the $({m^{\text{m}}_{\alpha,ab}})^{2}$ symbolizes the 
meson curvature masses calculated from the second derivatives of the pure mesonic potential U($x,y$).
The mass modifications $({\delta m^{\text{v}}_{\alpha,ab}})^{2}$ due to the 
fermionic vacuum correction are shown in the right half. Symbols used in the expressions are defined as 
 $ X=\left\{1+4\log\left( \frac{g \ x}{2\Lambda}\right)\right\} $ and $Y=\left\{1+4\log\left( \frac{g \ y}{\sqrt{2}\Lambda}\right)\right\} $.}
\label{tab*:MesonVM}
\resizebox{1.0\hsize}{!}
{\begin{tabular}{{p{0.08\textwidth} p{0.51\textwidth} p{0.10\textwidth} p{0.36\textwidth}}}
 \hline
    & Meson masses calculated from pure mesonic potential  & & Fermionic vacuum correction in meson masses \\ \hline
    $(m^{\text{m}}_{a_{0}})^2$ & $m^2 +\lambda_1(x^2+y^2)+\frac{3\lambda_2}{2} x^2+
    \frac{\sqrt{2}c}{2}y $ & $({\delta m^{\text{v}}_{\text{s},11}})^{2}$ & $-\frac{N_cg^4}{64\pi^2} \ \ x^2(4+3X)$
    \\ 
    $(m^{\text{m}}_{\kappa})^{2}$&$m^2+\lambda_1(x^2+y^2)+\frac{\lambda_{2}}{2}(x^2+\sqrt{2}xy+2y^2)+\frac{c}{2} x $ & $({\delta m^{\text{v}}_{\text{s},44}})^{2}$
    & $-\frac{N_cg^4}{64\pi^2} \ \ \left(\frac{x+\sqrt{2}y}{x^2-2y^2}\right) \left( x^3X-2\sqrt{2}y^3Y\right)$
    \\
    $(m^{\text{m}}_{\text{s},00})^2$ &  $m^2+\frac{\lambda_1}{3}(7x^2+4\sqrt{2}xy+5y^2)+\lambda_2(x^2 + y^2)-\frac{\sqrt{2}c}{3} (\sqrt{2} x +y)$ &
    $({\delta m^{\text{v}}_{\text{s},00}})^{2}$&$-\frac{N_cg^4}{96\pi^2} \ \ \left\{3\left( x^2X+y^2Y\right)+4\left( x^2+y^2\right)\right\}$ \\
    $(m^{\text{m}}_{\text{s},88})^{2}$ & $m^2 +\frac{\lambda_1}{3}(5x^2-4\sqrt{2}xy +7y^2)+\lambda_2(\frac{x^2}{2} +2y^2)+\frac{\sqrt{2}c}{3} (\sqrt{2}x-\frac{y}{2})$ & 
    $({\delta m^{\text{v}}_{\text{s},88}})^{2}$&  
    $-\frac{N_cg^4}{96\pi^2} \ \ \left\{ \frac{3}{2}\left( x^2X+4y^2Y\right)+2\left( x^2+4y^2\right)\right\}$
    \\
   $(m^{\text{m}}_{\text{s},08})^{2}$ &  $\frac{2\lambda_1}{3}(\sqrt{2}x^2 -xy -\sqrt{2}y^2) +\sqrt{2}\lambda_2(\frac{x^2}{2}-y^2) +\frac{c}{3\sqrt{2}}(x- \sqrt{2}y)$ &
  $({\delta m^{\text{v}}_{\text{s},08}})^{2}$&$-\frac{N_cg^4}{8\sqrt{2}\pi^2} \ \ \left\{\frac{1}{4}\left( x^2X-2y^2Y\right)+\frac{1}{3}\left( x^2-2y^2\right)\right\}$   
   \\
$ (m^{\text{m}}_{\pi})^{2}$ & $m^2 + \lambda_1 (x^2 + y^2) +\frac{\lambda_2}{2} x^2 -\frac{\sqrt{2} c}{2} y$& $({\delta m^{\text{v}}_{\text{p},11}})^{2}$  
&$-\frac{N_cg^4}{64\pi^2} \ \ x^2X$  \\ 
$(m^{\text{m}}_{K})^{2}$ &$m^2 + \lambda_1 (x^2 + y^2) +\frac{\lambda_2}{2} (x^2 - \sqrt{2} x y +2 y^2) - \frac{c}{2} x$&$({\delta m^{\text{v}}_{\text{p},44}})^{2}$
&$-\frac{N_cg^4}{64\pi^2} \ \ \left(\frac{x-\sqrt{2}y}{x^2-2y^2}\right) \left( x^3X+2\sqrt{2}y^3Y\right)$  \\ 
$(m^{\text{m}}_{\text{p},00})^{2}$ & $m^2 + \lambda_1(x^2 +y^2) + \frac{\lambda_2}{3}(x^2 +y^2) + \frac{c}{3} (2x + \sqrt{2} y)$&$({\delta m^{\text{v}}_{\text{p},00}})^{2}$
&$-\frac{N_cg^4}{96\pi^2} \ \ \left( x^2X+y^2Y\right)$ \\
$(m^{\text{m}}_{\text{p},88})^{2}$ & $m^2 +\lambda_1(x^2 +y^2) +\frac{\lambda_2}{6}(x^2 +4y^2)-\frac{c}{6}(4x -\sqrt{2}y)$& $({\delta m^{\text{v}}_{\text{p},88}})^{2}$
&$-\frac{N_cg^4}{192\pi^2} \ \ \left( x^2X+4y^2Y\right)$ \\
$(m^{\text{m}}_{\text{p},08})^{2}$ & $\frac{\sqrt{2}\lambda_2}{6}(x^2-2y^2)-\frac{c}{6}(\sqrt{2}x -2y)$& $({\delta m^{\text{v}}_{\text{p},08}})^{2}$
&$-\frac{N_cg^4}{96\sqrt{2}\pi^2} \ \ \left( x^2X-2y^2Y\right)$ \\ 
\hline
\end{tabular}}
\end{table*}   

The $f_{\pi}$ and $f_K$ give vacuum condensates according to the partially conserved axial vector current (PCAC) relation. The
$x=f_{\pi}$ and $y=\left( \frac{2f_K-f_{\pi}}{\sqrt{2}}\right)$ at $T=0$. The parameters $\lambda_2$ and $ c$ in vacuum are obtained as:
\bqa
\label{lam2c}
\lambda_2&=&\frac{3\left(2f_K-f_{\pi}\right)({m^{\text{m}}_K})^{2}-\left(2f_K+f_{\pi}\right)({m^{\text{m}}_{\pi}})^{2}
-2\left(({m^{\text{m}}_{\eta}})^{2}+({m^{\text{m}}_{\eta'}})^{2}\right) 
\left( f_K-f_{\pi}\right)}{\left(3f^2_{\pi}+8f_K\left( f_K-f_{\pi}\right)\right)\left( f_K-f_{\pi}\right)}\;, 
\\
\label{eqc}
c&=&\frac{({m^{\text{m}}_K})^{2}-({m^{\text{m}}_{\pi}})^{2}}{f_K-f_{\pi}}-\lambda_2\left(2f_K-f_{\pi}\right)\;.
\eqa
When expressions of $({m^{\text{m}}_{\pi}})^2$, $({m^{\text{m}}_K})^2$ and $\left( ({m^{\text{m}}_{\eta}})^2+
({m^{\text{m}}_{\eta'}})^2\right)$ from the Eq.(\ref{PIKETA}) are substituted in the Eq.(\ref{lam2c}), the Eq.(\ref{eqc}) 
and the vacuum value of the condensates are used, the final rearrangement of terms yields:
\bqa
\label{lam2VT}
\lambda_2&=&\lambda_{2s}+n+\lambda_{2+}+\lambda_{2\Lambda}\ \text{where} \ \lambda_{2s}=\frac{3\left(2f_K-f_{\pi}\right) m^{2}_K-\left(2f_K+f_{\pi}\right) m^{2}_{\pi}-2\left( m^{2}_{\eta}+m^{2}_{\eta'}\right) \left( f_K-f_{\pi}\right)}{\left(3f^2_{\pi}+8f_K\left( f_K-f_{\pi}\right)\right)\left( f_K-f_{\pi}\right)}\ ,\\ \nonumber
n&=&\frac{N_cg^4}{32\pi^2}, 
\ \lambda_{2+}=\frac{n{f_{\pi}}^2}{f_K \left( f_K-f_{\pi}\right)}\log\left( 
\frac{2\ f_K-f_{\pi}}{f_{\pi}}\right)
\text{and scale dependent part} \ \lambda_{2\Lambda}= 4n\log\left( \frac{g\left( 2f_K-f_{\pi}\right)}{2 \Lambda}\right),\\ 
\label{cfinal}
c&=&\frac{{m_K}^{2}-{m_{\pi}}^{2}}{f_K-f_{\pi}}-\lambda_{2s}\left(2f_K-f_{\pi}\right).
\eqa
Note that the $\lambda_{2s}$ is the old 
$\lambda_2$ parameter fixed in the QM/PQM model calculations in 
Ref.~\cite{Rischke:00,Rischke:001,Schaefer:09,TiPQM3F}. Here, the curvature mass parameterization
in the QMVT model, generates an addition of ($n+\lambda_{2+}$) to $\lambda_{2s}$
 and further, one gets a  scale $\Lambda$ dependent addition $\lambda_{2\Lambda}$ to the expression of the $\lambda_2$ in the  Eq.~(\ref{lam2VT}). The $\Lambda$ dependence completely cancels in the evaluation of $c$ and its value remains the same as in the QM model. The parameter $m^2$ can be expressed in terms of $\lambda_1$ using the scale $\Lambda$ independent formula of $m^2_{\pi}$ (given in the appendix B of the Ref.~\cite{vkkt13}) and putting  $x=f_{\pi}$ and $y=(\frac{2f_K-f_{\pi}}{\sqrt{2}})$, one gets,
\bqa
\nonumber
m^2&=&m^{2}_{\pi}-\lambda_1\{f_{\pi}^2+\frac{(2f_K-f_{\pi})^2}{2}\}-\frac{f_{\pi}^2}{2} \ \biggl[\ \lambda_{2\text{v}} \qquad \\ 
&& -4\log \ \{\ \frac{f_{\pi}}{\left(2f_K-f_{\pi}\right)} \}\biggl] +\frac{c}{2} \ (\ 2f_K-f_{\pi})\;.
\eqa
When the formula of $m^2_{\sigma}$ (given in the Table \ref{tab:table1}) is used with the vacuum values of the masses $m^2_{\text{s},00}$, $m^2_{\text{s},88}$, $m^2_{\text{s},08}$ and the above expression of $m^2$ is substituted in it, one gets the numerical value of $\lambda_1$ for different values of $m_\sigma$. The fermionic vacuum correction does not change the explicit symmetry breaking parameters $h_x$ and $h_y$. 
\section{THE TADPOLE TERMS}
\label{appenB}
The expressions of tadpole contributions in the self-energies are,

\begin{align}
\nonumber
&\Sigma^{tad}_{\text{s},00}=4\sqrt{\frac{2}{3}}N_cg\left[\frac{1}{m^2_{s,00}}\left\{3\left(\lambda_1+\frac{\lambda_2}{3}\right)\bar{\sigma_0}-\frac{c}{2}\sqrt{\frac{2}{3}}\right\}+\frac{1}{3m^2_{s,08}}(\lambda_1+\lambda_2)\bar(\sigma_8)\right]\mathcal{F}(m_u,m_s) \\ 
\label{appenBeq1}
&+\frac{8}{\sqrt{3}}N_cg\left[\frac{1}{m^2_{s,08}}\left\{3\left(\lambda_1+\frac{\lambda_2}{3}\right)\bar{\sigma_0}-\frac{c}{2}\sqrt{\frac{2}{3}}\right\}+\frac{1}{3m^2_{s,88}}(\lambda_1+\lambda_2)\bar{\sigma_8}\right]\mathcal{G}(m_u,m_s)\\
\nonumber
&\Sigma^{tad}_{\text{s},88}=4\sqrt{\frac{2}{3}}N_cg\left[\frac{1}{3 \ m^2_{s,00}}\left\{\left(\lambda_1+\lambda_2\right)\bar{\sigma_0}-\frac{\lambda_2}{\sqrt{2}}\bar{\sigma_8}+\frac{c}{4}\sqrt{\frac{2}{3}}\right\}+\frac{1}{m^2_{s,08}}\left\{\frac{-\lambda_2}{\sqrt{2}}\bar{\sigma_0}+3\left(\lambda_1+\frac{\lambda_2}{2}\right)\bar{\sigma_8}+\frac{c}{2\sqrt{3}}\right\}\right]\\ \nonumber
&{\hskip -0.5 cm}\mathcal{F}(m_u,m_s)+\frac{8}{\sqrt{3}}N_cg\left[\frac{1}{3 \ m^2_{s,08}}\left\{\left(\lambda_1+\lambda_2\right)\bar{\sigma_0}-\frac{\lambda_2}{\sqrt{2}}\bar{\sigma_8}+\frac{c}{4}\sqrt{\frac{2}{3}}\right\}+\frac{1}{m^2_{s,88}}\left\{\frac{-\lambda_2}{\sqrt{2}}\bar{\sigma_0}+3\left(\lambda_1+\frac{\lambda_2}{2}\right)\bar{\sigma_8}+\frac{c}{2\sqrt{3}}\right\}\right]\\ 
&\mathcal{G}(m_u,m_s), \\
 \nonumber
&\Sigma^{tad}_{\text{s},08}=4\sqrt{\frac{2}{3}}N_cg\left[\frac{1}{3 \ m^2_{s,00}}(\lambda_1+\lambda_2)\bar{\sigma_8}+\frac{1}{3 \ m^2_{s,08}}\left\{(\lambda_1+\lambda_2)\bar{\sigma_0}-\frac{\lambda_2}{\sqrt{2}}\bar{\sigma_8}+\frac{c}{4}\sqrt{\frac{2}{3}}\right\}\right]\mathcal{F}(m_u,m_s)\\ 
&\qquad \quad+\frac{8}{\sqrt{3}}N_cg\left[\frac{1}{3 \ m^2_{s,08}}(\lambda_1+\lambda_2)\bar{\sigma_8}+\frac{1}{3 \ m^2_{s,88}}\left\{(\lambda_1+\lambda_2)\bar{\sigma_0}-\frac{\lambda_2}{\sqrt{2}}\bar{\sigma_8}+\frac{c}{4}\sqrt{\frac{2}{3}}\right\}\right]\mathcal{G}(m_u,m_s)\;,
\nonumber
&\Sigma^{tad}_{\text{p},00}=4\sqrt{\frac{2}{3}}N_cg\left[\frac{1}{m^2_{s,00}}\left\{\left(\lambda_1+\frac{\lambda_2}{3}\right)\bar{\sigma_0}+\frac{c}{2}\sqrt{\frac{2}{3}}\right\}+\frac{1}{m^2_{s,08}}\left(\lambda_1+\frac{\lambda_2}{3}\right)\bar{\sigma_8}\right]\mathcal{F}(m_u,m_s)\\ 
&\qquad \quad+\frac{8}{\sqrt{3}}N_cg\left[\frac{1}{m^2_{s,08}}\left\{\left(\lambda_1+\frac{\lambda_2}{3}\right)\bar{\sigma_0}+\frac{c}{2}\sqrt{\frac{2}{3}}\right\}+\frac{1}{m^2_{s,88}}\left(\lambda_1+\frac{\lambda_2}{3}\right)\bar{\sigma_8}\right]\mathcal{G}(m_u,m_s)\;,
\\
\nonumber
&\Sigma^{tad}_{\text{p},88}=4\sqrt{\frac{2}{3}}N_cg\left[\frac{1}{m^2_{s,00}}\left\{\left(\lambda_1+\frac{\lambda_2}{3}\right)\bar{\sigma_0}-\frac{\lambda_2}{3\sqrt{2}}\bar{\sigma_8}-\frac{c}{4}\sqrt{\frac{2}{3}}\right\}+\frac{1}{m^2_{s,08}}\left\{\left(\lambda_1+\frac{\lambda_2}{2}\right)\bar{\sigma_8}-\frac{\lambda_2}{3\sqrt{2}}\bar{\sigma_0}
-\frac{c}{2\sqrt{3}}\right\}\right]\\ \nonumber
&{\hskip -0.4 cm} \mathcal{F}(m_u,m_s) +\frac{8}{\sqrt{3}}N_cg\left[\frac{1}{m^2_{s,08}}\left\{\left(\lambda_1+\frac{\lambda_2}{3}\right)\bar{\sigma_0}-\frac{\lambda_2}{3\sqrt{2}}\bar{\sigma_8}-\frac{c}{4}\sqrt{\frac{2}{3}}\right\}+\frac{1}{m^2_{s,88}}\left\{\left(\lambda_1+\frac{\lambda_2}{2}\right)\bar{\sigma_8}-\frac{\lambda_2}{3\sqrt{2}}\bar{\sigma_0}-\frac{c}{2\sqrt{3}}\right\}\right],\\ 
&\mathcal{G}(m_u,m_s)\\
\nonumber  
&\Sigma^{tad}_{\text{p},08}=4\sqrt{\frac{2}{3}}N_cg\left[\frac{1}{m^2_{s,00}}\left(\lambda_1+\frac{\lambda_2}{3}\right)\bar{\sigma_8}+\frac{1}{m^2_{s,08}}\left\{\left(\lambda_1+\frac{\lambda_2}{3}\right)\bar{\sigma_0}-\frac{\lambda_2}{3\sqrt{2}}\bar{\sigma_8}
-\frac{c}{4}\sqrt{\frac{2}{3}}\right\}\right]\mathcal{F}(m_u,m_s)\\
& \qquad \quad +\frac{8}{\sqrt{3}}N_cg\left[\frac{1}{m^2_{s,08}}\left(\lambda_1+\frac{\lambda_2}{3}\right)\bar{\sigma_8}+\frac{1}{m^2_{s,88}}\left\{\left(\lambda_1+\frac{\lambda_2}{3}\right)\bar{\sigma_0}-\frac{\lambda_2}{3\sqrt{2}}\bar{\sigma_8}
-\frac{c}{4}\sqrt{\frac{2}{3}}\right\}\right]\mathcal{G}(m_u,m_s)\;,\\ \nonumber
\nonumber 
&{\hskip -1.0 cm}\Sigma^{tad}_{\text{p},11}=4\sqrt{\frac{2}{3}}N_cg\left[\frac{1}{m^2_{s,00}}\left\{\left(\lambda_1+\frac{\lambda_2}{3}\right)\bar{\sigma_0}+\frac{\sqrt{2}}{6}\lambda_2\bar{\sigma_8}-\frac{c}{4}\sqrt{\frac{2}{3}}\right\}+\frac{1}{m^2_{s,08}}\left\{\frac{\sqrt{2}}{6}\lambda_2\bar{\sigma_0}+\left(\lambda_1+\frac{\lambda_2}{6}\right)\bar{\sigma_8}
+\frac{c}{2\sqrt{3}}\right\}\right] \\ \nonumber
&{\hskip -1.0 cm}\mathcal{F}(m_u,m_s)+\frac{8}{\sqrt{3}}N_cg\left[\frac{1}{m^2_{s,08}}\left\{\left(\lambda_1+\frac{\lambda_2}{3}\right)\bar{\sigma_0}+\frac{\sqrt{2}}{6}\lambda_2\bar{\sigma_8}-\frac{c}{4}\sqrt{\frac{2}{3}}\right\}+\frac{1}{m^2_{s,88}}\left\{\frac{\sqrt{2}}{6}\lambda_2\bar{\sigma_0}+\left(\lambda_1+\frac{\lambda_2}{6}\right)\bar{\sigma_8}
+\frac{c}{2\sqrt{3}}\right\}\right] \\
& \mathcal{G}(m_u,m_s) \\
\nonumber
&{\hskip -1.0 cm}\Sigma^{tad}_{\text{p},44}=4\sqrt{\frac{2}{3}}N_cg\left[\frac{1}{m^2_{s,00}}\left\{\left(\lambda_1+\frac{\lambda_2}{3}\right)\bar{\sigma_0}-\frac{\lambda_2}{6\sqrt{2}}\bar{\sigma_8}-\frac{c}{4}\sqrt{\frac{2}{3}}\right\}+\frac{1}{m^2_{s,08}}\left\{-\frac{\lambda_2}{6\sqrt{2}}\bar{\sigma_0}+\left(\lambda_1+\frac{7\lambda_2}{6}\right)\bar{\sigma_8}
-\frac{c}{4\sqrt{3}}\right\}\right]\\ \nonumber
\label{appenBeq8}
&{\hskip -1.0 cm}\mathcal{F}(m_u,m_s)+\frac{8}{\sqrt{3}}N_cg\left[\frac{1}{m^2_{s,08}}\left\{\left(\lambda_1+\frac{\lambda_2}{3}\right)\bar{\sigma_0}-\frac{\lambda_2}{6\sqrt{2}}\bar{\sigma_8}-\frac{c}{4}\sqrt{\frac{2}{3}}\right\}+\frac{1}{m^2_{s,88}}\left\{-\frac{\lambda_2}{6\sqrt{2}}\bar{\sigma_0}+\left(\lambda_1+\frac{7\lambda_2}{6}\right)\bar{\sigma_8}
-\frac{c}{4\sqrt{3}}\right\}\right]\\ \nonumber
&\mathcal{G}(m_u,m_s) \\ 
\end{align}
where $\mathcal{F}(m_u,m_s)=2m_u\mathcal{A}(m_u^2)+m_s\mathcal{A}(m_s^2)$ and $\mathcal{G}(m_u,m_s)=m_u\mathcal{A}(m_u^2)-m_s\mathcal{A}(m_s^2)$. Substituting $\bar{\sigma_0}= \frac{\sqrt{2} \ x+y}{\sqrt{3}}$ and $\bar{\sigma_8}= \frac{x-\sqrt{2} \ y}{\sqrt{3}}$ in the  Eqs.~(\ref{appenBeq1}) to (\ref{appenBeq8}) and rearranging the terms, one gets the following $x$ and $y$ dependent                                                       expressions of self energies for the tadpole terms.
\begin{align}
\nonumber
&\Sigma^{tad}_{\text{s},00}=\frac{8N_{c} \ g  \ m_{u} \  \mathcal{A}(m_{u}^2) }{3} \Biggl[-c\biggl(\frac{1}{m^2_{s,00}}+\frac{1}{\sqrt{2}\ m^2_{s,88}}\biggr)+x\left\{\frac{2\left(3\lambda_1+\lambda_2\right)}{m^2_{s,00}}+\frac{\left(\lambda_1+\lambda_2\right)}{3\ m^2_{s,88}}+\frac{2\sqrt{2}\left(2\lambda_1+\lambda_2\right)}{3\ m^2_{s,08}}\right\} \\ \nonumber 
&\qquad \quad +y   \left\{\sqrt{2}\left(\frac{1}{m^2_{s,00}}-\frac{1}{3\ m^2_{s,88}}\right) \biggl(3\lambda_1+\lambda_2\biggr)+\frac{\left(\lambda_1-\lambda_2\right)}{3\ m^2_{s,08}}\right\}\Biggr] 
+\frac{4N_{c} \ g   m_{s}  \mathcal{A}(m_{s}^2) }{3} \Biggl[-c\biggl(\frac{1}{m^2_{s,00}}-\frac{\sqrt{2}}{m^2_{s,08}}\biggr) \\ 
&{\hskip -0.6 cm} +x\left\{\frac{2\left(3\lambda_1+\lambda_2\right)}{m^2_{s,00}}-\frac{2\left(\lambda_1+\lambda_2\right)}{3\ m^2_{s,88}}-\frac{\sqrt{2}\left(17\lambda_1+5\lambda_2\right)}{3\ m^2_{s,08}}\right\}+y\left\{\frac{\sqrt{2}\left(3\lambda_1+\lambda_2\right)}{m^2_{s,00}}-\frac{2\sqrt{2}\left(\lambda_1+\lambda_2\right)}{3\ m^2_{s,88}}-\frac{2\left(5\lambda_1+2\lambda_2\right)}{3\ m^2_{s,08}}\right\}\Biggr]\;,\\
\nonumber
&\Sigma^{tad}_{\text{s},88}=\frac{8N_{c} g m_{u} \mathcal{A}(m_{u}^2) }{3} \Biggl[c\biggl(\frac{1}{6 m^2_{s,00}}+\frac{7}{6\sqrt{2} m^2_{s,08}}+\frac{1}{2 \ m^2_{s,88}}\biggr)+x\left\{\frac{\left(2\lambda_1+\lambda_2\right)}{3 \ m^2_{s,00}}+\frac{4\left(5\lambda_1+\lambda_2\right)}{3\sqrt{2}\ m^2_{s,08}}+\frac{\left(6\lambda_1+\lambda_2\right)}{2\ m^2_{s,88}}\right\} \\ \nonumber 
&\qquad \quad +y \left\{\frac{\sqrt{2} (\lambda_1+2\lambda_2)}{3 \ m^2_{s,00}}-\frac{\sqrt{2}}{ m^2_{s,88}} (3\lambda_1+2\lambda_2)-\frac{\left(17\lambda_1+10\lambda_2\right)}{3\ m^2_{s,08}}\right\}\Biggr]+\frac{4N_{c}  g   m_{s}  \mathcal{A}(m_{s}^2) }{3} \\ \nonumber
&\Biggl[ c \biggl(\frac{1}{6  m^2_{s,00}}+\frac{\sqrt{2}}{3  m^2_{s,08}}-\frac{1}{  m^2_{s,88}}\biggr) +x\left\{\frac{(2\lambda_1+\lambda_2)}{3 \ m^2_{s,00}}+\frac{\sqrt{2}(14 \lambda_1+\lambda_2 )}{6 \ m^2_{s,08}}-\frac{(6\lambda_1+\lambda_2)}{\ m^2_{s,88}}\right\}\\
&+y\left\{\frac{\sqrt{2}(\lambda_1+2\lambda_2)}{3 \ m^2_{s,00}}-\frac{4 \  (5\lambda_1+4\lambda_2)}{3\ m^2_{s,08}}+\frac{2\sqrt{2} \left(3\lambda_1+2\lambda_2\right)}{\ m^2_{s,88}}\right\}\Biggr] \\ 
\nonumber
&\Sigma^{tad}_{\text{s},08}=\frac{8N_{c} \ g  \ m_{u} \  \mathcal{A}(m_{u}^2) }{9} \Biggl[c\biggl(\frac{1}{2m^2_{s,08}}+\frac{1}{2\sqrt{2}\ m^2_{s,88}}\biggr)+x\left\{\frac{\sqrt{2}\left(\lambda_1+\lambda_2\right)}{m^2_{s,00}}+\frac{\left(3\lambda_1+2\lambda_2\right)}{\ m^2_{s,08}}+\frac{\left(2\lambda_1+\lambda_2\right)}{\sqrt{2} \ m^2_{s,88}}\right\}\; \\
\nonumber 
&\qquad \quad +y   \left\{\frac{-2(\lambda_1+\lambda_2)}{m^2_{s,00}}+\frac{\sqrt{2} \lambda_{2}}{m^2_{s,08}} +\frac{\left(\lambda_1+2\lambda_2\right)}{ m^2_{s,88}}\right\}\Biggr] 
+\frac{4N_{c} \ g \  m_{s}  \mathcal{A}(m_{s}^2) }{9} \Biggl[c\biggl(\frac{1}{2 \ m^2_{s,08}}-\frac{1}{\sqrt{2} \ m^2_{s,88}}\biggr) \\ 
& +x\left\{\frac{\sqrt{2}\left(\lambda_1+\lambda_2\right)}{m^2_{s,00}}-\frac{\lambda_2}{ m^2_{s,08}}-\frac{\sqrt{2}\left(2\lambda_1+\lambda_2\right)}{ m^2_{s,88}}\right\}+y   \left\{\frac{-2(\lambda_1+\lambda_2)}{m^2_{s,00}}+\frac{\sqrt{2}(3\lambda_1+ 4\lambda_{2})}{m^2_{s,08}} -\frac{2\left(\lambda_1+2\lambda_2\right)}{ m^2_{s,88}}\right\} \Biggr]
\;, \\  \nonumber
&\Sigma^{tad}_{\text{p},00}=\frac{8N_{c} \ g  \ m_{u} \  \mathcal{A}(m_{u}^2) }{3} \Biggl[c\biggl(\frac{1}{m^2_{s,00}}+\frac{1}{\sqrt{2}\ m^2_{s,08}}\biggr)+x\biggl(\lambda_1+\frac{\lambda_2}{3}\biggr)\left\{\frac{2}{m^2_{s,00}}+\frac{2\sqrt{2}}{m^2_{s,08}}+\frac{1}{m^2_{s,88}}\right\} \\ \nonumber 
&\qquad \quad +y \biggl(\lambda_1+\frac{\lambda_2}{3}\biggr)\left\{\frac{\sqrt{2}}{m^2_{s,00}}-\frac{1}{m^2_{s,08}}-\frac{\sqrt{2}}{m^2_{s,88}}\right\}  \Biggr] 
+\frac{4N_{c} \ g   m_{s}  \mathcal{A}(m_{s}^2) }{3} \Biggl[c\biggl(\frac{1}{m^2_{s,00}}-\frac{\sqrt{2}}{ m^2_{s,08}}\biggr) \\ 
&\qquad \quad +x\biggl(\lambda_1+\frac{\lambda_2}{3}\biggr)\left\{\frac{2}{m^2_{s,00}}-\frac{\sqrt{2}}{m^2_{s,08}}-\frac{2}{m^2_{s,88}}\right\}+y \biggl(\lambda_1+\frac{\lambda_2}{3}\biggr)\left\{\frac{\sqrt{2}}{m^2_{s,00}}-\frac{4}{m^2_{s,08}}+\frac{2\sqrt{2}}{m^2_{s,88}}\right\}\Biggr]\;, \\  \nonumber
\end{align}
\begin{align}
\nonumber
&{\hskip -1 cm}\Sigma^{tad}_{\text{p},88}=\frac{8N_{c} g m_{u}  \mathcal{A}(m_{u}^2) }{3} \Biggl[
-\frac{c}{2}\biggl(\frac{1}{m^2_{s,00}}+\frac{3}{\sqrt{2} m^2_{s,08}}+\frac{1}{\ m^2_{s,88}}\biggr)+x\left\{\frac{\left(2\lambda_1+\frac{\lambda_2}{3}\right)}{ \ m^2_{s,00}}+\frac{\sqrt{2}    
\left(2\lambda_1+\frac{\lambda_2}{3}\right)}{ m^2_{s,08}}+\frac{\left(\lambda_1+\frac{\lambda_2}{6}\right)}{ m^2_{s,88}}\right\} \\ \nonumber 
& +y   \left\{\frac{\sqrt{2}(\lambda_1+\frac{2\lambda_2}{3})}{\ m^2_{s,00}}-\frac{1}
{ m^2_{s,08}} (\lambda_1+\frac{2\lambda_2}{3})
-\frac{\sqrt{2}(\lambda_1+\frac{\lambda_2}{3})}{\ m^2_{s,88}}\right\}\Biggr] 
+\frac{4N_{c}  g   m_{s}  \mathcal{A}(m_{s}^2) }{3} \Biggl[-c \biggl(\frac{1}{2  m^2_{s,00}}-\frac{1}{  m^2_{s,88}}\biggr) \\ 
&{\hskip -1 cm}+x\left\{\frac{\left(2\lambda_1+\frac{\lambda_2}{3}\right)}{ \ m^2_{s,00}}-\frac{\sqrt{2}    
\left(\lambda_1-\frac{\lambda_2}{6}\right)}{ m^2_{s,08}}-\frac{2\left(\lambda_1+\frac{\lambda_2}{6}\right)}{ m^2_{s,88}}\right\} +y   \left\{\frac{\sqrt{2}(\lambda_1+\frac{2\lambda_2}{3})}{\ m^2_{s,00}}-\frac{4}{ m^2_{s,08}} (\lambda_1+\frac{2\lambda_2}{3})
+\frac{2\sqrt{2}(\lambda_1+\frac{2\lambda_2}{3})}{\ m^2_{s,88}}\right\}\Biggr]\;, \\ \nonumber
&\Sigma^{tad}_{\text{p},08}=\frac{8N_{c} \ g  \ m_{u} \  \mathcal{A}(m_{u}^2) }{3} \Biggl[
-\frac{c}{2}\biggl(\frac{1}{ \ m^2_{s,08}}+\frac{\sqrt{2}}{\ m^2_{s,88}}\biggr)+x\left\{\frac
{\sqrt{2}\left(\lambda_1+\frac{\lambda_2}{3}\right)}{ \ m^2_{s,00}}+\frac{   
\left(3\lambda_1+\frac{2\lambda_2}{3}\right)}{ m^2_{s,08}}+\frac{\sqrt{2}\left(\lambda_1+\frac{\lambda_2}{6}\right)}{ m^2_{s,88}}\right\} \\ \nonumber 
&\qquad \quad +y   \left\{-2\frac{(\lambda_1+\frac{2\lambda_2}{3})}{\ m^2_{s,00}}+\frac{\sqrt{2}}
{ m^2_{s,08}} (\frac{\lambda_2}{3})
+\frac{(\lambda_1+\frac{2\lambda_2}{3})}{\ m^2_{s,88}}\right\}\Biggr] 
+\frac{4N_{c}  g   m_{s}  \mathcal{A}(m_{s}^2) }{3} \Biggl[\frac{-c}{2} \biggl(\frac{1}{  m^2_{s,08}}-\frac{\sqrt{2}}{m^2_{s,88}}\biggr)
\\
&{\hskip -1 cm}+x\left\{\frac{\sqrt{2}\left(\lambda_1+\frac{\lambda_2}{3}\right)}{ \ m^2_{s,00}}-\frac{  \lambda_2}{3m^2_{s,08}}+\frac{-2\sqrt{2}\left(\lambda_1+\frac{\lambda_2}{6}\right)}{ m^2_{s,88}}\right\} +y   \left\{\frac{-2(\lambda_1+\frac{\lambda_2}{3})}{\ m^2_{s,00}}+\frac{\sqrt{2}}{ m^2_{s,08}} (3
\lambda_1+\frac{4\lambda_2}{3})
-\frac{2(\lambda_1+\frac{2\lambda_2}{3})}{\ m^2_{s,88}}\right\}\Biggr]\;,
\\
\nonumber
&\Sigma^{tad}_{\text{p},11}=\frac{8N_{c} \ g  \ m_{u} \  \mathcal{A}(m_{u}^2) }{3} \Biggl[
\frac{c}{2}\biggl(\frac{-1}{m^2_{s,00}}+\frac{1}{\sqrt{2}\ m^2_{s,08}}+\frac{1}{ m^2_{s,88}}\biggr)+x\biggl(\lambda_1+\frac{\lambda_2}{2}\biggr)\left\{\frac{2}{m^2_{s,00}}+\frac{2\sqrt{2}}{m^2_{s,08}}+\frac{1}{m^2_{s,88}}\right\} \\ \nonumber 
&\qquad \quad +y \ \lambda_1\left\{\frac{\sqrt{2}}{m^2_{s,00}}-\frac{1}{m^2_{s,08}}-\frac{\sqrt{2}}{m^2_{s,88}}\right\}\Biggr] 
+\frac{4N_{c} \ g   m_{s}  \mathcal{A}(m_{s}^2) }{3} \Biggl[\frac{c}{2}\biggl(\frac{-1}{m^2_{s,00}}+\frac{2\sqrt{2}}{ m^2_{s,08}}-\frac{2}{ m^2_{s,88}}\biggr)\\ 
&\qquad \quad +x\biggl(\lambda_1+\frac{\lambda_2}{2}\biggr)\left\{\frac{2}{m^2_{s,00}}-\frac{\sqrt{2}}{m^2_{s,08}}-\frac{2}{m^2_{s,88}}\right\}+y \  \lambda_1\left\{\frac{\sqrt{2}}{m^2_{s,00}}-\frac{4}{m^2_{s,08}}+\frac{2\sqrt{2}}{m^2_{s,88}}\right\}\Biggr]\;, \\
\nonumber
&{\hskip -1 cm}\Sigma^{tad}_{\text{p},44}=\frac{8N_{c} \ g  \ m_{u} \  \mathcal{A}(m_{u}^2) }{3} \Biggl[
-\frac{c}{4}\biggl(\frac{2}{ \ m^2_{s,00}}+\frac{2\sqrt{2}}{\ m^2_{s,08}}+\frac{1}{\ m^2_{s,88}}\biggr)+x\left\{\frac{2\left(\lambda_1+\frac{\lambda_2}{4}\right)}{ \ m^2_{s,00}}+\frac{\sqrt{2}    
\left(2\lambda_1+\frac{5\lambda_2}{4}\right)}{ m^2_{s,08}}+\frac{\left(\lambda_1+\lambda_2\right)}{ m^2_{s,88}}\right\}\\
\nonumber 
&{\hskip -0.5 cm} +y   \left\{\frac{\sqrt{2}(\lambda_1+\frac{\lambda_2}{2})}{\ m^2_{s,00}}-\frac{1}
{ m^2_{s,08}} (\lambda_1+2\lambda_2)
-\frac{\sqrt{2}(\lambda_1+\frac{5\lambda_2}{4})}{\ m^2_{s,88}}\right\}\Biggr] 
+\frac{4N_{c}  g   m_{s}  \mathcal{A}(m_{s}^2) }{3} \Biggl[-\frac{c}{4}\biggl(\frac{2}{ \ m^2_{s,00}}-\frac{\sqrt{2}}{\ m^2_{s,08}}-\frac{2}{\ m^2_{s,88}}\biggr)\\
&{\hskip -1 cm}+x\left\{\frac{2\left(\lambda_1+\frac{\lambda_2}{4}\right)}{ \ m^2_{s,00}}-\frac{\sqrt{2}    
\left(\lambda_1-\frac{\lambda_2}{2}\right)}{ m^2_{s,08}}-\frac{2\left(\lambda_1+\lambda_2\right)}{ m^2_{s,88}}\right\} +y   \left\{\frac{\sqrt{2}(\lambda_1+\frac{\lambda_2}{2})}{\ m^2_{s,00}}-\frac{2}{ m^2_{s,08}} (2\lambda_1+\frac{7\lambda_2}{4})
+\frac{2\sqrt{2}(\lambda_1+\frac{5\lambda_2}{4})}{\ m^2_{s,88}}\right\}\Biggr]\;.\\ \nonumber 
\end{align}

\section{INTEGRALS AND SUM INTEGRALS}
\label{appenC}
The divergent loop integrals are regularized by incorporating dimensional regularization.
\bqa
\int_p=\left(\frac{e^{\gamma_E}\Lambda^2}{4\pi}\right)^\epsilon\int \frac{d^dp}{(2\pi)^d}\;,
\eqa
where $d=4-2\epsilon$ , $\gamma_E$ is the Euler-Mascheroni constant, and $\Lambda$ is renormalization scale associated with the $\overline{\text{MS}}$.

\bqa
\nonumber
\mathcal{A}(m^2_f)&=&\int_p \frac{1}{p^2-m^2_f}=\frac{i m^2_f}{(4\pi)^2}\left[\frac{1}{\epsilon}+1\right. \\
\nonumber
&&\left.+\ln(4\pi e^{-\gamma_E})+\ln\left(\frac{\Lambda^2}{m^2_f}\right)\right]\;,
\eqa

we rewrite this after redefining $\Lambda^2\longrightarrow \Lambda^2\frac{e^{\gamma_E}}{4\pi}$.

\bqa
\label{aint1}
\mathcal{A}(m^2_f)&=&\frac{i m^2_f}{(4\pi)^2}\left[\frac{1}{\epsilon}+1+\ln\left(\frac{\Lambda^2}{m^2_f}\right)\right]\;,
\eqa

\bqa
\label{bint1}
\nonumber
\mathcal{B}(p^2,m_f)&=&\int_k \frac{1}{(k^2-m^2_f)[(k+p)^2-m^2_f)]} \\
&=&\frac{i}{(4\pi)^2}\left[\frac{1}{\epsilon}+\ln\left(\frac{\Lambda^2}{m^2_f}\right)+\mathcal{C}(p^2,m_f)\right]\;,
\eqa

\bqa
\label{bprimeint1}
\mathcal{B}^\prime(p^2,m_f)&=&\frac{i}{(4\pi)^2}\mathcal{C}^\prime(p^2,m_f)\;,
\eqa

\begin{equation}
\label{eq:cp1}
{\hskip -1 cm}\mathcal{C}(p^2,m_f)=2-2\sqrt{\dfrac{4 m^2_f}{p^2}-1}\arctan\left(\dfrac{1}{\sqrt{\dfrac{4 m^2_f}{p^2}-1}}\right); \quad  \mathcal{C}^{\prime}(p^2,m_f)=\frac{4 m^2_f}{p^4\sqrt{\dfrac{4 m^2_f}{p^2}-1}}\arctan\left(\dfrac{1}{\sqrt{\dfrac{4 m^2_f}{p^2}-1}}\right)-\frac{1}{p^2}\;,
\end{equation}
\begin{equation}
\label{eq:cp2}
{\hskip -1 cm}\mathcal{C}(p^2,m_f)=2+\sqrt{1-\dfrac{4 m^2_f}{p^2}}\ln\left(\dfrac{1-\sqrt{1-\dfrac{4 m^2_f}{p^2}}}{1+\sqrt{1-\dfrac{4 m^2_f}{p^2}}}\right); \quad \mathcal{C}^{\prime}(p^2,m_f)=\frac{2 m^2_f}{p^4\sqrt{\dfrac{4 m^2_f}{p^2}-1}}\ln\left(\dfrac{1-\sqrt{1-\dfrac{4 m^2_f}{p^2}}}{1+\sqrt{1-\dfrac{4 m^2_f}{p^2}}}\right)-\frac{1}{p^2}\;,
\end{equation}
The Eqs.(\ref{eq:cp1}) and (\ref{eq:cp2}) are valid with the constraints  ($p^2<4m^2_f$) and ($p^2>4m^2_f$) respectively. 
\begin{align}
\label{bint1}
&\mathcal{B}(p^2,m_u,m_s)=\int_k \frac{1}{(k^2-m^2_s)[(k+p)^2-m^2_u)]}=\frac{i}{(4\pi)^2}\left[\frac{1}{\epsilon}+\ln\left(\frac{\Lambda^2}{m^2_u}\right)+\mathcal{C}(p^2,m_u,m_s)\right]\;,\\
&{\hskip -1 cm}\mathcal{C}(p^2,m_u,m_s)=2-\frac{1}{2}\Biggl[ 1+\frac{m_s^2-m_u^2}{p^2}\Biggr]\ln\left(\frac{m_s^2}{m_u^2}\right)-\frac{\mathcal{G}(p^2)}{p^2}\Biggl[\arctan\left(\frac{p^2-m^2_s+m^2_u}{\mathcal{G}(p^2)}\right)+\arctan\left(\frac{p^2+m^2_s-m^2_u}{\mathcal{G}(p^2)}\right)\Biggr]\;, \quad \quad\\
&\mathcal{G}(p^2)=\sqrt{\{(m_s+m_u)^2-p^2\}\{p^2-(m_s-m_u)^2\}}\;,\\ \nonumber
&\mathcal{C}^\prime(p^2,m_u,m_s)=\frac{m_s^2-m_u^2}{2p^4}\ln\Biggl(\frac{m^2_s}{m^2_u}\Biggr)+\frac{p^2(m_s^2+m_u^2)-(m_s^2-m_u^2)^2}{p^4\mathcal{G}(p^2)}\Biggl[\arctan\Biggl(\frac{(p^2-m^2_s+m^2_u)}{\mathcal{G}(p^2)}\Biggr)\\
&\qquad \qquad \qquad \quad+\arctan\Biggl(\frac{(p^2+m^2_s-m^2_u)}{\mathcal{G}(p^2)}\Biggr)\Biggr]-\frac{1}{p^2}\;.
\end{align}

\acknowledgments
Computational support of the computing facility which has been developed by the Nuclear Particle Physics group of the Department of Physics, University of Allahabad (UOA) under the Center of Advanced Studies (CAS) funding of UGC, India, is acknowledged. Department of Science and Technology, Government of India, DST-PURSE program Phase 2/43(C), financial support to the science faculty of the UOA is also acknowledged. I acknowledge the support of Dr. Pramod Kumar Shukla and Mr. Suraj Kumar Rai for making some figures, reading the manuscript and suggesting corrections. Dr. Swatantra Kumar Tiwari is also thanked for reading the manuscript and suggesting corrections.

\end{document}